\documentclass[a4paper, fleqn, usenatbib]{mnras}

\usepackage[utf8]{inputenc}
\usepackage[english]{babel}
\usepackage{newtxtext,newtxmath}

\usepackage[T1]{fontenc}
\usepackage{ae, aecompl}

\usepackage{layouts}

\usepackage{graphicx}	
\usepackage{amsmath}	

\usepackage{amssymb}	
\usepackage{subfigure}
\usepackage[usenames,dvipsnames]{xcolor}
\usepackage{letltxmacro}
\usepackage{todonotes}

\hypersetup{%
colorlinks,
pdfborder = {0 0 0},
breaklinks,
linkcolor=black,
urlcolor=NavyBlue,
citecolor=NavyBlue
}

\LetLtxMacro{\originaleqref}{\eqref}
\renewcommand{\eqref}{Eq.~\originaleqref}

\renewcommand{\vec}[1]{\ensuremath{\boldsymbol{#1}}}

\newcommand{\expVal}[1]{\ensuremath{\left \langle #1 \right \rangle}}

\defcitealias{2015MNRAS.451.2663H}{H15}
\defcitealias{Guo:2010ap}{G11}

\title[Galaxy formation with lensing and clustering]{Joint galaxy-galaxy lensing and clustering constraints on galaxy formation}

\author[Renneby et al.]{Malin Renneby$^{1, \, 2}$\thanks{E-mail: mrenneby@gmail.com}, Bruno M. B. Henriques$^{3, \, 4}$, Stefan Hilbert$^{5, \, 6}$, Dylan Nelson$^{4}$,
\newauthor Mark Vogelsberger$^{7}$, Raúl E. Angulo$^{8, \,9}$, Volker Springel$^{4}$, Lars Hernquist$^{10}$\\
$^{1}$HEP/ALCF Divisions, Argonne National Laboratory, 9700 S. Cass Avenue, Lemont, IL 60439, USA\\
$^{2}$Kavli Institute for Cosmological Physics, University of Chicago, Chicago, IL 60637, USA\\
$^{3}$ETH Zürich, Departement Physik, Institut für Teilchenphysik und Astrophysik, Wolfang-Pauli-Straße 27, CH-8093 Zürich, Switzerland\\
$^{4}$Max-Planck-Institut für Astrophysik, Karl-Schwarzschild-Str. 1, 85741 Garching b. München, Germany\\
$^{5}$Excellence Cluster Universe, Boltzmannstraße 2, 85748 Garching b. München, Germany\\
$^{6}$Ludwig-Maximilians-Universität, Fakultät für Physik, Universitäts-Sternwarte, Scheinerstraße 1, 81679 Munich, Germany\\
$^{7}$Department of Physics, Kavli Institute for Astrophysics and Space Research, MIT, Cambridge, MA 02139, USA\\
$^{8}$Donostia International Physics Centre (DIPC), Paseo Manuel de Lardizabal 4, 20018 Donostia-San Sebastian, Spain\\
$^{9}$IKERBASQUE, Basque Foundation for Science, E-48013, Bilbao, Spain\\
$^{10}$Harvard-Smithsonian Center for Astrophysics, 60 Garden Street, Cambridge, MA 02138, USA
}

\date{Accepted XXX. Received YYY; in original form ZZZ}

\pubyear{2020}

\def\figrelpath{}

\begin{document}
\label{firstpage}
\pagerange{\pageref{firstpage}--\pageref{lastpage}}
\maketitle

\begin{abstract}
We compare predictions for galaxy-galaxy lensing profiles and clustering from the Henriques et al. (2015) public version of the Munich semi-analytical model of galaxy formation (SAM) and the IllustrisTNG suite, primarily TNG300, with observations from KiDS+GAMA and SDSS-DR7 using four different selection functions for the lenses (stellar mass, stellar mass and group membership, stellar mass and isolation criteria, stellar mass and colour). We find that this version of the SAM does not agree well with the current data for stellar mass-only lenses with $M_\ast > 10^{11}\,M_\odot$. By decreasing the merger time for satellite galaxies as well as reducing the radio-mode AGN accretion efficiency in the SAM, we obtain better agreement, both for the lensing and the clustering, at the high mass end. We show that the new model is consistent with the signals for central galaxies presented in Velliscig et al. (2017). Turning to the hydrodynamical simulation, TNG300 produces good lensing predictions, both for stellar mass-only ($\chi^2 = 1.81$ compared to $\chi^2 = 7.79$ for the SAM), and locally brightest galaxies samples ($\chi^2 = 3.80$ compared to $\chi^2 = 5.01$). With added dust corrections to the colours it matches the SDSS clustering signal well for red low mass galaxies. We find that both the SAMs and TNG300 predict $\sim 50\,\%$ excessive lensing signals for intermediate mass red galaxies with $10.2 < \log_{10} M_\ast [ M_\odot ] < 11.2$ at $r \approx 0.6\,h^{-1}\,\text{Mpc}$, which require further theoretical development.
\end{abstract}

\begin{keywords}
gravitational lensing: weak -- galaxies: evolution -- galaxies: haloes -- cosmology: theory -- methods: numerical
\end{keywords}



\section{Introduction}

The next generation of large scale structure surveys, such as \emph{Euclid}
\citep{2011arXiv1110.3193L}, WFIRST \citep{2015arXiv150303757S} and
LSST \citep{2008arXiv0805.2366I}, will cover a wide range of scales in
the cosmic web with unprecedented precision. Weak gravitational
lensing, specifically galaxy-galaxy lensing \citep[GGL, see
  e.g.][]{Bartelmann:1999yn}, and galaxy clustering are two promising
diagnostics of structure growth that in combination can be used to
constrain the matter fraction $\Omega_\text{m}$, the amplitude of
matter density fluctuations $\sigma_8$, and the galaxy bias
$b_\text{g}$, which are all of interest to cosmologists. From the
perspective of astrophysicists, these probes offer the
opportunity to constrain galaxy evolution processes that determine
which classes of galaxies reside in what types of dark matter haloes
and the spatial distribution of the halo material.

Modelling the signals on small scales beyond the validity limit of perturbation
theory requires empirical or computational approaches. Examples of the
former are halo occupation models (HODs)
\citep[e.g.][]{2000MNRAS.318.1144P, 2000MNRAS.318..203S,
  2002ApJ...575..587B, 2002PhR...372....1C, 2011ApJ...738...45L,
  2012ApJ...744..159L, 2015MNRAS.454.1161Z, 2016MNRAS.457.4360Z},
which give the probability distribution of galaxies satisfying some
criteria, such as a stellar mass cut, conditioned on a property of the
host haloes, like their masses. Advances have made possible
construction of HODs using additional secondary properties such as
halo concentration \citep[e.g.][]{2016MNRAS.460.2552H} as well as
boosting their statistical input by accounting for the
incompleteness of stellar mass selected samples
\citep{2015MNRAS.454.1161Z, 2016MNRAS.457.4360Z}. This has allowed the construction of fast engines for 2-pt statistics predictions over a wide redshift range \citep[e.g.][]{2019MNRAS.488.3143B, 2019ApJ...884...29N}. However, these
approaches have difficulty in including many secondary parameters and
lack the connection between these and the governing physical
processes.

Semi-analytical models (SAMs) \citep{1991ApJ...379...52W,1999MNRAS.303..188K,
  2001MNRAS.328..726S,2006MNRAS.370..645B,2007MNRAS.375....2D,Guo:2010ap,
  2013MNRAS.431.3373H, 2015MNRAS.451.2663H} and hydrodynamical
simulations such as Illustris \citep{2014MNRAS.444.1518V,
  2014Natur.509..177V, 2014MNRAS.445..175G}, EAGLE
\citep{2015MNRAS.446..521S, 2015MNRAS.450.1937C} and IllustrisTNG
\citep[see e.g.][for methods and introductory publications]{2017MNRAS.465.3291W, 2018MNRAS.473.4077P, 2018MNRAS.475..676S, 2018MNRAS.477.1206N, 2018MNRAS.475..648P, 2018MNRAS.480.5113M, 2018MNRAS.475..624N} are examples of
methods in which haloes are instead populated with galaxies through modelling of the relevant physical mechanisms. Hydrodynamical simulations such as IllustrisTNG invest effort in consistently modelling and tracking the evolution of gas cells with subgrid recipes for star formation and regulating feedback. Thanks to its large volume, one is able to compute cosmological statistics, such as galaxy clustering, in TNG300 as was done in \citet{2018MNRAS.475..676S} out to comparably large radial scales with similar statistics as for the SAMs. This is in contrast to previous studies of clustering \citep[e.g.][]{2017MNRAS.470.1771A} restricted to EAGLE and Illustris, which have smaller volumes, and with higher resolution than simulations run in even larger volumes such as the BAHAMAS suite \citep{2017MNRAS.465.2936M, 2018MNRAS.476.2999M}. This large scale analysis is not restricted to the clustering of galaxies; one can also probe the spatial distribution of neutral hydrogen to gain insight into the physics of reionisation \citep[e.g.][]{2018ApJ...866..135V}. The small-scale lensing predictions for these types of simulations have previously been partly explored for different datasets \citep[see e.g.][]{2017MNRAS.467.3024L, 2017MNRAS.471.2856V, 2019A&A...626A..72G}. The stellar mass functions and colour distributions for the different box sizes of the IllustrisTNG suite have been presented in \citet{2018MNRAS.475..648P, 2018MNRAS.475..624N}, and halo-occupation distribution prescriptions for the galaxy-halo relation in \citet{2019MNRAS.490.5693B, 2020arXiv200804913H}. This Paper continues to address this issue, e.g. if state-of-the-art SAMs and hydrodynamical simulations yield consistent predictions when compared to the best current observational constraints. We use the TNG suite to probe the impact of baryons, expanding on the work of \citet{2019MNRAS.488.5771L} who compared Illustris and TNG300, and are thus able to answer how the signal from the SAM galaxies should be altered to account for this.

Thanks to the low computational cost of the SAMs, it is possible to
explore the parameter space of the underlying physical models using
Monte Carlo Markov chains (MCMC) \citep{2009MNRAS.396..535H,
  2013MNRAS.431.3373H, 2015MNRAS.451.2663H}, with observational
constraints such as the stellar mass function (SMF) and the red fraction
of galaxies ($f_\text{red}$) (1-pt functions). The parameters of a hydrodynamical simulation are usually calibrated in small test boxes and are then fixed at runtime, which does not allow for the same flexibility. In
\citet{2016MNRAS.458..934V} it was shown that the introduction of
galaxy clustering constraints (2-pt functions) in the SAM MCMC sampling provide
additional insights into the formation physics. In this Paper, we
focus on the public version of the Munich semi-analytical model
\textsc{L-Galaxies} released in 2015 \citep{2015MNRAS.451.2663H}, henceforth
\citetalias{2015MNRAS.451.2663H}, for a Planck 2014 cosmology
\citep{2014A&A...571A..16P}, and show how lensing complemented by clustering signals can inform on the parameter choices for the feedback processes. Recently, a newer version of the model, presented in \citet{2020MNRAS.491.5795H}, became available. The improvements introduced there primarily concerns the ability to provide predictions for discretised gas distributions in radial rings, which have negligible to small impact on galaxy clustering and lensing signals. Moreover, \citet{2020arXiv200414390A} have conducted a study where they compare general properties of galaxies in TNG100 and TNG300 with \textsc{L-Galaxies}, apart from lensing and clustering predictions, which this analysis complements. \citet{2016MNRAS.456.2301W} found that the \citetalias{2015MNRAS.451.2663H} model predicts an excessive 
lensing signal around Sloan Digital Sky Survey (SDSS)
locally brightest galaxies, LBGs \citep{2013A&A...557A..52P,
  2015MNRAS.449.3806A}. By enforcing a stellar mass correction based
on abundance matching to SDSS via the fitting function in
\citet{2009MNRAS.398.2177L}, a better agreement was reached. The motivation for this correction was that if the model stellar masses were adjusted so that the SMFs traced the observed data, the lensing signals should agree better with the observed lensing data by default. The
version with the smallest necessary abundance correction was the
\cite{Guo:2010ap} model, henceforth \citetalias{Guo:2010ap}, adapted
for the Planck 2014 cosmology, owing to the MCMC tuning to low
redshift observations. This model also passed a more stringent test in
\citet{2016MNRAS.457.3200M} with a separation of the lensing signal
for red and blue LBGs. Still, due to the low redshift tuning, this
version has difficulties in making predictions for future deep
surveys, such as the Hyper Suprime-Cam SSP Survey (HSC)
\citep{2018PASJ...70S...4A} and the \emph{Euclid} mission, where the signal
will be measured for lens systems beyond $z = 1$. In addition, it does
not feature recent developments to improve the modelling of low mass
galaxies, where \citetalias{2015MNRAS.451.2663H} has reduced the
over-abundance of $8.0 < \log_{10} M_\ast/M_\odot < 9.5$ systems at $z
\geqslant 1$ as well as the excessive fraction of red dwarf galaxies
at low redshift. Our task here is to see if we can alter the \citetalias{2015MNRAS.451.2663H} model sufficiently to match the low redshift lensing signals while retaining the higher-$z$ SMF agreement, and see how it fares against other datasets.

We focus on selections based on stellar
mass, joint stellar mass and colour and joint stellar mass and
isolation/group membership criteria. The latter is especially
important for upcoming group and cluster finders, where lensing can be
used to validate models of feedback from active galactic nuclei (AGN)
\citep[e.g.][]{2010MNRAS.406..822M, 2015MNRAS.452.3529V}. Colour
bimodality can inform models for quenching mechanisms of star
formation and their relations to the host halo mass
\citep[e.g.][]{2016MNRAS.457.4360Z, 2016MNRAS.457.3200M}. With respect
to \citet{2016MNRAS.456.2301W, 2016MNRAS.457.3200M}, we both consider
the locally brightest galaxies and the full galaxy distribution. For lensing observations, we consider a deeper field from the equatorial overlap of
the KiDS+GAMA surveys
\citep{2015MNRAS.452.2087L, 2015MNRAS.454.3500K} with data from
\citet{2016MNRAS.459.3251V} and \citet{2017MNRAS.471.2856V} and a
shallow field (SDSS-DR7) \citep{2016MNRAS.456.2301W,
  2016MNRAS.457.3200M, 2016MNRAS.457.4360Z} to illustrate how
different surveys and redshifts affect the lensing profiles. We also
compare predictions from the HOD models of \citet[iHODS,
][]{2016MNRAS.457.4360Z} to
illustrate how well the different frameworks with increasingly
granular levels of model sophistication can capture the signal. For
the SAM, we use the LBG and stellar mass only samples to constrain the
model parameters and then use the group lens samples from
\citet{2017MNRAS.471.2856V} as validation cases for the new models.

The purpose of our study is three-fold: (i) Investigate if
\textsc{L-Galaxies} fits current observational constraints from
galaxy-galaxy lensing and clustering, and (ii) Examine if
modest changes to a few model parameters can bring about better
agreement. (iii) Assess the agreement of the IllustrisTNG hydrodynamical simulation with observations and explore differences with respect to the SAM.

This Paper is organised as follows: We introduce our observables in Section~\ref{sec:theory}, the simulations as well as 
review the physical recipes of the feedback processes in \textsc{L-Galaxies} in Section~\ref{sec:simulations}. The different datasets we use
to gauge the performance of the models, as well as their colour
distributions, are given in Section~\ref{sec:data}. In Section~\ref{sec:method} we present our
methodology to match the simulations with the different datasets. In
Section~\ref{sec:results}, we show our results for the modified galaxy
formation models for the stellar mass functions
(Section~\ref{sec:abundanceCorrections}), stellar mass-selected lenses
(Section~\ref{sec:mstarOnlyKiDSGAMA}) followed by the implications of
cosmology (Section~\ref{sec:cosmology}) and baryons (Section~\ref{sec:baryons}), colour-selected lenses
(Section~\ref{sec:sdssColour}), LBGs
(Section~\ref{sec:lbgLensingSAMs} and Section~\ref{sec:lbgLensingTNG}) and galaxy clustering
(Section~\ref{sec:clustering}). Finally, we conclude with computing
the predictions for a few of our models for the KiDS+GAMA group lens
sample in Section~\ref{sec:groupCriteria}.

\section{Galaxy-galaxy lensing and clustering}\label{sec:theory}

Under the assumption of statistical isotropy, the spatial two-point correlation functions $\xi_\text{gi}$
\begin{equation}
\xi_\text{gi} (\vert \vec{r} - \vec{r^\prime} \vert) = \expVal{\delta_\text{g}(\vec{r})\delta_\text{i}(\vec{r^\prime})},
\end{equation}
\noindent with i\,$=$\,g for the galaxy field with itself (galaxy clustering) and i\,$=$\,m for the galaxy field with the matter field (galaxy-galaxy lensing) can be inferred by their projections integrated along the line-of-sight $\text{d}l$,
\begin{align}\label{eq:projectedAutocorrelation}
\omega_\text{p}(r) &= \int^{\infty}_{-\infty} \xi_\text{gg} \left (\sqrt{r^2 + l^2} \right ) \,\text{d} l,\\
\Sigma (r) &= \bar{\rho} \int^{\infty}_{0}  \xi_\text{gm} \left (\sqrt{r^2 + l^2} \right ) \,\text{d} l,\label{eq:projectedCrosscorrelation}
\end{align}
\noindent where $\bar{\rho}$ is the average matter density, evaluated at the projected radius $r$ with $\omega_\text{p}(r)$ as the projected clustering correlation function and $\Sigma (r)$ as the projected surface mass density. This latter quantity can be used to construct a differential excess surface mass density $\Delta \Sigma (r)$ related to the observed tangential shear $\gamma_\text{t}$ of background galaxies around foreground matter overdensities as the difference between the average projected mass inside a circular aperture $\bar{\Sigma}(r)$ with radius $r$ and a boundary term evaluated in a thin cylindrical shell $\Sigma(r)$ by
\begin{equation}\label{eq:deltaSigmaDef}
\Delta \Sigma (r) = \gamma_\text{t} \Sigma_\text{crit.}^{-1} = \bar{\Sigma}(r) - \Sigma (r),
\end{equation}
\noindent where $\Sigma_\text{crit.} = c^2/4\pi G \cdot D_\text{s}/(D_\text{l}D_\text{ls})$ is a geometric pre-factor containing the angular diameter distances of the lenses $D_\text{l}$, sources $D_\text{s}$ and the relative distance between them $D_\text{ls}$, and the gravitational constant $G$ and the speed of light $c$. We estimate the autocorrelation function $\hat{\xi}_\text{gg}(r)$ using pair counts according to the standard definition as
\begin{equation}
\hat{\xi}_\text{gg}(r) = \frac{V}{\expVal{N_\text{gal}}^2V(r)}N_\text{gal}(r) - 1,
\end{equation}
\noindent where $N_\text{gal}$ as the total number of galaxies in the snapshot, $V$ the total volume, and $V(r)$ and $N_\text{gal}(r)$ the volume and number of galaxies per cylindrical shell with radius $r$ around each galaxy. Effectively, the integration for $\omega_\text{p}(r)$ in \eqref{eq:projectedAutocorrelation} is carried out to a maximal distance $l_\text{max}^\pi$ to account for the uncertainty in determining galaxy redshifts. We set $l_\text{max}^\pi = 60 \, h^{-1} \, \text{Mpc}$ following \citet{2016MNRAS.457.4360Z}. However, this choice primarily influences the clustering 2-halo term. For the lensing signal we integrate along the entire simulation box length $L$.

We denote the central galaxy lensing signal as $\Delta \Sigma_\text{cent}$, taken to be the same as the friends-of-friends group signal, and $\Delta \Sigma_\text{sat}$ as the satellite signal. The joint central-satellite signal is calculated as 
a sum where these contributions are weighted with $1 - f_\text{sat}$ and $f_\text{sat}$, the satellite fraction, respectively. The central signal is effectively the lensing of the host haloes, whereas the satellite signal features a central sharpening from the presence of the subhalo which decreases radially until the contribution from the central host halo kicks in as a central bump. The radial distance between these two features reflects the average projected distance between the satellites and their centrals. 

\section{Simulations}\label{sec:simulations}

We list the different simulations used in this study below, all with
flat $\Lambda$CDM universes. Subhaloes are identified
using \textsc{Subfind} \citep{2001MNRAS.328..726S} in every
friends-of-friends (FOF) group constructed with a halo finder
\citep{1985ApJ...292..371D}. For the merger trees for the galaxy
formation models, subhaloes with more than twenty bound particles are
linked uniquely to descendants in the subsequent snapshots following
\citet{Springel:2005nw} with merger trees built with the \textsc{LHaloTree} algorithm.

\subsection{IllustrisTNG}

IllustrisTNG is the next generation of the Illustris simulation suite,
also run with the moving mesh-code \textsc{AREPO} \citep{2010MNRAS.401..791S} with an updated
galaxy formation model \citep{2017MNRAS.465.3291W,
  2018MNRAS.473.4077P} extending the original Illustris model
\citep{2013MNRAS.436.3031V, 2014MNRAS.438.1985T}, assuming a Planck
2016 cosmology $\{\Omega_\text{m},\, \Omega_\text{b},\,\sigma_8,\,
n_\text{s},\, h\} = \{0.3089,\, 0.0486, \, 0.8159, \, 0.9667, \,
0.6774\}$ \citep{2016A&A...594A..13P}. The two main changes from the fiducial Illustris implementation concern black holes and supernova-driven winds \citep{2017MNRAS.465.3291W, 2018MNRAS.475..648P}, which include a new AGN feedback model for the low accretion state of the black holes, and changes to the stellar feedback winds. This decreases the stellar-to-halo mass ratio for massive central galaxies while
retaining more gas in the inner parts of the haloes and significantly improves the stellar masses and colours of galaxies below the knee. Box lengths and particle numbers
are $75\, h^{-1}\,\text{Mpc}$ with $2\times1820^3$ particles (TNG100,
with the same phases as Illustris in the initial conditions, which
enables object-by-object comparisons) and $205\,h^{-1}\,\text{Mpc}$
with $2 \times 2500^3$ particles (TNG300) for the highest resolution
runs. Particle masses are $m_\text{b} = 9.44 \times 10^5\,h^{-1} \,
M_\odot$ and $m_\text{dm} = 5.06 \times 10^6 \,h^{-1}\, M_\odot$
(TNG100) and $m_\text{b} = 7.44 \times 10^6 \,h^{-1}\, M_\odot$ and
$m_\text{dm} = 3.98 \times 10^7 \,h^{-1}\, M_\odot$ (TNG300). For the
gravity-only runs, the corresponding particle masses are $m_\text{dm}
= 6.00 \times 10^6 \, h^{-1}\, M_\odot$ (TNG100-DMO) and $m_\text{dm} =
4.73 \times 10^7\, h^{-1}\, M_\odot$ (TNG300-DMO). The maximum
softening lengths are $\epsilon = 0.5 \, h^{-1}\,\text{kpc}$ (TNG100)
and $\epsilon = 1.0 \,h^{-1}\, \text{kpc}$ (TNG300) for the dark
matter and stars, with a minimum adaptive gas cell softening of $184\,
\text{pc}$ (TNG100) and $370\,\text{pc}$ (TNG300). Results for the stellar and halo mass functions, galaxy colours, clustering and matter power spectra, magnetic fields and chemical evolution have been presented in \citet{2018MNRAS.475..648P, 2018MNRAS.475..624N, 2018MNRAS.475..676S, 2018MNRAS.480.5113M, 2018MNRAS.477.1206N}. These two simulations have recently been publicly released\footnote{Available at: \url{www.tng-project.org}.}, as described in \citet{2019ComAC...6....2N}. We primarily use TNG300
to obtain comparable statistics as for the Millennium simulation. We also enforce the resolution
correction from the appendices of \citet{2018MNRAS.475..648P} for TNG300, called
'rTNG300' for some comparisons. This correction brings the stellar-to-halo mass relation, as well as the stellar mass function, in line with that of TNG100 and observations, as numerical convergence results in higher stellar masses and star formation rates with increasing resolution. Specifically, one uses the differences between the higher resolved TNG100-1 simulation and its lower resolution companion TNG100-2, which has the same resolution as TNG300-1 (which we refer to as TNG300), to determine the correction.

\subsection{Millennium and Millennium-II}

For the \textsc{L-Galaxies} comparisons to observations we primarily use the Millennium suite simulations. Millennium (MR) \citep{Springel:2005nw} and Millennium-II (MRII)
\citep{BoylanKolchin:2009nc} are cold dark matter-only simulations
performed using \textsc{Gadget-2} and \textsc{Gadget-3}
\citep{2005MNRAS.364.1105S}, respectively, with $2160^3$ particles
with masses $8.61 \times 10^8 \, h^{-1}\, M_\odot$ and $6.88 \times
10^6\,h^{-1}\,M_\odot$, respectively, with a WMAP1 cosmology
$\{\Omega_\text{m},\, \Omega_\text{b},\,\sigma_8,\, n_\text{s},\, h\}
= \{0.25,\,0.045,\,0.90,\, 1.0,\, 0.73\}$ \citep{2003ApJS..148..175S}
and box lengths of $500\, h^{-1}\,\text{Mpc}$ and $100\,
h^{-1}\,\text{Mpc}$. The Plummer-equivalent softening lengths
$\epsilon$ are $5\, h^{-1}\, \text{kpc}$ and $1\, h^{-1}\,
\text{kpc}$, respectively. We primarily use rescaled versions of these
simulations with a Planck 2014 cosmology \citep{2014A&A...571A..16P}
applying the techniques of \citet{Angulo:2009rc, Angulo:2014gza} with
$\{\Omega_\text{m},\, \Omega_\text{b},\,\sigma_8,\, n_\text{s},\, h\}
= \{0.315,\, 0.049, \, 0.826, \, 0.961, \, 0.673\}$ and box lengths
of $480.279\, h^{-1}\,\text{Mpc}$ and $96.0558\, h^{-1}\,\text{Mpc}$, and
particle masses $9.61 \times 10^8 \, h^{-1}\, M_\odot$ (MRscPlanck)
and $7.69 \times 10^6\,h^{-1}\,M_\odot$ (MRIIscPlanck). Cosmological rescaling is an established technique to match the linear growth and the fluctuations of the matter power spectrum scales over scales corresponding to a range of halo masses one seeks to match in a target cosmology using a simulation with a different fiducial cosmology \citep[see e.g.][]{Angulo:2009rc, Angulo:2014gza, 2018MNRAS.479.1100R, 2019MNRAS.489.5938Z}. \citet{2018MNRAS.479.1100R} showed that it is possible to predict and correct for the bias in the lensing signal in such rescaled cosmologies using linear theory and fits to the concentration-mass-redshift relation. Here the correction is negligible and thus we ignore it. SAM lensing comparisons to direct simulations with different
cosmological parameters were already carried out in
\citet{2016MNRAS.456.2301W}. Some of these galaxy
formation models as well as the merger trees and halo catalogues are
accessible through the Virgo Millennium database
\citep{2006astro.ph..8019L}.

\subsection{L-Galaxies}

The seventeen free parameters in \textsc{L-Galaxies} have been
calibrated against the stellar mass function (SMF) at $z = 0, \, 0.4,
\, 1, \, 2, \, 3$ and the red fraction of galaxies $z = 0, \, 0.4, \, 1,
\, 2$. These parameters cover star formation, feedback from supernovae and active galactic nuclei, metal yields and galaxy merger criteria. The \citetalias{2015MNRAS.451.2663H} model is described in full in the Supplementary Material of that publication. The choices from the MCMC fit do not necessarily 
provide the best match to the SMF at low redshifts, since stress is put on obtaining good predictions at higher redshifts as well. As there are many free parameters, as well as degeneracies between the impact of different physical processes in the observables one attempts to match, the model is the output of an exploration of a very high-dimensional parameter space. In this Paper, we investigate the \citetalias{2015MNRAS.451.2663H} model\footnote{Public release available at: \url{http://galformod.mpa-garching.mpg.de/public/LGalaxies/index.html}.} and a subsection of model alterations to see if they can provide better fits to 2-pt statistics. The idea is that clustering and lensing observations could break some of the model degeneracies, and possibly be able to rule out the model in certain regimes.

We investigate the 2-pt statistics predictions from the existing \citetalias{2015MNRAS.451.2663H} model and alterations of it where we restrict the modifications to values of three parameters, $k_\text{AGN}$, $\epsilon_\text{reheat}$ and $\alpha_\text{dyn}$, which govern the stellar-to-halo mass relation and the satellite fraction. The benefit here is to see whether it is possible to change the model marginally with the variables which we deem most liable in determining the lensing signal, while keeping the other model variables fixed, avoiding a full new MCMC search. Lensing predictions are more computationally intense to obtain than clustering signals, and this analysis serves to prove whether such constraints are useful, or if all relevant information is already contained in the SMF. The clustering observations primary purpose is to illustrate that the new models work for them as well, i.e. that they provide consistent predictions for the galaxy and matter fields. This is in contrast to a constraint analysis performed by \citet{2016MNRAS.458..934V}, which focuses on galaxy clustering observations using an older version of the model. Below we review the parts of the \citetalias{2015MNRAS.451.2663H} model where the relevant parameters for the lensing signals occur.

\subsubsection{$k_\text{AGN}$ - AGN feedback efficiency regulator}

From the peak of star formation efficiency for Milky Way class galaxies, the lower mass end is regulated by supernovae (SN) and galactic wind feedback and the high mass end by AGN feedback \citep[see abundance matching results in e.g.][]{2010ApJ...710..903M, 2010ApJ...717..379B}; although recent studies hint that AGN feedback could also play an important role for less massive systems \citep[e.g.][]{2019MNRAS.tmpL.103K}. Hence, these two processes are a natural starting point for modifications to alter the lensing predictions. Since the lensing signals for the \citetalias{2015MNRAS.451.2663H} model in \citet{2016MNRAS.456.2301W} were too high, it means that one could attempt lower each or both efficiencies for these processes to increase the stellar-to-halo mass ratio for the galaxies. In \citetalias{2015MNRAS.451.2663H}, AGN feedback is implemented with a radio mode accretion model \citep{2006MNRAS.365...11C} normalised to the expansion rate of the Universe $H(z)$, which increases the accretion at lower redshifts,
\begin{equation}\label{eq:radioModeAGNAccretion}
\dot{M}_\text{BH} = k_\text{AGN} \left ( \frac{M_\text{hot}}{10^{11}M_\odot}\right ) \left ( \frac{M_\text{BH}}{10^8 M_\odot}\right ),
\end{equation}
\noindent where $\dot{M}_\text{BH}$ is the accretion rate, $k_\text{AGN}$ is a free parameter that regulates the efficiency of the accretion (in units of $M_\odot \text{yr}^{-1}$), and $M_\text{hot}$ and $M_\text{BH}$ are the masses of the hot gas halo and the supermassive black hole (SMBH), respectively. This accretion then impedes the cooling flow onto the cold disc as it is accompanied by depositing energy into the hot gas halo. With respect to previous versions of the SAM, $k_\text{AGN}$ is assumed to be fixed across all redshifts. This change was introduced to make certain that galaxies with stellar masses around the knee of the SMF were sufficiently quenched at $z = 0$.

\subsubsection{$\epsilon_\text{reheat}$ - supernovae gas reheating efficiency}

For SN feedback, the \citetalias{2015MNRAS.451.2663H} model has two regulators. The one which is relevant here sets the fraction of this energy for the reheating of cold gas and the subsequent injection into the hot gas atmosphere. The mass of cold gas reheated due to star formation $\Delta M_\text{reheat}$ is set to be proportional to the amount of stars formed \citep[see e.g.][]{1999ApJ...513..156M}
\begin{equation}
\Delta M_\text{reheat} = \epsilon_\text{disc} M_\text{disc},
\end{equation}
\noindent where $M_\text{disc}$ is the mass of stars in the galaxy disc and $\epsilon_\text{disc}$ is 
\begin{equation}
\epsilon_\text{disc} = \epsilon_\text{reheat} \left ( 0.5 + \left ( \frac{V_\text{max}}{V_\text{reheat}}\right )^{-\beta}\right ),
\end{equation}
\noindent where $\epsilon_\text{reheat}$ is the efficiency, $V_\text{max}$ the maximum circular velocity and $V_\text{reheat}$ and $\beta$ parameters determining the normalisation and slope of the feedback, respectively. In this study we keep these two parameters fixed to the fiducial \citetalias{2015MNRAS.451.2663H} values.

\subsubsection{$\alpha_\text{dyn}$ - dynamic friction multiplier}

Another way to increase the stellar masses is to modify processes governing the merging of systems. In SAMs, a subhalo of a satellite galaxy can be disrupted and the satellite shortly lives on as an orphan galaxy before falling into the central galaxy due to dynamical friction. The time between disruption and accretion, $t_\text{friction}$, is fixed by a merging timescale \citep[see e.g.][]{1987gady.book.....B} as
\begin{equation}
t_\text{friction} = \alpha_\text{dyn} \frac{V_{200\text{c}}r_\text{sat}^2}{G M_\text{sat} \ln (1 + M_\text{sat}/M_{200\text{c}})},
\end{equation}
\noindent where $M_\text{sat}$ is the total mass of the satellite, $r_\text{sat}$ the radius of the satellite orbit, $M_{200\text{c}}$ and $V_{200\text{c}}$ the mass and circular velocity of the friends-of-friends host halo, $G$ the gravitational constant and $\alpha_\text{dyn}$ a merger time multiplier. This value was set to $\alpha_\text{dyn} = 2.4$ by \citet{2007MNRAS.375....2D} to conform with the bright end of the luminosity function at $z = 0$. This choice was later found to be consistent with direct numerical simulation inferences \citep{2008MNRAS.383...93B, 2010MNRAS.406.1533D}. Intuitively, decreasing $\alpha_\text{dyn}$ lowers $f_\text{sat}$ and boosts the stellar mass of central galaxies which dominate the high mass end of the SMF as mergers are associated with starbursts. However, a short merger timescale implies that one overall ends up with fewer massive systems. One can decrease the efficiency of the feedback process to increase this number, which means that these two simultaneous modifications produce indistinguishable\footnote{This does not guarantee that other observables, such as radial profiles, agree, which influence the lensing and clustering predictions under certain selection functions.} SMFs.

\begin{table}
	\centering
	\begin{tabular}{r r r r}
	\hline
		{\bf Model} & ${\bf k_\text{\bf AGN}} \left[M_\odot\,\text{year}^{-1} \right]$ & ${\bf \epsilon_\text{\bf reheat}}$ & ${\bf \alpha_\text{\bf dyn.}}$\\
		\hline
\citetalias{Guo:2010ap} & $1.5 \times 10^{-3}$ & 6.5 & 2.0 \\
		\citetalias{2015MNRAS.451.2663H} & $5.3 \times 10^{-3}$ & $ 2.6$ & $2.5$\\
		\hline
	\end{tabular}
	\caption{The fiducial SAM model parameters. \citetalias{Guo:2010ap} has a different implementation of the AGN feedback, neglecting the normalisation with $H(z)$.}
	\label{tab:samModelsFiducialParameters}
\end{table}

\begin{table}
	\centering
	\begin{tabular}{ r r r r}
	\hline
		${\bf k_\text{\bf AGN}/ k_\text{\bf AGN}^\text{\bf fid.}}$ & ${\bf \epsilon_\text{\bf reheat}}/\epsilon_\text{\bf reheat}^\text{\bf fid.}$ & ${\bf \alpha_\text{\bf dyn.}}/\alpha_\text{\bf dyn.}^\text{\bf fid}$ & $\chi^2$\\
		\hline
		 $0.1$ & $1$ & $1$ & 1.67\\
		 $1$ & $0.1$ & $1$ & 2.62\\
		$1$ & $1$ & $0.1 $ & 38.89\\
		 $0.1$ & $1$ & $0.1$ & 3.14\\
		 $0.5$ & $1$ & $0.1$ & 18.91\\
		 $0.1$ & $1$ & $0.3$ & 2.21\\
		 $0.2$ & $1$ & $0.3$ & 4.23\\
		 $0.3$ & $1$ & $0.3$ & 6.44\\
		 $0.1$ & $1$ & $0.4 $ & 2.05\\
		 $0.2$ & $1$ & $0.4$ & 3.72\\
		 $0.3$ & $1$ & $0.4$ & 5.53\\
		 $0.1$ & $1$ & $0.5$ & 1.90\\
		 $0.2$ & $1$ & $0.5$ & 3.32\\
		 $0.3$ & $1$ & $0.5$ & 4.86\\
		$0.4$ & $1$ & $0.5$ & 6.48\\
		$0.5$ & $1$ & $0.5$ & 7.91\\
		$0.5$ & $0.5$ & $0.5$ & 3.96\\
		$0.5$ & $1.5$ & $0.5$ & 9.58\\
		\hline
	\end{tabular}
	\caption{The different SAM configurations compared in this Paper, derivatives of the \citetalias{2015MNRAS.451.2663H} model. 'fid' refers to the values in the \citetalias{2015MNRAS.451.2663H} model. We list their score, see definition in \eqref{eq:fom}, on the first lensing stellar mass-only comparison with the \citet{2016MNRAS.459.3251V} dataset, see Sections~\ref{sec:dataKiDSGAMA} and~\ref{sec:mstarOnlyKiDSGAMA}. The \citetalias{2015MNRAS.451.2663H} model ($\chi^2 =7.79$) and \citetalias{Guo:2010ap} model ($\chi^2 = 14.53$), as well as the \citetalias{Guo:2010ap} parameter values on the \citetalias{2015MNRAS.451.2663H} model ($\chi^2 = 7.71$), are less favoured by the data than some model variations.}
	\label{tab:samModelsParameters}
\end{table}

\subsubsection{Model variations and picking the best SAM}\label{sec:fomSAM}

We list the fiducial values of these parameters in Table~\ref{tab:samModelsFiducialParameters} and the variations in Table~\ref{tab:samModelsParameters}. The extreme models with 10\,\% of the fiducial model parameter values mainly serve as test cases. In the \citetalias{Guo:2010ap} version of the model, $\alpha_\text{dyn}$ has a marginally lower value and in \citet{2016MNRAS.458..934V}, a 40\,\% to 50\,\% lower value was required to match clustering observations. Hence we are focusing on derivative models with a lower $\alpha_\text{dyn}$ and lower $k_\text{AGN}$ than in the fiducial \citetalias{2015MNRAS.451.2663H} model. \citetalias{2015MNRAS.451.2663H} also found that boosting $V_\text{reheat}$ was necessary for a better clustering agreement. In the \citetalias{2015MNRAS.451.2663H} model this value is already fixed to a much higher value so we just modify the efficiency. With respect to observations, this SN mass loading factor was found to be on the high end in \citetalias{2015MNRAS.451.2663H} and this motivates the decrease.

We quantify the best SAMs under each selection function for the lensing and clustering observables through a figure-of-merit:

\begin{equation}\label{eq:fom}
\chi^2 = \frac{1}{N} \sum_i \frac{1}{\sigma^2}\left (\xi_\text{model}(r) - \xi_\text{data} (r) \right )^2,
\end{equation}
\noindent where $i$ goes over all overlapping data points $N$, where we
linearly interpolate the model between the bins and $\sigma$ is the
reported variance of the observations. To reduce this to a scalar for the different mass bins for a given dataset, we effectively compute the average $\expVal{\chi^2}$ for the dataset, but write $\chi^2$ for brevity.

\section{Data}\label{sec:data}

In this Section we describe the different lensing and clustering datasets
used in this study.

\subsection{KiDS+GAMA: Data selections}\label{sec:dataKiDSGAMA}

We compare the predicted lensing signals to observational results from
the KiDS shear catalogues and GAMA foreground lens sample in the
equatorial regions (fields G09, G12 and G15) using the published data
in \citet{2016MNRAS.459.3251V} for the partly overlapping region (75.1
deg$^2$) with an effective source density of 5.98\,arcmin$^{-2}$
\citep{2015MNRAS.454.3500K}. For the sample, we consider all galaxies
which satisfy the stellar mass criteria, based on the stellar mass
information in \citet{2011MNRAS.418.1587T}. Error bars incorporate
the effect of cosmic variance. This dataset serves as our principal lensing observations, since it has the simplest selection function. The average redshifts range from $\expVal{z} = 0.17$ to $\expVal{z} = 0.35$ from the lowest to the highest mass bins.

We also make use of observations presented in \citet{2017MNRAS.471.2856V},
which were compared to the EAGLE hydrodynamical simulation with
satisfactory agreement in the same publication. This study considered measurements which
satisfy stellar mass criteria and consist of galaxy
groups with at least five members $(N_\text{FOF} \geqslant 5)$ from
the GAMA group catalogue G3Cv7 \citep{2011MNRAS.416.2640R}. Galaxies
in this group catalogue are linked via friends-of-friends based on
their line-of-sight and projected distances and the catalogue has been
calibrated against the Millennium simulation with SAMs
\citep{2006MNRAS.370..645B}. For groups with more than five members,
galaxies are reliably classified as centrals/satellites above the
completeness limit of GAMA which is $\sim \log_{10}(M_\ast/M_\odot) =
8$. The field overlap is 180\,deg$^2$.

\subsection{SDSS: Data selections}\label{sec:sdssDataSelections}

Locally brightest galaxies, LBGs, are found with the following procedure: A cylinder with
radius $1\, \text{Mpc}$ in physical coordinates spanning $\pm\, 1000
\, \text{km/s}$ in redshift is constructed around each galaxy and if the galaxy has
the brightest absolute $r-$band magnitude with dust extinction in this cylinder it is
considered an LBG. For the observational LBGs, we use the lensing measurements in
\citet{2016MNRAS.456.2301W} (all) and in \citet{2016MNRAS.457.3200M}
(red and blue). The source catalogue is described in
\citet{2012MNRAS.425.2610R} and the effective source density is
1.2\,arcmin$^{-2}$.

For the stellar mass only clustering, we use the observations from
\citetalias{Guo:2010ap}, which have appeared for comparisons with
\citetalias{2015MNRAS.451.2663H} in \citet{2017MNRAS.469.2626H} and
with TNG100 and TNG300 in \citet{2018MNRAS.475..676S}.

We use the SDSS-DR7 \citep{2009ApJS..182..543A} lens and clustering
sample from \citet{2016MNRAS.457.4360Z} as well as the all main SDSS-DR7 lensing sample from \citet{2016MNRAS.457.3200M}. Red and blue galaxies are separated according to the following
$^{0.1}(g - r)$ colour cut (with filter magnitudes computed in rest-frame
wavebands blueshifted to $z = 0.1$),

\begin{equation}\label{eq:zuMandelbaum2016ColorCut}
^{0.1}(g - r)_\text{cut} = 0.8 \left ( \frac{\log_{10} M_\ast}{10.5} \right )^{0.6}.
\end{equation}
\noindent with $M_\ast$ as the galaxy stellar mass. For the SDSS LBGs in \citet{2016MNRAS.457.3200M}, we separate red and blue according to

\begin{equation}\label{eq:mandelbaum2016ColorCut}
^{0.1}(g - r)_\text{cut} = 0.8.
\end{equation}

\noindent We first $K$-correct our magnitudes and convert this cut
into a separation for magnitudes in rest-frame wavebands at $z = 0$
using the empirical filter conversion formulae of
\citet{2007AJ....133..734B}. Transformed to the $^{0}(g-r)$ filters,
this cut is similar to the one used by \citet{2018MNRAS.475..676S} and
\citet{2017MNRAS.469.2626H} and it reasonably follows the depth of the
green valley in \textsc{L-Galaxies} as well as in TNG300. We see negligible differences in the colour
distributions at the high mass end between the $^{0}(g-r)$ colours
with and without dust extinction added, but there is a shift for low
mass galaxies around $10^{9.5} \, h^{-2} \, M_\odot$ with blue
galaxies being misclassified as red, which leads to a slight blurring
of the green valley. The \citet{2018MNRAS.475..648P} stellar
mass resolution correction for TNG300, which produces rTNG300, does not take into account
the differences in the colour distributions between TNG100 and
TNG300, which primarily affects galaxies with $9.0 < \log_{10} M_\ast \left
[ M_\odot \right ] < 10.5$ in the range of stellar masses we are
probing. The consequence of this correction is that the red sequence in rTNG300 is shifted into the blue by
about 0.1 mag for $9.5 < \log_{10} M_\ast \left [ M_\odot \right ] < 10.0$,
with a slightly smaller shift for higher masses. This is one example of a relation which the resolution correction affects, implying that one should proceed with caution as consistency conditions of the simulation are altered. However, we are mainly interested in the stellar mass-halo mass relation for lensing, and since the correction brings (r)TNG300 in line with observations, it is still a meriting case to examine. We discuss how satellite fractions are affected in Section~\ref{sec:baryons}.

\section{Method}\label{sec:method}

In this Section we outline how we model the different lensing and clustering datasets under different selection functions.

\subsection{Background cosmology, box size and hydrodynamics}\label{sec:cosmologicalRescalingImpact}

Apart from differences in galaxy formation physics, the galaxy-galaxy lensing
and clustering signals are also influenced by cosmological
parameters. To illustrate this we also compute the SAM lensing
predictions for the fiducial \citetalias{2015MNRAS.451.2663H} model in the fiducial Millennium WMAP1 cosmology, which has a lower matter fraction $\Omega_\text{m}$
and greater $\sigma_8$ than the Planck 2014 cosmology. We have run the \textsc{L-Galaxies} SAM on TNG100-DMO and TNG300-DMO, the gravity-only
versions of IllustrisTNG boxes. In these
simulations, the background cosmology, Planck 2016
\citep{2016A&A...594A..13P}, is close to the adopted Planck 2014
cosmology, which means that the model parameters chosen should be
fairly optimal.

We have compared the halo mass functions, which are what the rescaling
algorithm is designed to match, for the central galaxies for the
\citetalias{2015MNRAS.451.2663H} model run on top of the rescaled MR
and MRII runs as well as the gravity-only runs of
TNG100 and TNG300 and note negligible differences. This, however, does
not necessarily translate to good agreement in the SMFs, for the
rescaled MRII whose SMF deviates from the TNG100-DMO results above
$10^{10.2}\,h^{-2}\,M_\odot$. We attribute this mass bias to
small number statistics and potential biases in how the SAM assigns
galaxies to the rescaled merger trees. Because of this issue, and in
order to conform with the number of objects and simulation volume for EAGLE in the
\citet{2017MNRAS.471.2856V} study, we carry out the group lensing
comparison with the SAM derivatives run on the
gravity-only TNG100-DMO simulation.

In the IllustrisTNG suite, the lensing signal is affected by the hydrodynamics. To gauge the impact we compare the lensing signal around matched subhaloes in the full physics and gravity-only runs. This matching is bijective and based on the particle IDs in the structures. We also compare the TNG results to its predecessor Illustris and the EAGLE simulation. The matched EAGLE catalogues we use are built using a matching of the 50 most bound particles of the substructures \citep[see e.g.][]{2015MNRAS.451.1247S}. For the central (sub-)haloes studied in this analysis, there should be negligible differences between the matching techniques, but we note potential differences for satellite galaxies (see Section~\ref{sec:groupCriteriaTNG}).

\subsection{Simulation mocks}\label{sec:simulationMocks}

To produce lensing predictions from simulations, we bin particles and tessellation elements in concentric cylindrical shells around the structures of interest using the full distributions in a given simulation snapshot with average redshift close to the average lens redshift. We use this method rather than computing full lightcones as it is simpler and yields comparable results. The signal is computed as the average of the projection along the three spatial axes. Theoretical
error bars for the lensing are computed using hundred bootstrap
samplings of the signal, where we treat each component in the full physics run separately, with replacements with the 95\,\% percentiles
shown. We do not account for this model spread in the $\chi^2$-computations, as it is typically small. For galaxy clustering we measure the signal in 40
log-equidistant bins between $20\,h^{-1}\, \text{kpc}$ and
$20\,h^{-1}\,\text{Mpc}$ and for lensing 40 log-equidistant bins
between $20\,h^{-1}\, \text{kpc}$ and $2\,h^{-1}\,\text{Mpc}$
(KiDS+GAMA) and $30\,h^{-1}\, \text{kpc}$ and $3\,h^{-1}\,\text{Mpc}$
(SDSS-DR7). Hence we probe the 1-halo and 2-halo terms for the
clustering and mainly the 1-halo term for galaxy-galaxy lensing. For the clustering data points
without error bars, we use $\pm 15\,\%$ estimates for the
variance which correspond to the smallest quoted errors. We make some of these profiles publicly available at the TNG website, including a library of object-by-object component-wise lensing profiles for $> 400\,000$ galaxies with $M_\ast > 10^{8.2}\,h^{-2}\,M_\odot$ from TNG300 and their dark matter-only counterparts in TNG300-DMO at $z = 0$. We also make profiles for $\sim 35\,000$ galaxies from TNG100 and TNG100-DMO passing this stellar mass cut at $z = 0$ available. 

For the \citet{2016MNRAS.459.3251V} stellar mass-only comparison, our baseline model test data, we measure the
signal\footnote{We have performed the same analysis at $z = 0.11$ for a few of the SAMs and note negligible
  differences. Hence we use the same snapshot for all mass bins.} at
$z = 0.31$ in the rescaled MR and MRII runs which is reasonably close to the average redshift of $z = 0.28$. We also show the
corresponding predictions from the TNG300, TNG100 and Illustris simulations at $z = 0.30$. For the EAGLE simulation we use the $z = 0.26$ snapshot. For the SDSS comparisons, we use the $z = 0.0$ snapshot for TNG300
to boost the statistical signal and the $z = 0.11$ snapshot for the MR
run, but we have checked that there are negligible differences at such
low redshifts. For the galaxy group lensing sample from \citet{2017MNRAS.471.2856V},
we show predictions from the different SAMs run on TNG100-DMO at $z = 0.18$. We also compute the corresponding predictions from TNG100, TNG300 and Illustris at this redshift.

The samples in \citet{2016MNRAS.457.4360Z} and
\citet{2015MNRAS.454.1161Z} are approximately volume complete until an
imposed limit in stellar mass $M^\text{mix}_\ast$ which gives the
maximum redshift a galaxy with a given stellar mass could be observed at as a sensitivity function. This sensitivity function can be incorporated into the
differential comoving volume element $ \text{d} V(z)$, which can be
used to set the relative weight of the different simulation snapshots
for each stellar mass bin $i$. Such a setup effectively down-weights
the contribution from the highest redshift snapshots. Below the mass
limit $M_\ast^\text{mix}$, the sensitivity is considered to be full
and we use the ordinary differential comoving volume element to define
that volume. Each individual stellar mass lensing sample is thus
constructed from the list of available snapshots with individual
weights set by their fractional contribution to the total comoving
volume. We have verified that there are negligible differences in the host halo masses for
centrals and satellites for samples defined using this technique with
respect to using a single snapshot at $z = 0.11$ to define the sample,
although the satellite fraction changes on the order of $\sim 1\,\%$
for the \citetalias{2015MNRAS.451.2663H} model. Hence, we use the $z =
0.11$ snapshot for our mocks.

For the KiDS+GAMA group lens sample, we introduce a minimal stellar mass
$M_\ast^\text{lim}$ following \citet{2017MNRAS.471.2856V} from which
we start counting group members. This mass is set such that the
satellite fraction for galaxies in the GAMA fields is matched for a
given stellar mass bin. Increasing this mass leads to an almost
monotonic increase in $f_\text{sat}$, depending on the sample size,
as the number of group central galaxies decreases whereas the number
of satellite galaxies is almost constant for a given stellar mass bin.

\subsection{Locally brightest galaxies - LBGs}

We select the mock LBGs, which were investigated in previous comparisons \citep[see e.g.][]{2016MNRAS.456.2301W,
  2016MNRAS.457.3200M} by matching the observational criteria as
adopted by \citet{2013A&A...557A..52P}. At the high mass end, identification rates for LBGs exceed $90\, \%$ for
central galaxies -- i.e. the fraction of central galaxies which are also
LBGs -- but less luminous red galaxies are excluded to a greater extent
than blue centrals since red centrals live in denser and thus more
clustered environments,
although this effect is of the order of $5-10\,\%$. All distances quoted are in comoving units, except for the LBG
selection cylinder which has a radius in physical Mpc.

\subsection{Galaxy classification and stellar masses}

Galaxies in the SAMs are classified as centrals, satellites or orphans
in their host haloes depending on whether their associated subhaloes
are central, satellite, or stripped. In IllustrisTNG there are only
central and satellite subhalo hosts. In the SAM the stellar mass is given as
the combined mass in stars in the bulge and disk, where the
intracluster light (ICL) component is neglected (this primarily affects the high
mass end). For IllustrisTNG and the other hydrodynamical simulations we use the bound stellar masses to be in line with EAGLE \citep{2017MNRAS.471.2856V}. We have also conducted the analysis with 30 pkpc (physical kpc) aperture masses to conform with previous studies which have compared SAMs and results from the EAGLE simulation \citep[][]{2016MNRAS.461.3457G, 2018MNRAS.474..492M}, and we achieve comparable results. For the clustering signals we show results with these 30 pkpc aperture masses; the bound mass results have already been presented in \citet{2018MNRAS.475..676S}. We use the simulation specific
$h$ values to convert between stellar masses.

As we are primarily interested in looking at predictions from the largest boxes, this
limits the lowest allowed stellar masses due to resolution
effects. Hence we consider only galaxies with $M_\ast >
10^{9.39}\,h^{-2}\,M_\odot$ in accordance with
\citet{2017MNRAS.469.2626H}.

\subsection{Colours and dust}

The division of galaxies into red and blue can be affected by the
dust model used, especially for dusty star-forming
galaxies. This in turn can influence the predicted clustering and
lensing signals. To illustrate this we perform the analysis with and
without dust extinction for the derivative
\citetalias{2015MNRAS.451.2663H} models as well as the IllustrisTNG
suite. The main difference in the dust treatment between the
\citetalias{Guo:2010ap} and \citetalias{2015MNRAS.451.2663H} versions
is a stronger scaling with redshift in the latter for the extinction
by the interstellar medium in galactic discs\footnote{The total dust
  model is separated into an ISM treatment and one for the molecular
  birth clouds of stars following \citet{2007MNRAS.375....2D}.}. This
should have a minor impact since we only probe colours
at $z = 0.11$.

For IllustrisTNG, we use the fiducial dust model from \citet{2018MNRAS.475..624N},
which describes technical details, with resolved dust attenuation
following the simulated distribution of neutral gas and metals. This
model depends on the viewing angle and we use the median magnitudes of
the twelve angles provided. The dust attenuation in this model has a weak redshift scaling and is almost negligible close to $z = 0$. More recent models can have stronger
redshift dependencies \citep[as investigated in e.g.][]{2019MNRAS.487.4870V, 2020MNRAS.492.5167V}, with the new model described in \citet{2016MNRAS.457.3775M, 2017MNRAS.468.1505M, 2018MNRAS.478.2851M}, but this is of limited importance in the local Universe and thus we do not investigate the impact of different dust models. For
the uncorrected colours we sum the magnitudes of all the individual bound
star particles per subhalo, which are assigned using the
\citet{2003MNRAS.344.1000B} stellar synthesis models assuming a
Chabrier IMF.

\section{Results}\label{sec:results}

In the following Sections we present and discuss our results for the different datasets,
starting with the SMFs and the predictions for the different
galaxy-galaxy lensing datasets, followed by the galaxy clustering
results and lastly by the galaxy group lensing signals.

\subsection{Stellar mass functions}\label{sec:abundanceCorrections}

\begin{figure}
\includegraphics[width=1.02\columnwidth]{\figrelpath 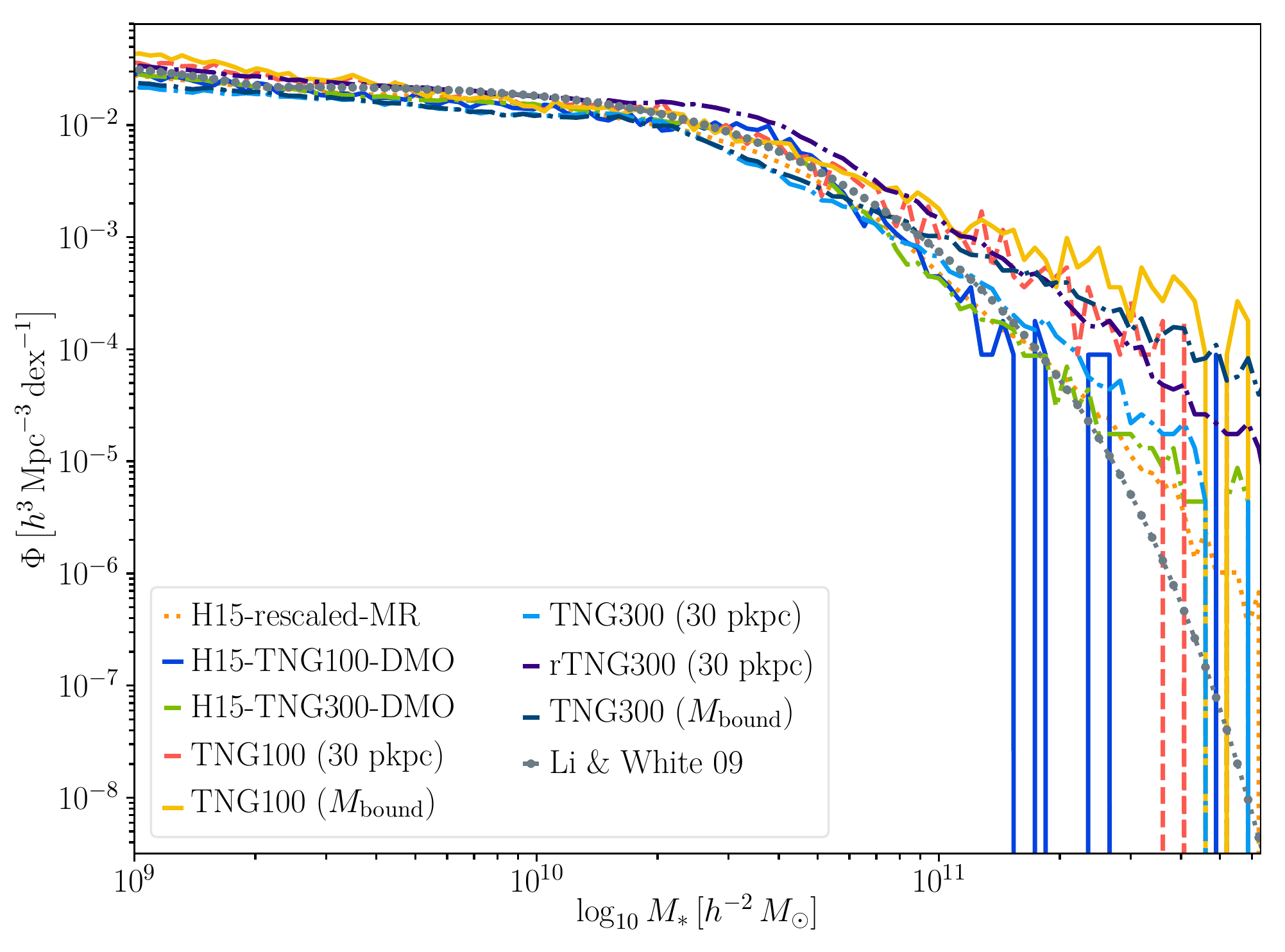}
\caption{The stellar mass function at $z = 0.11$ for the SAM from \citetalias{2015MNRAS.451.2663H} run on top of the rescaled MR simulation as well as the gravity-only runs TNG100-DMO and TNG300-DMO compared to hydrodynamical results from the baryonic runs for TNG100 and TNG300, for 30 pkpc and bound subhalo stellar masses, and the SDSS fit from \citet{2009MNRAS.398.2177L}. If one resolution corrects the TNG300 stellar masses and arrive at rTNG300, those results conform with the smaller TNG100. We note that the SMFs for the SAM on top of TNG100-DMO and TNG300-DMO results are similar to the rescaled MR. The hydrodynamical TNG curves lie above the SAM curves beyond the knee of the SMF, regardless of the mass definition.}
\label{fig:smfLG15DifferentSimulations}
\end{figure}

The stellar mass functions and the stellar-to-halo mass relations for the IllustrisTNG suite have been extensively covered in \citet{2018MNRAS.475..648P}. For accessibility we plot the curves for TNG300, TNG100 and the \citetalias{2015MNRAS.451.2663H} model run on the gravity-only versions of these simulations as well as the rescaled MR simulation in Fig.~\ref{fig:smfLG15DifferentSimulations}. We see that the DMO-curves and the rescaled MR results are consistent with one another. Here we show the results for two different stellar mass definitions for the hydrodynamical simulations, that contained in a 30 pkpc aperture and the total bound subhalo stellar mass, $M_\text{bound}$. This choice especially matters beyond the knee of the SMF, as the apertures cannot capture all bound star particles. We note that both TNG100 and TNG300 favour a higher signal in Fig.~\ref{fig:smfLG15DifferentSimulations} in this regime than the \citetalias{2015MNRAS.451.2663H} model, especially if one considers bound stellar masses. We will show that this results in a better agreement with observational lensing data in the upcoming Sections. Note that the $M_\text{bound}$ SMFs for TNG100 and TNG300 are more similar at the high mass end than the 30 pkpc SMFs, where the resolution corrected rTNG300 SMF nicely traces the TNG100 curve.

Due to
difficulties in properly integrating the sizes of the galaxies, as
well as accounting for the ICL, and flux corrections, stellar masses at the massive end can be under-estimated by up to 0.3\,dex \citep{2015MNRAS.454.4027D}. Thus all these curves are in agreement with observations and there is still room for the modifications of the SAM parameters in Table~\ref{tab:samModelsParameters} while being consistent with the data.

\subsubsection{SMFs and abundance corrections for SAMs}

\begin{figure*}
\begin{centering}
\includegraphics[width=1.02\columnwidth]{\figrelpath 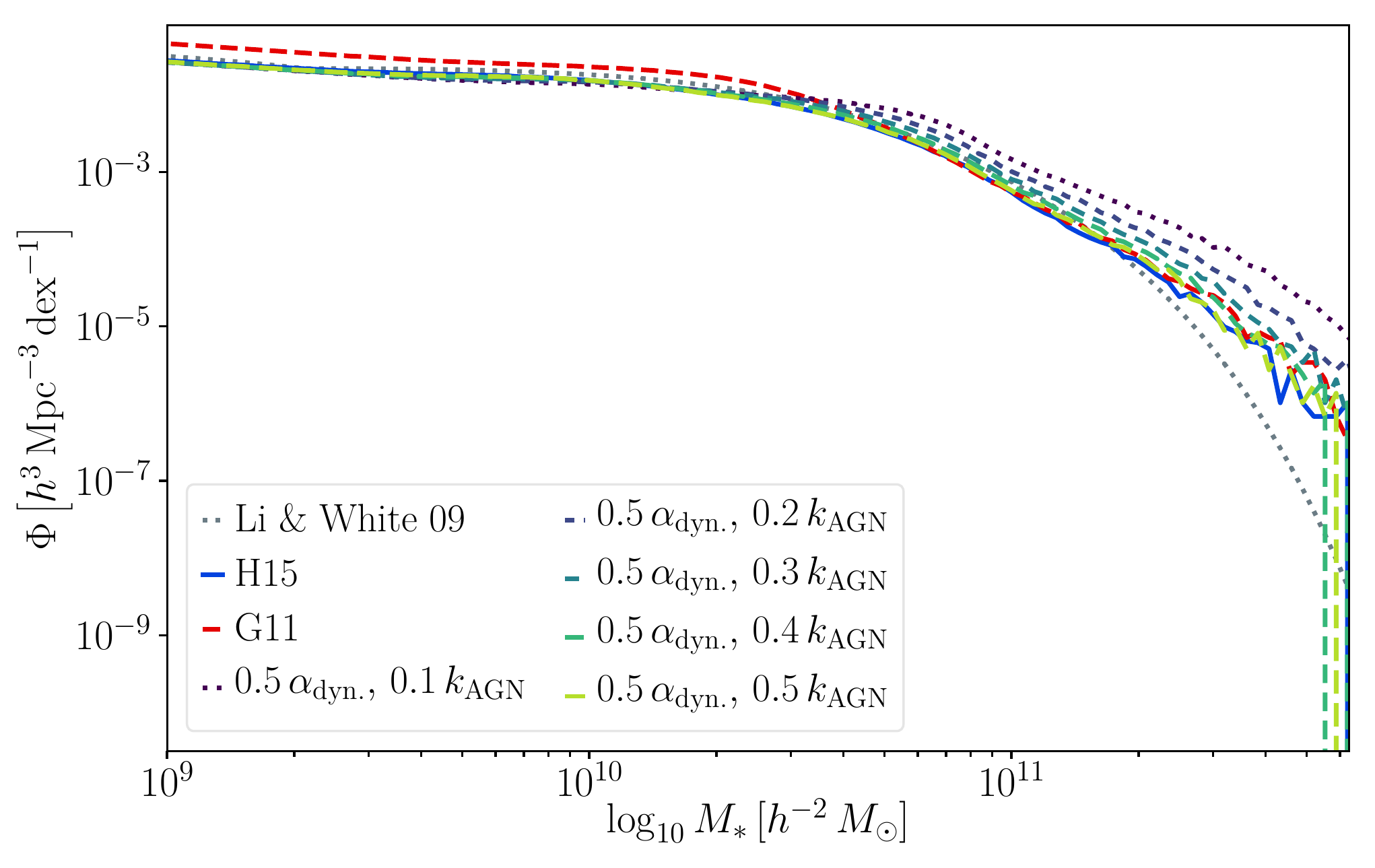}
\includegraphics[width=1.02\columnwidth]{\figrelpath 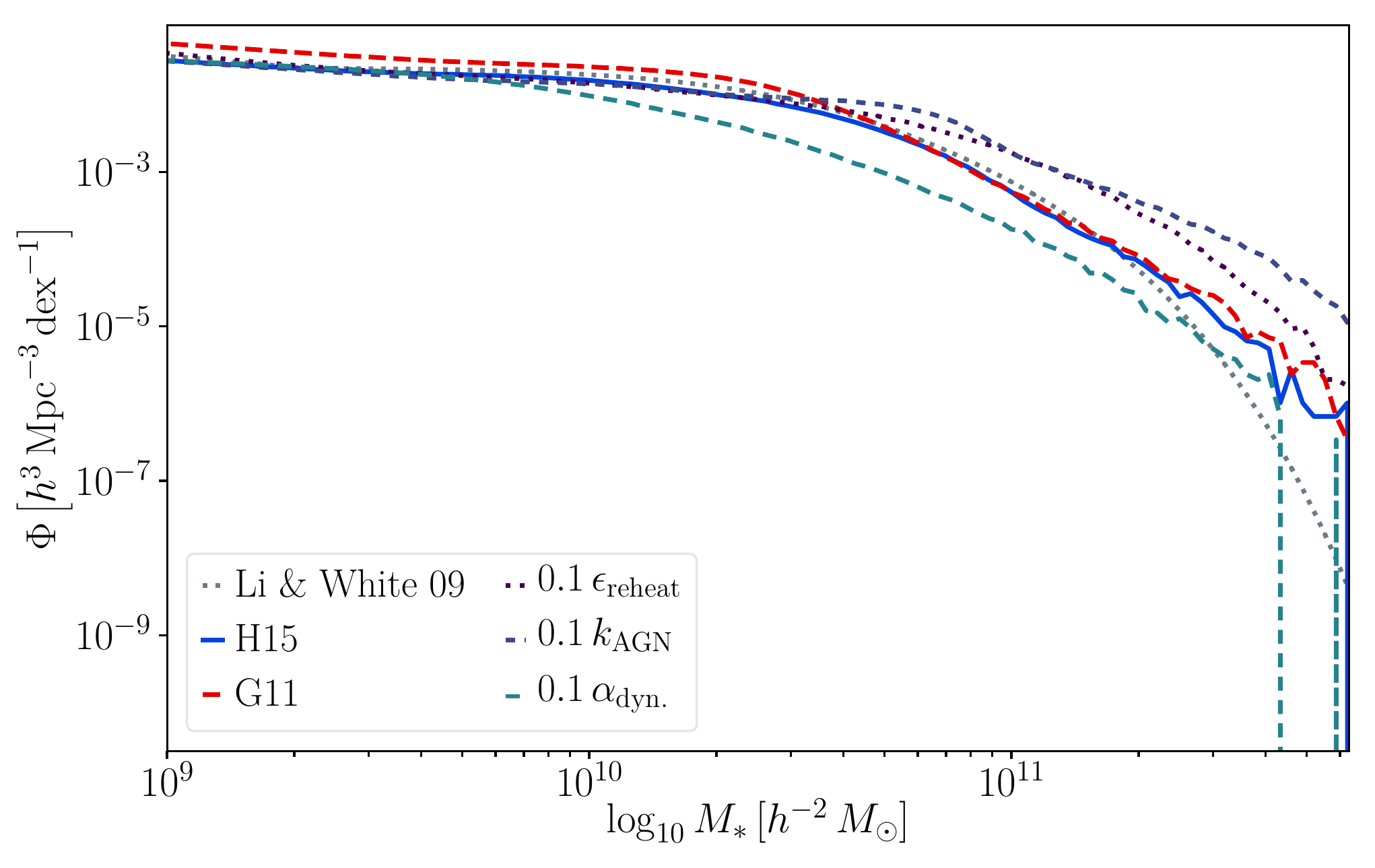}
\caption{Stellar mass functions at $z = 0.11$ for the \citetalias{2015MNRAS.451.2663H} and \citetalias{Guo:2010ap} fiducial models and model derivatives with different strengths of AGN feedback (\emph{left}). The $(0.5\, \alpha_\text{dyn},0.5\,k_\text{AGN})$ model traces the fiducial \citetalias{2015MNRAS.451.2663H} solution and the different AGN feedback strengths become noticeable above the knee of the SMF. In the right figure we illustrate the same situation with the fiducial models compared to the three most extreme parameter choices. Similarly as for TNG100 and TNG300, see Fig.~\ref{fig:smfLG15DifferentSimulations}, the weak feedback models $0.1\,k_\text{AGN}$ and $0.1\,\epsilon_\text{reheat}$ predict an excessive number of galaxies beyond the knee of the SMF. The $0.1\,\alpha_\text{dyn}$ model on the other hand has very few massive galaxies and the change of the SMF is opposite to the direction allowed by observations, leading us to discard this solution.}
\label{fig:smfDifferentkAGN}
\end{centering}
\end{figure*}

\begin{figure}
\begin{centering}
\includegraphics[width=1.02\columnwidth]{\figrelpath 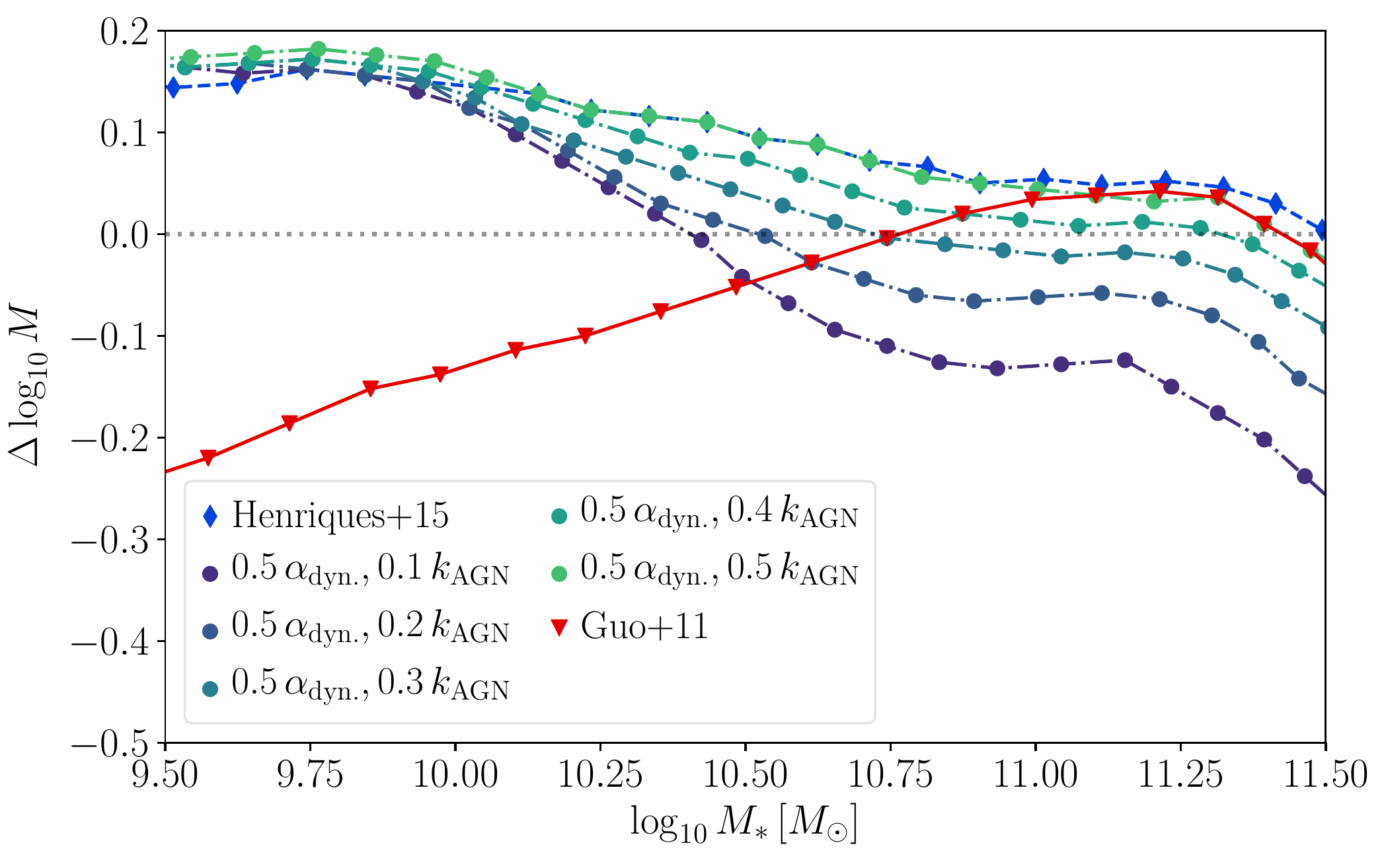}
\caption{Abundance corrections for models with the same $\alpha_\text{dyn}$, but different $k_\text{AGN}$ (\emph{left}), where the $y$-axis shows the correction to bring the model into agreement with SDSS and the $x$-axis the stellar mass after the correction. The $\left (0.5 \, \alpha_\text{dyn}, \, 0.5 \, k_\text{AGN} \right )$ model is almost degenerate with the fiducial \citetalias{2015MNRAS.451.2663H} model, and the $\left (0.5 \, \alpha_\text{dyn}, \, 0.4 \, k_\text{AGN} \right)$ and $\left (0.5 \, \alpha_\text{dyn}, \, 0.3 \, k_\text{AGN} \right)$ solutions have the smallest correction factors around the turnover point of the SMF at $10^{11} M_\odot$.}
\label{fig:smfSameAlphaDifferentkAGN}
\end{centering}
\end{figure}

In Fig.~\ref{fig:smfDifferentkAGN} we show the SMFs
at $z = 0.11$ compared to the fitting function in
\citet{2009MNRAS.398.2177L} for a few of the different $k_\text{AGN}$
SAMs at fixed $\alpha_\text{dyn}$ and the two fiducial models (\emph{left}), as
well as the effect of the most extreme parameter choices from
Table~\ref{tab:samModelsParameters} in (\emph{right}). Here we have not
convolved the masses with any observational error estimate but this
has a minor effect below $10^{11.2}\,h^{-2}\,M_\odot$ and only affects
the massive end. We find that the SMF of the
$(0.5\,\alpha_\text{dyn}, \, 0.5\,k_\text{AGN})$ model closely
resembles the \citetalias{2015MNRAS.451.2663H} result, indicating that
reducing the dynamical friction time while simultaneously reducing the
AGN efficiency introduces an SMF degeneracy with the model run with the fiducial parameter values. The more extreme AGN
feedback choices produce deviations away from the fitting function
starting at $10^{10.4}\,h^{-2}\,M_\odot$. Hence, we determine
that these modifications are allowed by the observational
constraints as discussed in Section~\ref{sec:abundanceCorrections}. The $0.1\, \epsilon_\text{reheat}$ and
$0.1\,k_\text{AGN}$ models lie on the extreme end of what is allowed
whereas the $0.1\,\alpha_\text{dyn}$ model is ruled out. Compared to
the TNG suite predictions in
Fig.~\ref{fig:smfLG15DifferentSimulations}, these model derivatives
reflect those results above $10^{10.2}\,h^{-2}\,M_\odot$.

We quantify the deviations of the SAMs from the 'local SMF' by
computing the necessary stellar mass correction to bring about
agreement with the \citet{2009MNRAS.398.2177L} fitting
formula for SDSS following \citet{2016MNRAS.456.2301W}, i.e. abundance correcting the stellar masses. These abundance corrections are illustrated in
Fig.~\ref{fig:smfSameAlphaDifferentkAGN} with the mass correction in
dex on the $y-$axis for a given stellar mass on the $x-$axis. All
derivative models of \citetalias{2015MNRAS.451.2663H} have a positive
correction for low stellar masses whereas it is negative for the
\citetalias{Guo:2010ap} model with approximately the same
magnitude. These two models have a similar correction for stellar
masses around $10^{11} M_\odot$. The model with reduced
$\alpha_\text{dyn}$ and AGN feedback efficiency $k_\text{AGN}$,
$(0.5 \, \alpha_\text{dyn}, \, 0.5 \, k_\text{AGN} )$,
needs a similar correction as \citetalias{2015MNRAS.451.2663H} as
seen in Fig.~\ref{fig:smfSameAlphaDifferentkAGN}. At fixed
$\alpha_\text{dyn}$, altering $k_\text{AGN}$ has the net effect of
gradually decreasing the correction for high stellar masses, but the
effect is small for dwarf galaxies with a congruence towards the
fiducial solution. As we shall see in the following Sections, the
$(0.5 \, \alpha_\text{dyn}, \, 0.2 \, k_\text{AGN})$
model will give the best LBG and good clustering results, and we find that
it comes with a small correction (\emph{left}). Fixing $k_\text{AGN}$ and changing
$\alpha_\text{dyn}$ gradually offsets the solution similarly across
the whole range of stellar masses, although the effect is slightly
larger around $10^{10.5} \, M_\odot$. Lastly, varying the SN feedback produces
concave $(0.5\, \alpha_\text{dyn}, \, 0.5 \, k_\text{AGN},\,1.5\,\epsilon_\text{reheat})$ and convex  $(0.5\, \alpha_\text{dyn}, \, 0.5 \, k_\text{AGN},\,0.5\,\epsilon_\text{reheat})$ curves around the fiducial valued
$\epsilon_\text{reheat}$ model for $10.0 < \log_{10} M_\ast [M_\odot] < 11.50$, with a congruence at $10^{10} \,
M_\odot$, with $(0.5\, \alpha_\text{dyn}, \, 0.5 \, k_\text{AGN},\,0.5\,\epsilon_\text{reheat})$ yielding a similar curve to $(0.5 \, \alpha_\text{dyn}, \, 0.2 \, k_\text{AGN})$. Simultaneously decreasing $k_\text{AGN}$ and $\epsilon_\text{reheat}$ produces a smoother transition around the knee than solely decreasing the AGN feedback efficiency. The extreme solutions with 10\,\% of the fiducial
\citetalias{2015MNRAS.451.2663H} values for the AGN feedback and SN
feedback are similar to the low $k_\text{AGN}$ solutions, where the
$0.1\,\epsilon_\text{reheat}$ model lacks the plateau feature around
$10^{11.25} \, M_\odot$. The $0.1\,\alpha_\text{dyn}$
solution is ruled out and remains positive across the whole mass range.

\subsection{Stellar mass selection: KiDS+GAMA}\label{sec:mstarOnlyKiDSGAMA}

We begin by investigating the lensing signal for \textsc{L-Galaxies} for stellar mass-only samples to match the observations from the KiDS+GAMA fields at $z = 0.31$ for the \citet{2016MNRAS.459.3251V} datasets, quantify the cosmological dependency, and then proceed to compute the same signals for TNG300 and use that simulation, as well as TNG100, Illustris and EAGLE, to measure the baryonic effects on the signal. As already mentioned, this dataset has the simplest selection function of those covered in this analysis, and thus it serves as the principal benchmark for the SAM and TNG predictions.

\subsubsection{L-Galaxies and variations}\label{sec:mstarOnlyKiDSGAMALG}

{\renewcommand{\arraystretch}{1.4}
\begin{table*}
	\centering
	\begin{tabular}{l l l l l}
	\hline
		{\bf Stellar mass lensing }& {\bf Fiducial} & $\chi^2$ & {\bf Abundance corrected} & $\chi^2$ \\
		\hline
First & $0.1\,k_{\mathrm{AGN}}$ & 1.67 & \citetalias{2015MNRAS.451.2663H} & 5.95 \\
		Second & $\left (0.5\, \alpha_\text{dyn}, \, 0.1\,k_\text{AGN} \right )$ & 1.90 & $0.1\,\epsilon_{\mathrm{reheat}}$ & 6.32\\
		Third &  $\left (0.4\, \alpha_\text{dyn} , \, 0.1\,k_\text{AGN}\right )$ & 2.05 & $\left (0.5\, \alpha_\text{dyn}, \, 0.5\,k_\text{AGN}, \,0.5\, \epsilon_\text{reheat} \right )$ & 6.66\\
		\ldots & & \ldots &  & \ldots \\
		- & \citetalias{Guo:2010ap} & 14.53 & $(0.1\,\alpha_\text{dyn}, \, 0.5 k_\text{AGN})$ & 10.23\\
		- & $(0.1\,\alpha_\text{dyn}, \, 0.5 k_\text{AGN})$ & 18.91 & $(0.1\,\alpha_\text{dyn}, \, 0.1 k_\text{AGN})$ & 11.45\\
		- & $0.1\,\alpha_\text{dyn}$ & 38.89 & \citetalias{Guo:2010ap} & 15.67\\
		\hline
	\end{tabular}
	\caption{The best and worst fit models according to stellar mass-only lensing without and with abundance corrected masses. Lensing prefers models with weaker AGN feedback and the \citetalias{2015MNRAS.451.2663H} model (drops from $\chi^2 = 7.79$) are competitive once the stellar masses have been altered to comply with SDSS abundances (in the case of \citetalias{2015MNRAS.451.2663H}, this means increasing the stellar masses for the range covered in Fig.~\ref{fig:smfSameAlphaDifferentkAGN}). After abundance corrections, the $0.1\,k_\text{AGN}$ slips to $\chi^2 = 10.05$, the $\left (0.5\, \alpha_\text{dyn}, \, 0.2\,k_\text{AGN} \right )$ model has $\chi^2 = 8.29$ and the \citetalias{2015MNRAS.451.2663H} model with \citetalias{Guo:2010ap} parameter values has $\chi^2 = 8.47$, which is not as good as the fiducial \citetalias{2015MNRAS.451.2663H} model.}
	\label{tab:fomLensingStellarMassOnly}
\end{table*}}

\begin{figure*}
\begin{centering}
\includegraphics[width=0.80\columnwidth]{\figrelpath 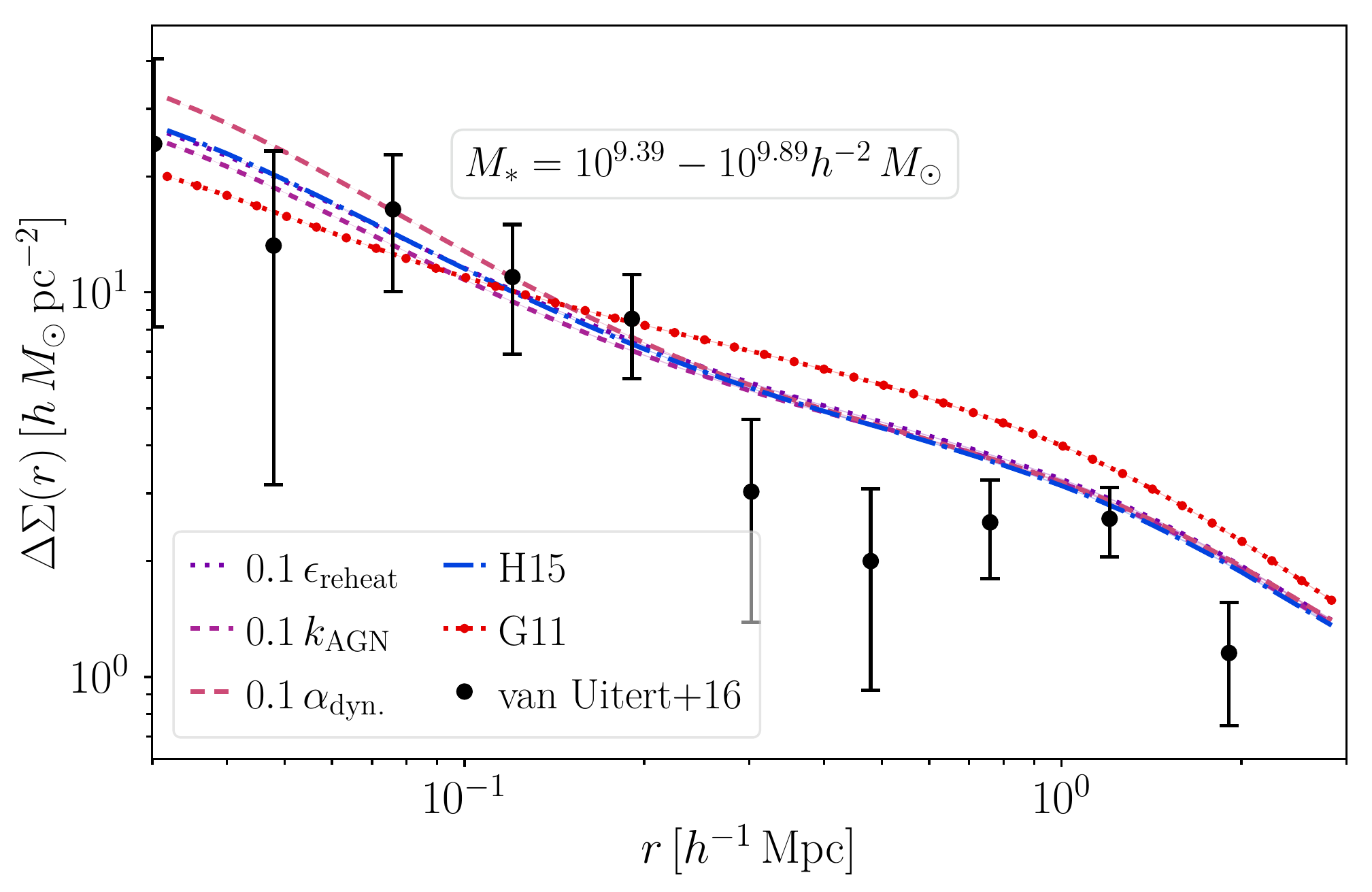}
\includegraphics[width=0.80\columnwidth]{\figrelpath 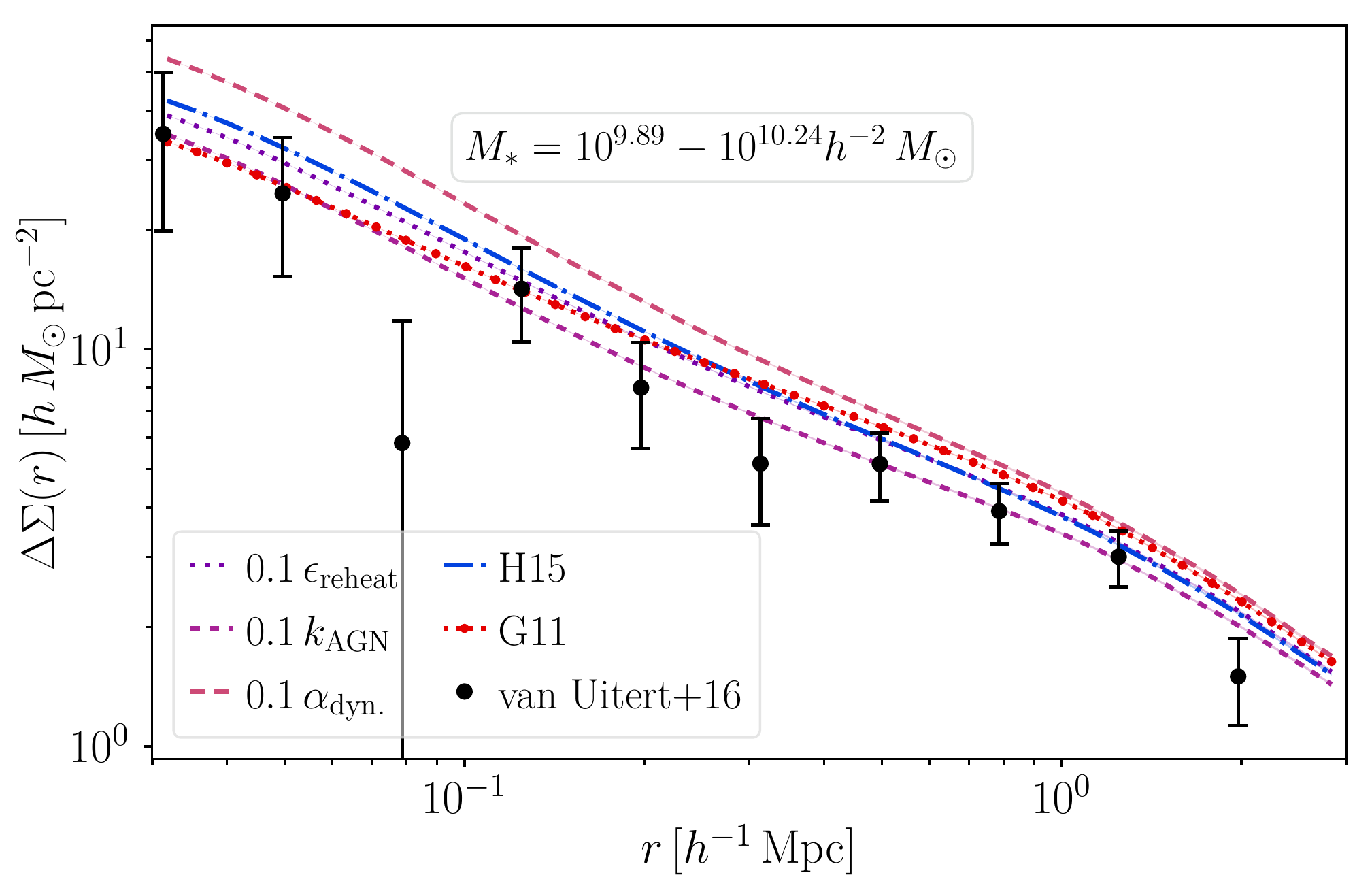}
\includegraphics[width=0.80\columnwidth]{\figrelpath 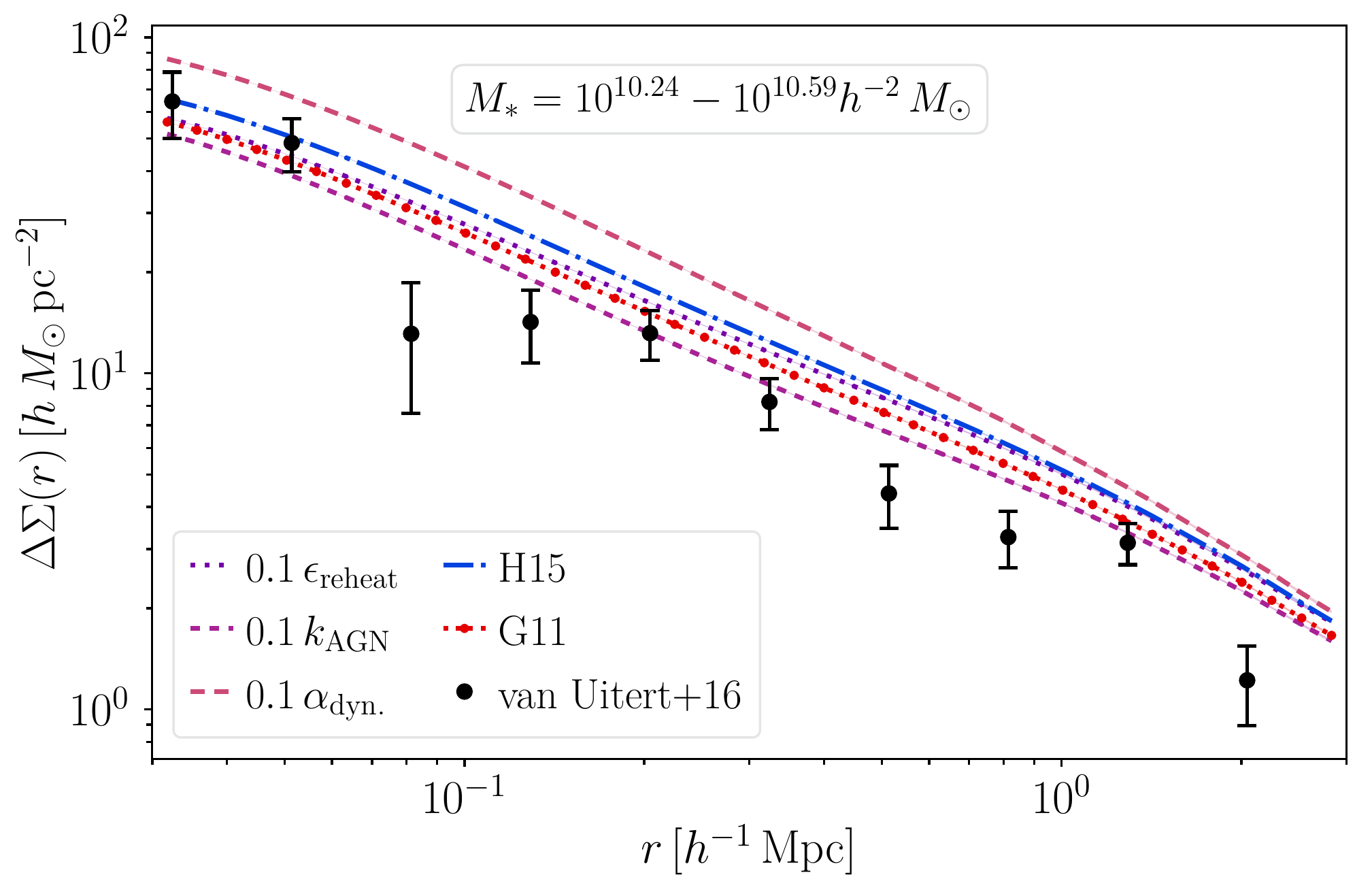}
\includegraphics[width=0.80\columnwidth]{\figrelpath 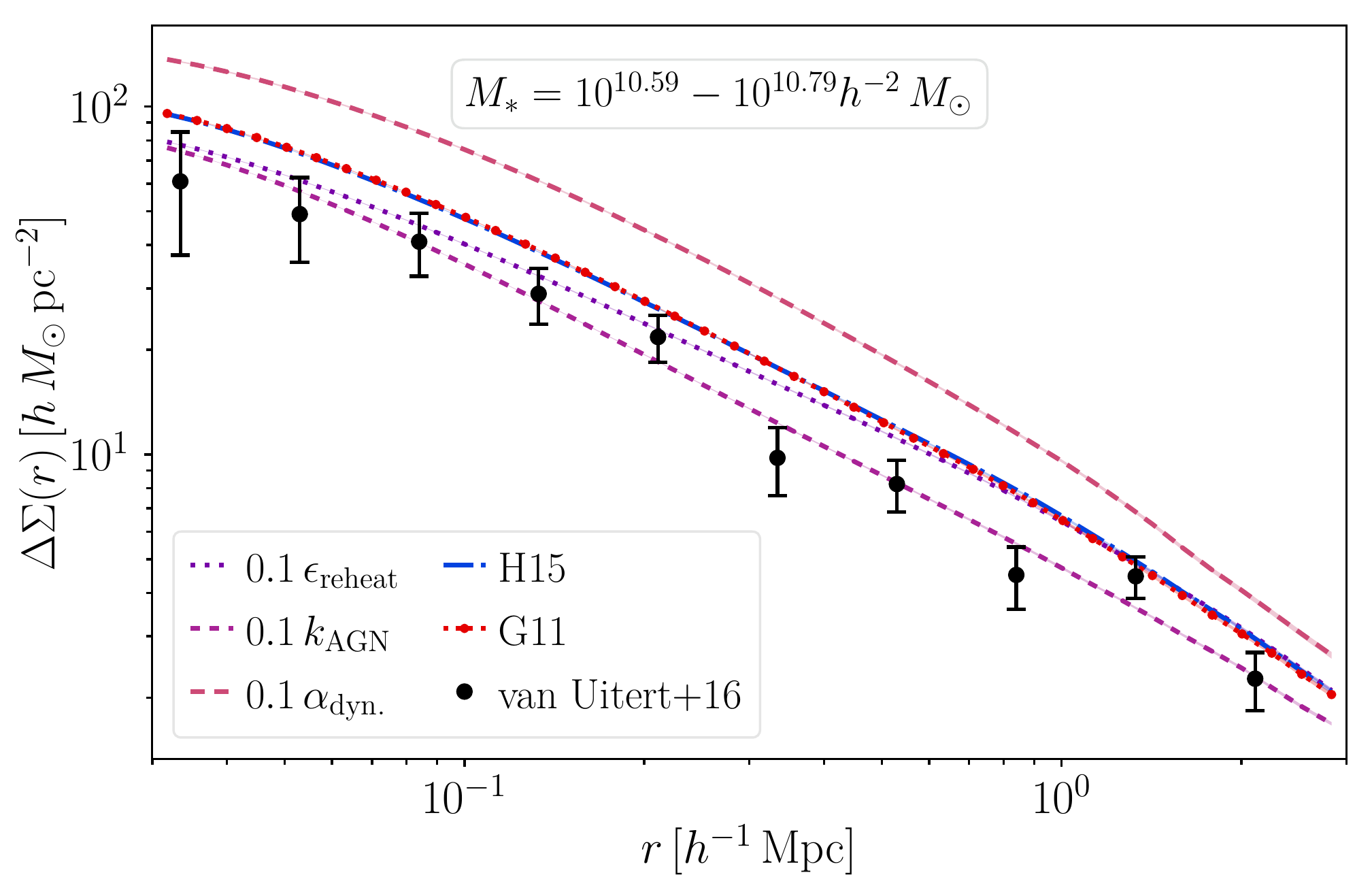}
\includegraphics[width=0.80\columnwidth]{\figrelpath 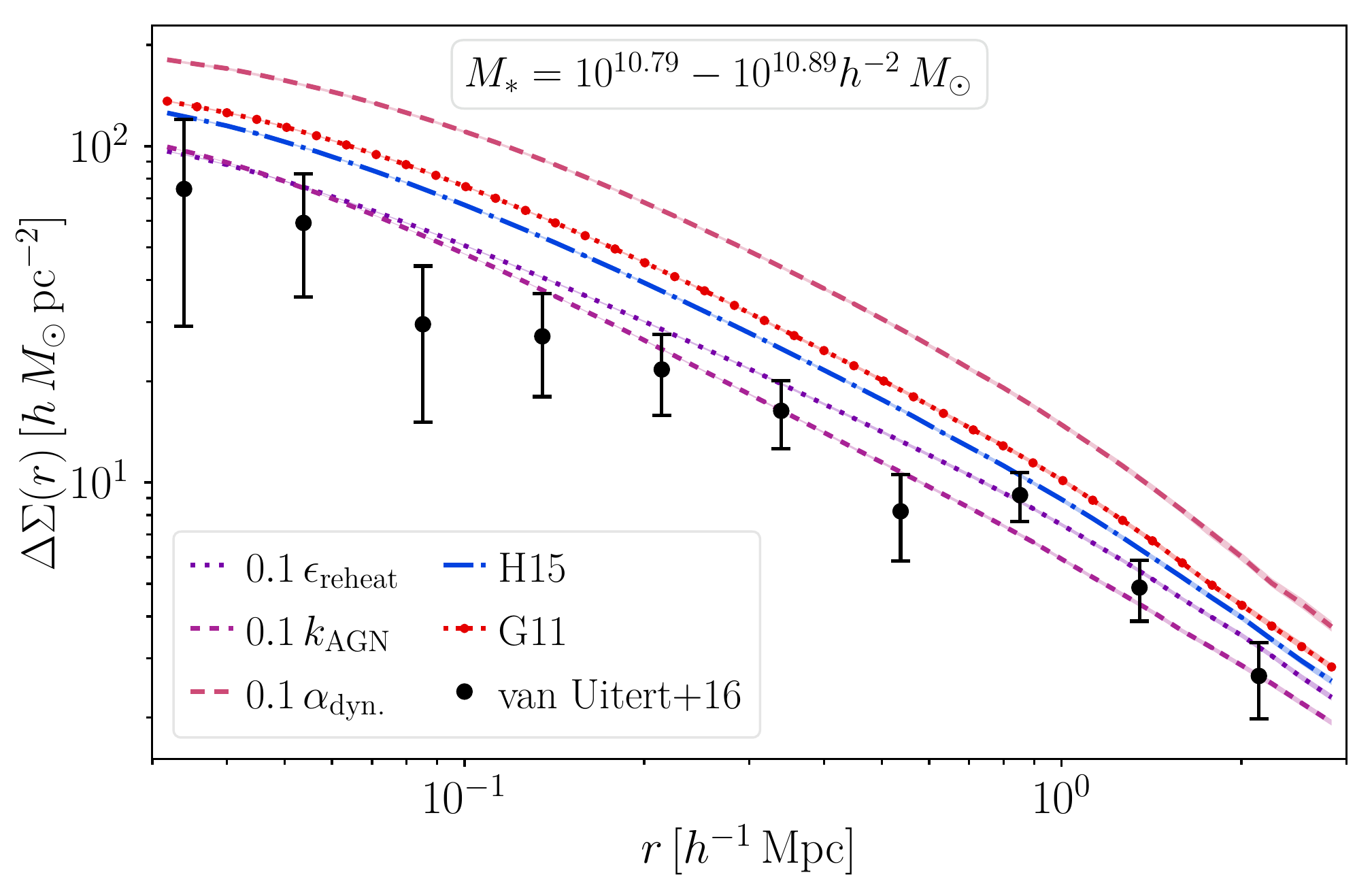}
\includegraphics[width=0.80\columnwidth]{\figrelpath 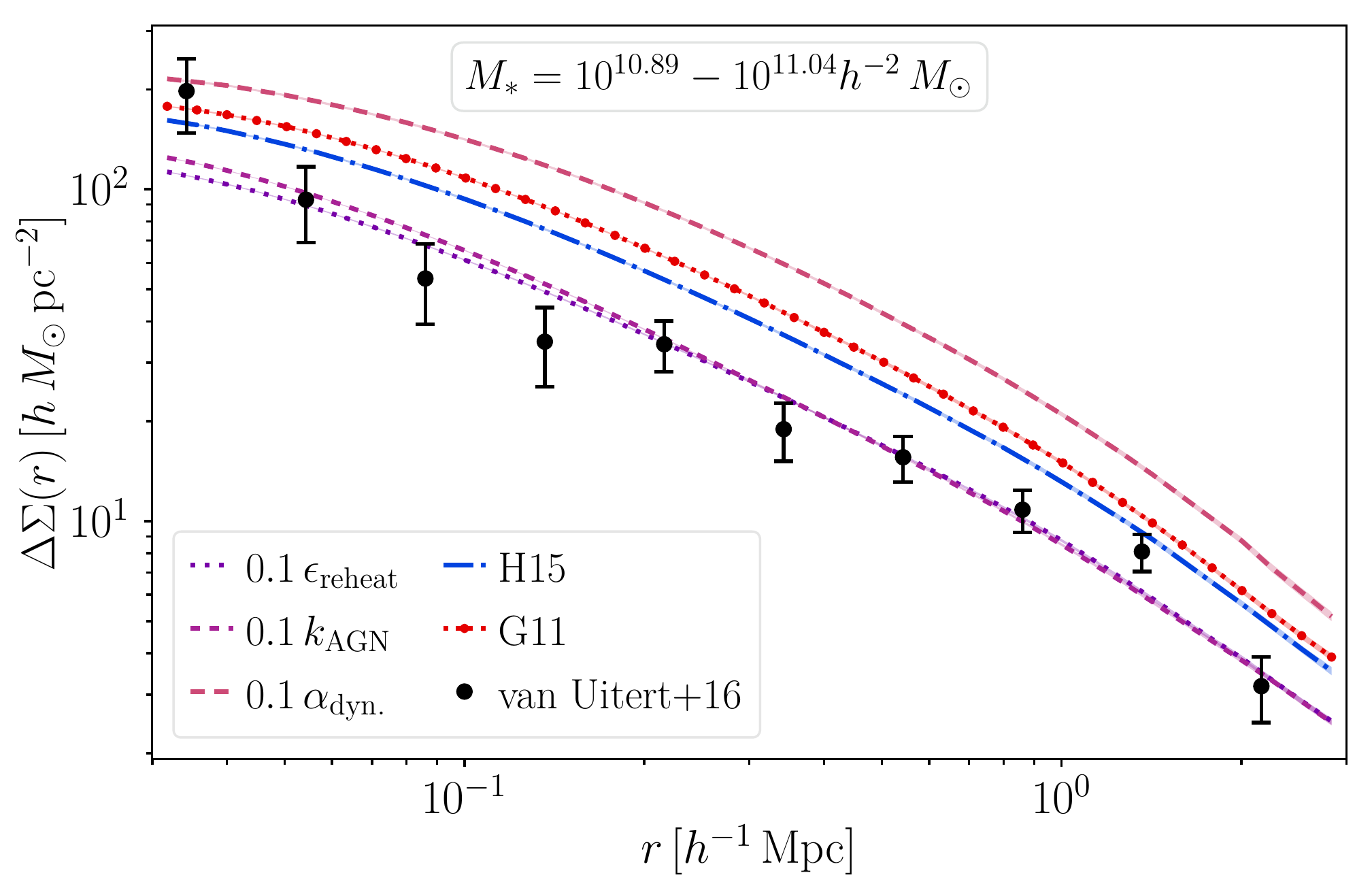}
\includegraphics[width=0.80\columnwidth]{\figrelpath 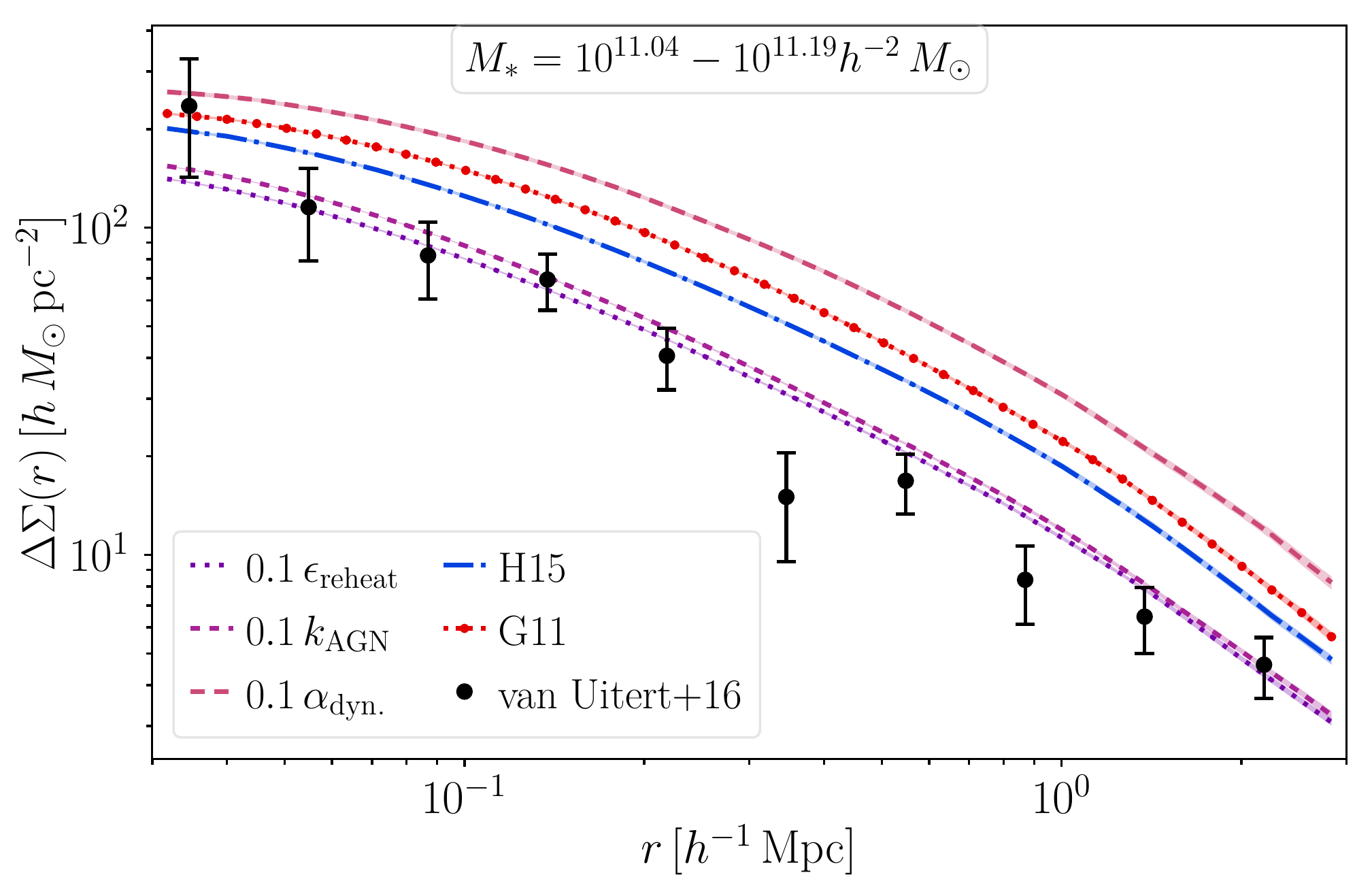}
\includegraphics[width=0.80\columnwidth]{\figrelpath 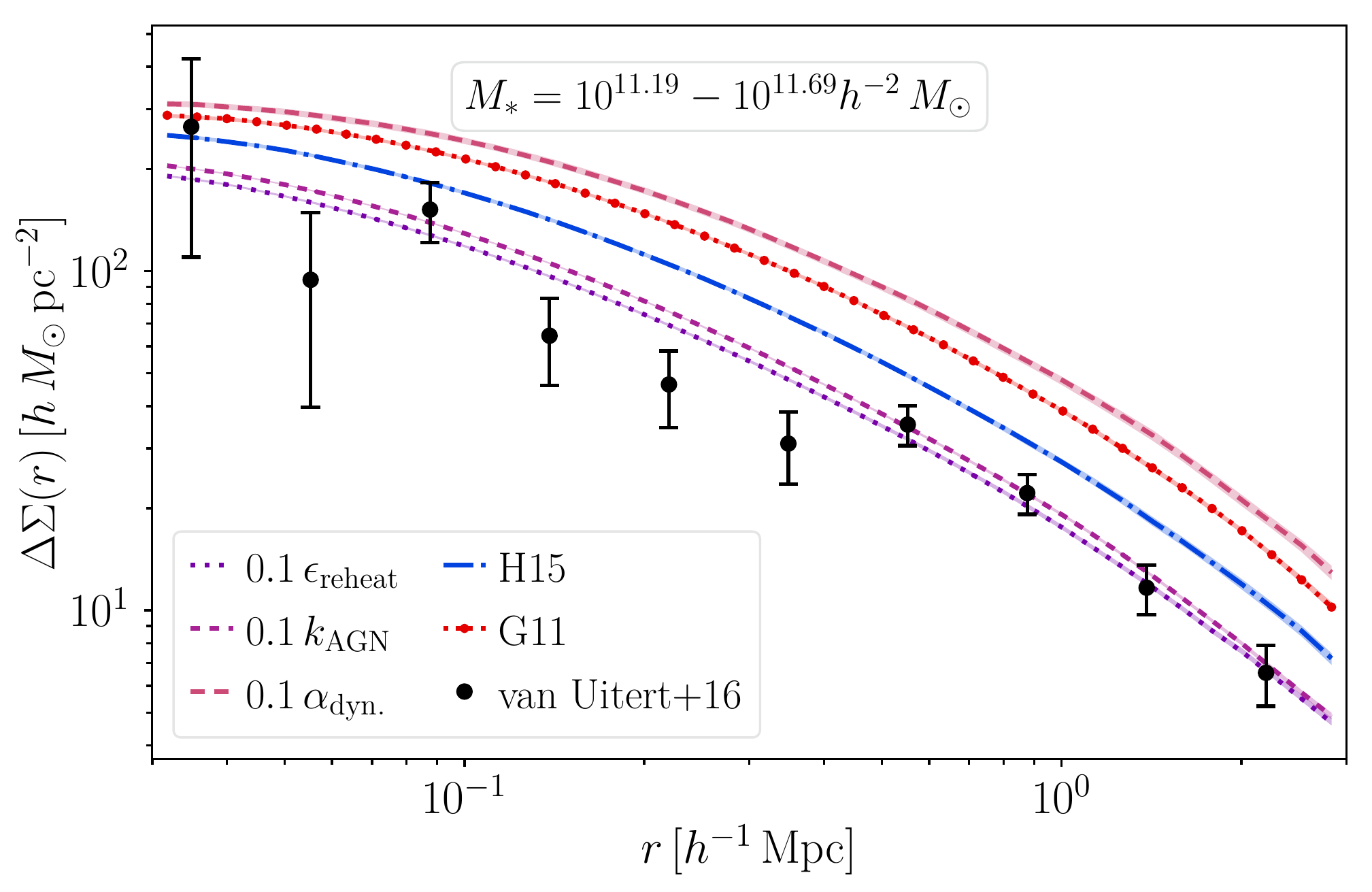}
\caption{Lensing signals for galaxies selected according to stellar mass at $z = 0.31$ compared to data from \citet{2016MNRAS.459.3251V} for the extreme models, and the two fiducial SAMs. Predictions from the \citetalias{Guo:2010ap} model exceed the \citetalias{2015MNRAS.451.2663H} model for the lowest mass bin and for mass bins with $M_\ast > 10^{10.79}\,h^{-2}\,M_\odot$. From this mass onwards, the two extreme SAMs with $0.1\,k_\text{AGN}$ and $0.1\,\epsilon_\text{reheat}$ perform the best.}
\label{fig:vanUitert16ExtremeModels}
\end{centering}
\end{figure*}

Starting with \textsc{L-Galaxies}, the fiducial SAMs, both the \citetalias{2015MNRAS.451.2663H} and
\citetalias{Guo:2010ap} models, predict an excessive signal around all
galaxies for masses $M_\ast > 10^{10.89}\, h^{-2} \, M_{\odot}$ approximately ranging from $50\,\%$ to a factor of two above the data points at  $r \approx 140\,h^{-1}\,\text{kpc}$ (and approximately the same discrepancy at $r \approx 0.9\,h^{-1}\,\text{Mpc}$ for the two most massive bins for the \citetalias{2015MNRAS.451.2663H} model) with a median excess exceeding 30\,\% from the upper quoted observational errors. The \citetalias{2015MNRAS.451.2663H} predictions also eclipse this 30\,\% bar for the $10.24 < \log_{10} M_\ast [h^{-2}\,M_\odot] < 10.59$ bin. However, the median excess for the \citetalias{2015MNRAS.451.2663H} model are within $50\,\%$ of the upper quoted errors for all mass bins. These profiles are shown in Fig.~\ref{fig:vanUitert16ExtremeModels} where we illustrate the
fiducial model predictions together with the results for the extreme
models. We have also investigated the effect of
gradually lowering the AGN feedback efficiency on the produced lensing profiles. At the high mass end, predictions for the $ (0.5\, \alpha_\text{dyn}, \, 0.1\,k_\text{AGN} )$ model are similar to the $0.1\,k_\text{AGN}$ results. It is the favoured solution from $M_\ast > 10^{10.79}\, h^{-2} \, M_{\odot}$ upwards, and the intermediate models do better for the $9.89 < \log_{10} M_\ast \left [ h^{-1} \, M_\odot
  \right ] < 10.24$ and $10.59 < \log_{10} M_\ast \left [ h^{-1} \, M_\odot
  \right ] < 10.79$ mass bins. Decreasing the feedback efficiency
lowers the signal step-by-step, except for the least massive bin where
there are only small differences between the models, which we could
also infer from the convergence of the abundance corrections in
Fig.~\ref{fig:smfSameAlphaDifferentkAGN}.
\citetalias{2015MNRAS.451.2663H} predicts a lower lensing signal than
\citetalias{Guo:2010ap} from this mass onward. For the least massive
bins, the \citetalias{Guo:2010ap} model yields a smaller signal in the
centre, but more pronounced central bumps owing to its high satellite
fractions. Such a signal is
disfavoured by the observations, leading us to conclude that the
\citetalias{2015MNRAS.451.2663H} model has the best fiducial
performance. We have also computed the result for moderate changes in $\epsilon_\text{reheat}$,
where simultaneously lowering $k_\text{AGN}$ and
$\epsilon_\text{reheat}$ and $\alpha_\text{dyn}$ help to mitigate the
tension with observations. This model performs well except in the two
most massive stellar mass bins, and comes with a
smaller discrepancy for the SMF at the high mass end than models with
lower $k_\text{AGN}$ only. For small variations in
$\alpha_\text{dyn}$ while $k_\text{AGN}$ is fixed, the resulting
lensing profiles change only marginally.

Previous studies \citep[e.g.][]{2017A&A...601A..98S} have not recorded a similar tension for the \citetalias{2015MNRAS.451.2663H} model using stellar mass-only selections. \citet{2017A&A...601A..98S} compared the model predictions for $\expVal{\gamma_\text{t}}$, cf. \eqref{eq:deltaSigmaDef}, to CFHTLenS observations, with photometric redshifts for the lenses. They have one mass bin in the $10.89 < \log_{10} M_\ast [h^{-2}\,M_\odot] < 11.19$ regime which is twice as broad as the two bins here with a good agreement between \citetalias{2015MNRAS.451.2663H} and the data. The reasons causing this conundrum could be multiple; the more precise spectroscopic redshifts for the lenses in the GAMA survey could play a role as well as the width of the mass bins and the background cosmology \citep[][assume the fiducial MR WMAP1 cosmology which can decrease the $\Delta \Sigma$ signal by $\sim 15\,\%$ shown in Fig.~\ref{fig:lg15WMAP1vsPlanck}]{2017A&A...601A..98S}.

In the lowest mass bin we have roughly 1 million galaxies in the
fiducial \citetalias{2015MNRAS.451.2663H} model and its derivatives
and $\sim 1.5$ million for the \citetalias{Guo:2010ap} model and approximately $1\,000-10\,000$ galaxies in the most massive bin, which
means that we are analysing robust statistical averages. For low
masses, all models perform approximately equally well, but the more
extreme choices with low supernovae and/or low AGN efficiency are able
to capture the signal across the whole mass range. As is visible in Fig.~\ref{fig:vanUitert16ExtremeModels}, we can lower either (or both, not shown)
of the AGN or supernovae efficiencies to obtain better agreement
with data. These two extremes produce
equivalent predictions for $M_\ast > 10^{10.89}\,h^{-2}\, M_\odot$,
but at lower masses the $0.1\,k_\text{AGN}$ model suggests a lower
lensing signal from $r \sim 100\,h^{-1}\,\text{kpc}$ outwards for
$10.59 < \log_{10} M_\ast \left [ h^{-1} \, M_\odot \right ] < 10.89$ which
starts already at the centre for lower mass bins. This difference
could be driven by the stronger relative strength of the AGN feedback
modification for the SMF and also the higher satellite fraction of the
$0.1\,\epsilon_\text{reheat}$ model (by about $5-10\,\%$ with respect to the fiducial \citetalias{2015MNRAS.451.2663H} model). The satellite fraction for this model
is higher as the lower SN feedback boosts star formation in centrals
and satellites alike, whereas the AGN feedback modification mainly
concerns the centrals.

However, these signals feature degenerate effects from the host halo
masses and the satellite fractions $f_\text{sat}$, which complicates
the modelling interpretations. Still, the discrepancies shown are too
large to be a product of these factors alone for the SAMs. We have computed the predicted satellite
fractions for the different mass bins for all models in our comparison and find that they lie within
the allowed range from the lensing observations and trace the GAMA
group $N_\text{FOF} > 2$ results well. Intuitively, the satellite
fractions are lower for the models with low $\alpha_\text{dyn}$ as
satellite galaxies merge faster. Most models trace a degenerate
solution close to the fiducial \citetalias{2015MNRAS.451.2663H} model (which starts at $f_\text{sat} = 40\,\%$ for low mass objects, drops at the knee of the SMF and ends up at $\sim 15\,\%$ at the high mass end) and the \citetalias{Guo:2010ap} model predicts more low mass
satellites by about $\sim 5\,\%$. Although the two extreme feedback models
$0.1\,k_\text{AGN}$ and $0.1\,\epsilon_\text{reheat}$ predict similar
lensing signals in Fig.~\ref{fig:vanUitert16ExtremeModels}, especially
at the high mass end, the $0.1\,\epsilon_\text{reheat}$ model predicts
more satellites. We shall see in Section~\ref{sec:clustering} that
this influences the clustering signal at $z = 0.11$. 

In Table~\ref{tab:fomLensingStellarMassOnly}, we list the SAMs which
perform best according to the mean figure-of-merit from all lensing
mass bins with and without abundance corrected stellar masses, as well as the worst. We present a full list for the $\chi^2$-values of the uncorrected profiles in Table~\ref{tab:samModelsParameters}. The
lensing data favour low AGN feedback, with a preference for the
fiducial dynamical friction parameter or large fractions of it. If we
perform the same test with post-abundance corrected stellar masses, the fiducial
\citetalias{2015MNRAS.451.2663H} model comes out on top followed by
the low SN feedback efficiency models. These models lie in the mid-range of the ranked uncorrected predictions. The corrected
\citetalias{2015MNRAS.451.2663H} profiles lie closer to the data than
the corrected and (uncorrected) \citetalias{Guo:2010ap} model
predictions, even if there is a very modest preference for the \citetalias{2015MNRAS.451.2663H} model with \citetalias{Guo:2010ap} parameter values for the uncorrected profiles ($\Delta \chi^2 = 0.08$) with respect to \citetalias{2015MNRAS.451.2663H} at $\chi^2 = 7.79$.

If we account for these abundance corrections, how much are the lensing
profiles altered? In \citet{2016MNRAS.456.2301W}, such corrections
were able to reconcile the discrepancies for the
\citetalias{2015MNRAS.451.2663H} model for LBG lenses. We investigate if these modifications are potent enough to mitigate
the large deviations observed in
Fig.~\ref{fig:vanUitert16ExtremeModels} for a more general lens
sample. Since we do not have a fitting function for the SMF at $z =
0.31$, we perform the corrections and measurements for the $z = 0.11$
sample and we assume that the GAMA SMF is similar to the SDSS SMF
which has been shown to be the case
\citep[e.g.][]{2016MNRAS.459.2150W}. While the
correction for the \citetalias{2015MNRAS.451.2663H} model serves to mitigate the tension, lowering the signal by $\sim 10\,\%$ for $10.24 < \log_{10} M_\ast [h^{-2}\,M_\odot] < 11.04$ with a slight percentage level boost for the most massive bin, it is not enough to solve
it, as is indicated by the relatively high $\chi^2$-value.

In addition, we convolve the stellar masses with a Gaussian in $\log_{10}
M_\ast$ with width $0.08 \times (1 + z)$ following
\citetalias{2015MNRAS.451.2663H}. We refer to
\citetalias{2015MNRAS.451.2663H} for the motivation of this choice in an
observational context. We have performed this comparison at $z = 0.31$ and $z =
0.11$ and note that the effect is slightly more pronounced at the
higher redshift due to the redshift dependence of the convolution. We have computed the result for the
\citetalias{2015MNRAS.451.2663H} model and find that the effect is
negligible for the low mass signal, but can amount to $\sim 15\,\%$ at
the high mass end, peaking at the knee of the SMF. The impact is model
specific, with $\sim 5\,\%$ effects for the
$0.1\,\epsilon_\text{reheat}$ and $0.1\,k_\text{AGN}$ derivatives,
whereas the result for the \citetalias{Guo:2010ap} model is similar to
\citetalias{2015MNRAS.451.2663H}. These errors lower the lensing
signal as abundant lower stellar mass galaxies, generally residing in
less massive host haloes, are upscattered to a more massive bin. Alone, it is not enough to explain the observed discrepancy. Moreover,
the observational error bars should already account for these stellar
mass errors, which means that this is a conservative estimate.

\subsubsection{Cosmological impact}\label{sec:cosmology}

\begin{figure}
\includegraphics[width=1.04\columnwidth]{\figrelpath 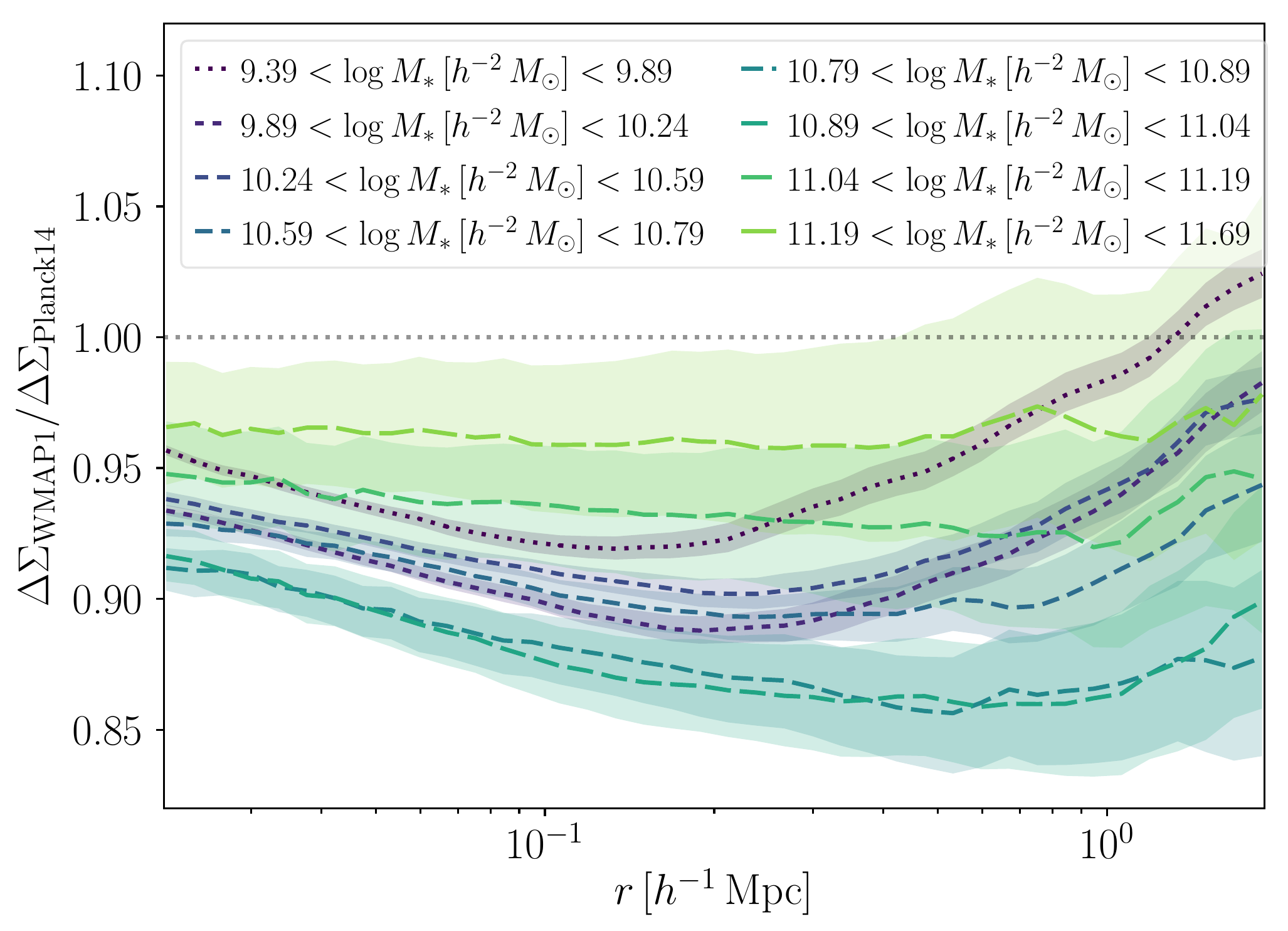}
\caption{Residuals for the \citetalias{2015MNRAS.451.2663H} model run on top of the fiducial Millennium run w.r.t. the rescaled simulation at $z = 0.31$. The signal is suppressed by approximately $10\,\%$ with the largest differences recorded around the knee of the SMF.}
\label{fig:lg15WMAP1vsPlanck}
\end{figure}

To gauge the impact of a different background cosmology we plot the predictions for \citetalias{2015MNRAS.451.2663H} model run
on top of the unscaled, fiducial MR simulation. We see in
Fig.~\ref{fig:lg15WMAP1vsPlanck} that the predictions are slightly
lower by about $\sim 10\,\%$ than for the Planck cosmology but not
sufficient to explain the observational difference. This suppression
has a flat radial evolution for the highest mass bins which are
central-galaxy dominated, whereas there is a radial difference for the satellite
population which dominates the lowest mass bins. The largest effect is
recorded around the knee of the SMF, which is to be expected since it
is most subject to calibration. A fairer comparison from the
perspective of the galaxy formation model, would be to retune a few model parameters to account for this change, which leads us to
conclude that the results in Fig.~\ref{fig:lg15WMAP1vsPlanck} are
upper conservative estimates of the cosmological impact. In
\citet{2016MNRAS.456.2301W}, predictions from the
\citetalias{Guo:2010ap} model were compared across three different
cosmologies (WMAP1, WMAP7 and Planck 2014) for LBG profiles and the
WMAP1 curves were notably higher for the two most massive bins
w.r.t. the other cosmologies, which means that one cannot draw a
general conclusion on the sign of the impact as a function of
background cosmology for all formation models.

\subsubsection{TNG and baryonic effects}\label{sec:baryons}

\begin{figure*}
\begin{centering}
\includegraphics[width=2.1\columnwidth]{\figrelpath 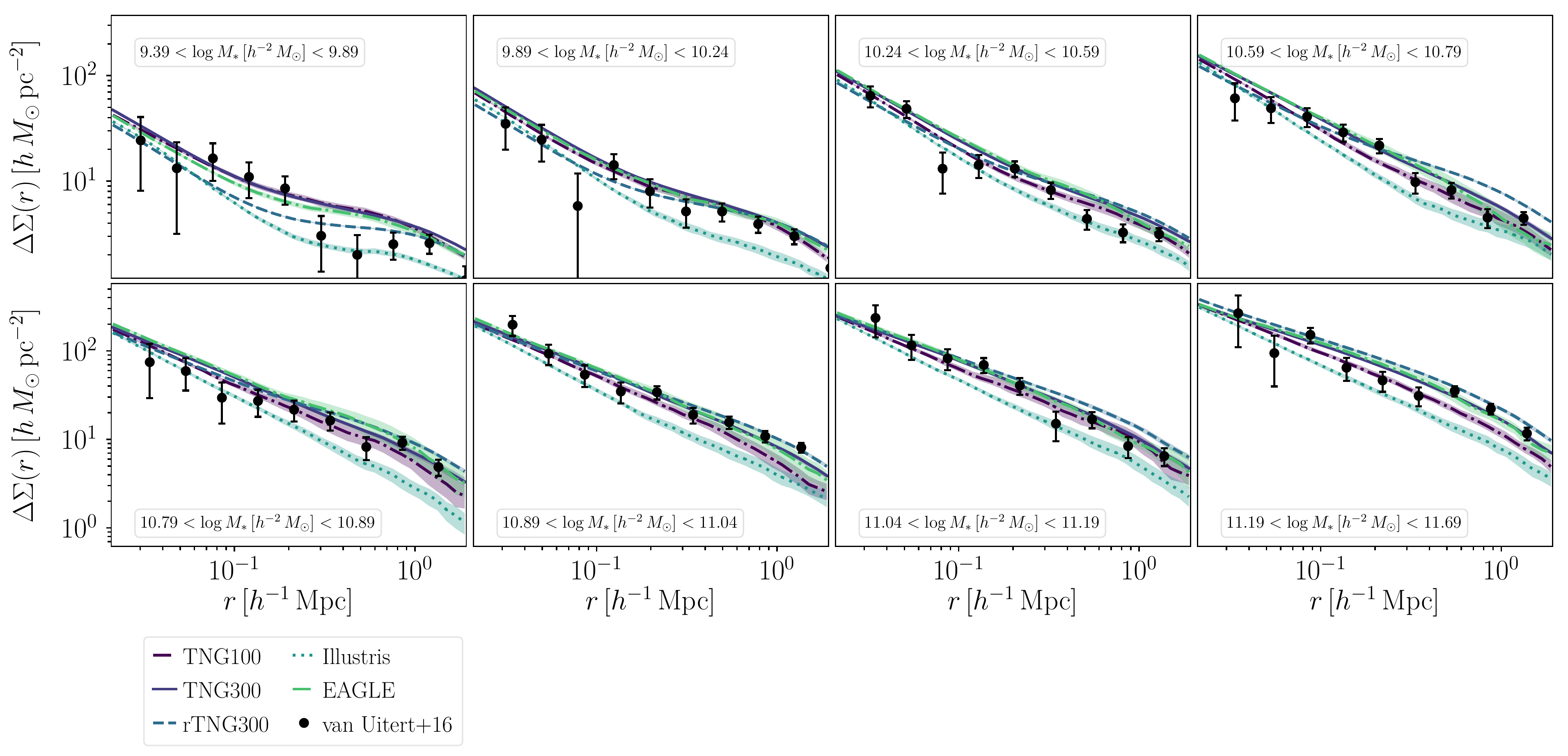}
\caption{Lensing measurements for different hydrodynamical simulations with respect to observations from \citet{2016MNRAS.459.3251V}. Note that we are using a 30 pkpc stellar mass definition for rTNG300, and bound subhalo stellar masses for the other profiles. Compared to the SAMs, the hydrodynamical predictions agree well with the data, with internal model variations of the order of the quoted errors. Illustris produces the lowest lensing signals, but is still in agreement with data, particularly for the lower mass bins. Overall, results from the EAGLE simulation and TNG300 agree well with one another, and the TNG300 curves are slightly boosted with respect to the TNG100 predictions.}
\label{fig:vanUitert16HydroComparison}
\end{centering}
\end{figure*}

\begin{figure*}
\begin{centering}
\includegraphics[width=1.02\columnwidth]{\figrelpath 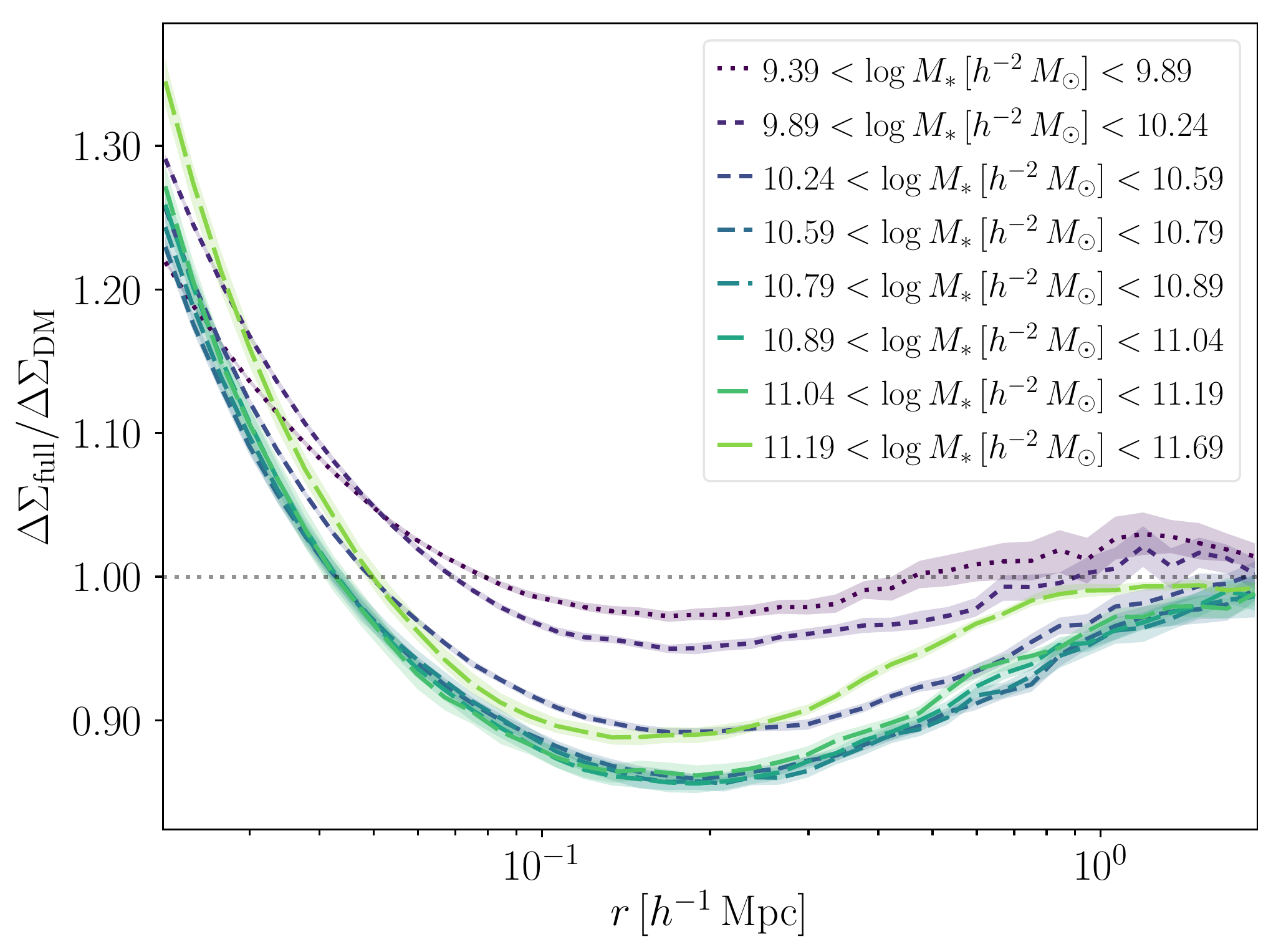}
\includegraphics[width=1.02\columnwidth]{\figrelpath 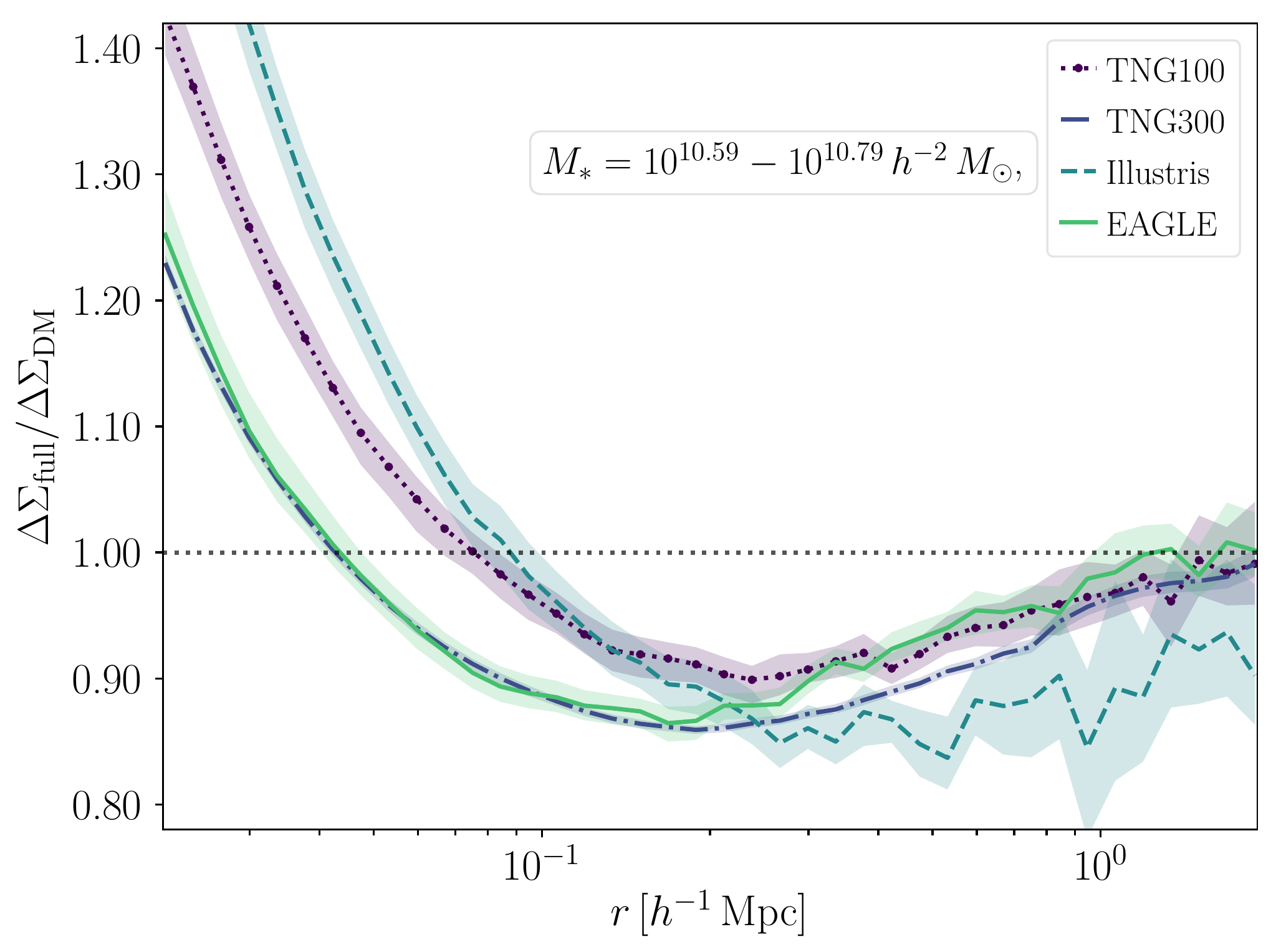}
\caption{Computed baryonic deformations using the residuals for TNG300 (\emph{left}) at $z = 0.30$ between the full physics run and the gravity-only run for matched centrals. The baryonic imprint is characterised by three features; a central enhancement, an intermediate scale suppression and a residual enhancement/suppression around $r\sim 1\,h^{-1}\,\text{Mpc}$, depending on the stellar mass of the bin. Baryonic deformations (\emph{right}) from TNG300, compared to TNG100, Illustris and EAGLE. The observed difference in the deformation between TNG100 and TNG300 suggests that the difference for the same galaxy formation model run in different volumes can be larger than the difference between different formation models.}
\label{fig:barEffectsIllustrisTNGEAGLE}
\end{centering}
\end{figure*}

In Fig.~\ref{fig:vanUitert16HydroComparison}, we show the predictions from the TNG300 simulation at $z = 0.30$ with respect to
the \citet{2016MNRAS.459.3251V} observations. We have also computed the same stellar mass only predictions for some other hydrodynamical simulations, which are shown for comparison. These simulations have already been compared with respect to the power spectrum \citep{2018MNRAS.475..676S, 2018MNRAS.480.3962C}. TNG100 and EAGLE perform equally well as TNG300 with respect to the data, whereas the Illustris signal is too low. Compared to the two
fiducial SAM models, the curves do not persistently lie above the data
points, with a similar excellent performance as the
$0.1\,k_\text{AGN}$ model for the most
massive stellar mass bins, and the results are overall more consistent
across the whole stellar mass range. Even though we obtain deficient statistics at the high mass end beyond $10^{11}\,h^{-2}\,M_\odot$, the models, apart from Illustris, agree well with the data. It is interesting and a milestone that EAGLE and IllustrisTNG produce very similar lensing predictions, despite the different physical prescriptions used, and that they also agree with observations. However, this very nice agreement, as we shall see in Section~\ref{sec:sdssColour}, does not guarantee conforming results for the colour-separated signal for TNG300 and TNG100 with respect to SDSS-DR7. If we turn to the $\chi^2$-values, again using \eqref{eq:fom}, we find that the data mildly prefer TNG100 ($\chi^2 = 1.80$) and TNG300 ($\chi^2 = 1.81$) over EAGLE ($\chi^2 = 1.92$), with values comparable to the low-feedback SAM variations. The resolution corrected TNG300, rTNG300 with 30 pkpc masses, fares slightly worse ($\chi^2 = 2.64$) and Illustris ($\chi^2 = 3.81$) is the worst hydrodynamical model, but still better than the \citetalias{2015MNRAS.451.2663H} model at $\chi^2 = 7.79$.

We can compare the TNG100 and TNG300 signals to get a handle on how the simulation volume affects the signal. In the centres, TNG300 is boosted by about 0-10\,\% where the increase is largest for the least massive bins compared to TNG100; and this increase can amount to $\sim 20-60\,\%$ for $r \in [1, 2] [h^{-1}\,\text{Mpc}]$. Also this increase is tightly connected to the chosen stellar masses, with the largest effects for the $10.89 < \log_{10} M_\ast [h^{-2}\,M_\odot] < 11.04$ bin. For the least massive bin with $9.39 < \log_{10} M_\ast [h^{-2}\,M_\odot] < 9.89$ there is a very slight suppression for TNG300 with respect to TNG100 for scales $r \sim 150 - 800\,h^{-1}\,\text{kpc}$. We also show results for the resolution corrected stellar masses, rTNG300, but with a 30 pkpc mass aperture. Its signals are similar to TNG300 with bound masses, but there is a suppression in the centres by about $25\,\%$ for the two least massive bins followed by a continuum upwards towards a slight boost for the two most massive bins with a few percent. At scales $r \sim 1\,h^{-1}\,\text{Mpc}$, rTNG300 lies around 25-30\,\% above TNG300 for $M_\ast > 10^{10.59}\,M_\odot$. Still, part of this increase should be attributed to the different mass definitions. We have computed this effect for TNG100, and there the 30 pkpc signal is boosted with respect to the bound mass signal with more than 50\,\% for the most massive bin and around 30-40\,\% for $10.79 < \log_{10} M_\ast [h^{-2}\,M_\odot] < 11.19$ (the effect is smaller for the less massive bins). Hence, there is considerable flexibility in the predicted signal depending on the stellar mass definition, but the variations are allowed within the current data error bars.

The resolution
corrected rTNG300 satellite fractions are excessive around the knee of
the SMF compared to TNG300 and TNG100, and the different SAMs ($f_\text{sat} \sim 50\,\%$ with respect to $\sim 30\,\%$ for the other models). This issue and proposed corrections have been covered by \citet{2020arXiv200211119E}. For TNG300, the satellite fractions trace the TNG100 solution.

By matching subhaloes in the baryonic runs with their dark matter-only
counterparts through their particle IDs using the publicly available catalogues \citep[see][]{2015A&C....13...12N, 2019ComAC...6....2N}, we can obtain an estimate of the
impact of baryonic feedback processes on the profiles. This works
particularly well for central galaxies and we will use this matched
central galaxy signal to estimate the baryonic deformation of the
profiles using the TNG300 simulation. We discuss the issues with matching satellite galaxies in Section~\ref{sec:groupCriteria}. The result for the
\emph{central} galaxies satisfying the stellar mass criteria of
\citet{2016MNRAS.459.3251V} is given in
Fig.~\ref{fig:barEffectsIllustrisTNGEAGLE}. As already found in the
literature \citep[e.g.][]{2015MNRAS.451.1247S, 2017MNRAS.467.3024L, 2019MNRAS.488.5771L},
the baryons enhance the profiles close to the central galaxy due to
the presence of additional cooling from the stellar component and the
associated adiabatic contraction of the dark matter, induce a suppression of the profiles
from $r \approx 100\,h^{-1}\,\text{kpc}$ to $r \approx
1\,h^{-1}\,\text{Mpc}$ which then converge (at $r$ where $\Delta\Sigma_\text{full} = \Delta\Sigma_\text{DM}$) at larger scales since the
same projected mass is contained inside the aperture. This is what we
observe in Fig.~\ref{fig:barEffectsIllustrisTNGEAGLE} (\emph{left}) where the
suppression increases with increasing stellar mass until $M_\ast >
10^{10.6}\, h^{-2} \, M_{\odot}$ and is self-similar for the four
subsequent mass bins with deviations for the most massive bin. Note
that the satellite fraction is high for the lower stellar mass bins
which means that the baryonic effect measured here is not a good proxy
for the observational signal. The maximum suppression amounts to $\sim
15\,\%$ attained at $r \approx 200\,h^{-1}\,\text{kpc}$ and the central
enhancement is roughly $\sim 20-40\,\%$ for the scales probed here, depending on the stellar mass
and radial bin. Except for the two least massive bins, a good
convergence is reached at $r \approx 1-2\,h^{-1}\,\text{Mpc}$ with the
dark matter-only run. We note that the maximum suppression at intermediate radial scales is reached for the two $10.59 < \log_{10} M_\ast \left[ h^{-2}\,M_\odot \right] < 10.89$ stellar mass bins, and that the signal gradually rises for the most massive bins. This is expected as the mean host halo mass increases and the deep gravitational potential wells of clusters are efficient at holding material, even in the presence of AGN feedback.

We also note that the baryonic imprint does not necessarily have to be the same as in TNG300 for the physical recipes used in the SAMs. Thus we also perform a comparison between Illustris, EAGLE and TNG300, as well as the smaller TNG100, with their gravity-only companion simulations across the different stellar mass bins. Here we also use the bound stellar masses (we have verified that the results hold for the 30 pkpc masses as well), restrict ourselves once more to central galaxies and present a representative mass bin in Fig.~\ref{fig:barEffectsIllustrisTNGEAGLE} (\emph{right}). We expect that the small redshift difference ($\Delta z = 0.04$) between the EAGLE and the other simulation snapshots has a negligible impact on the results. The same deformation trends already observed in the left figure  hold for all mass bins except for the least massive one, where the stellar term differs by $\sim 15\,\%$ for the different models (with Illustris up from the TNG300 values in the left figure of Fig.~\ref{fig:barEffectsIllustrisTNGEAGLE}, TNG100 agreeing with TNG300 and EAGLE down). In addition, the deformation in the EAGLE simulation crosses one at $r \approx 40\,h^{-1}\,\text{kpc}$, earlier than for the other simulations. This could be attributed to differences in the SN feedback implementation for low-mass centrals, but as the satellite fractions are $f_\text{sat} \approx 40 \pm 5\,\%$ for this low-mass sample, it is not clear how this would contribute to the joint total signal, and we do not have a representative group lens sample for such objects.

For the representative mass bin, EAGLE and TNG300 produce similar signals. The results from these two simulations are more similar than those from TNG100 and TNG300. Potentially the similar effective resolution could play a role, but it could also be incidental. This model conformance holds for $M_\ast > 10^{10.24}\,h^{-2}\,M_\odot$, with more scatter from $r \sim 200\,h^{-1}\,\text{kpc}$ for EAGLE with $3-4\,\%$ less deformation than TNG300 from $M_\ast > 10^{10.59}\,h^{-2}\,M_\odot$\footnote{There is a marginal shift between the two models for the second most massive bin, but they still agree within error bars.}. We also note that there are differences between TNG100 and TNG300, except for the least massive bin where they are in agreement. We have checked that this is independent of the resolution-correction (although this alters the TNG300 results slightly for the least massive and the two most massive bins), and is likely a consequence of the larger volume of the TNG300 simulation, with more massive host haloes present. Apart from differences in the stellar term, the biggest impact on scales where the lensing signal is usually measured is that the deformation in TNG300 amounts to $\sim 15\,\%$, whereas it is $\sim 10\,\%$ for TNG100. Convergence between the two models is typically attained at $r \approx 400-500\,h^{-1}\,\text{Mpc}$, with some scatter for TNG100. Hence, for future comparisons and calibrations of SAMs to account for baryons one must carefully take into account volume differences. One positive note is that the outer deformation convergence radius for TNG300 and TNG100, as well as EAGLE, is roughly the same across the different mass bins at $r \approx 1-2\,h^{-1}\,\text{Mpc}$, which bodes well for cosmological analyses, whereas the convergence radii for Illustris extends to $r \approx 5-6\,h^{-1}\,\text{Mpc}$, reflecting its stronger AGN feedback; and the deformation stays at $\sim 15 \pm 5\,\%$ till $r \approx 1-2\,h^{-1}\,\text{Mpc}$. In Section~\ref{sec:groupCriteria}, we show that this model does not produce good central lensing signals for galaxy groups. This model has the largest contribution from the stellar term across all mass bins covered here, followed by TNG100. Compared to TNG300, where the crossing from positive to negative deformation is more mass-dependent, the Illustris signal makes the crossing at roughly $r \approx 100\,h^{-1}\,\text{kpc}$ for bins with $M_\ast < 10^{11.04}\,h^{-2}\,\text{Mpc}$, and thereafter it gradually decreases to $r \approx 70\,h^{-1}\,\text{kpc}$ for the most massive bin. Regarding TNG100, its stellar term lies between Illustris and TNG300 for all mass bins.

We note that the baryonic effects are smaller than the
measured deviations for the two fiducial SAMs with respect to the data, in
e.g. Fig.~\ref{fig:vanUitert16ExtremeModels}, implying that a better
assignment scheme between galaxies and subhaloes is required and their parameters have to be re-tuned for better lensing agreement.

\subsection{SDSS lensing with colour split}\label{sec:sdssColour}

\begin{figure}
\includegraphics[width=1.02\columnwidth]{\figrelpath 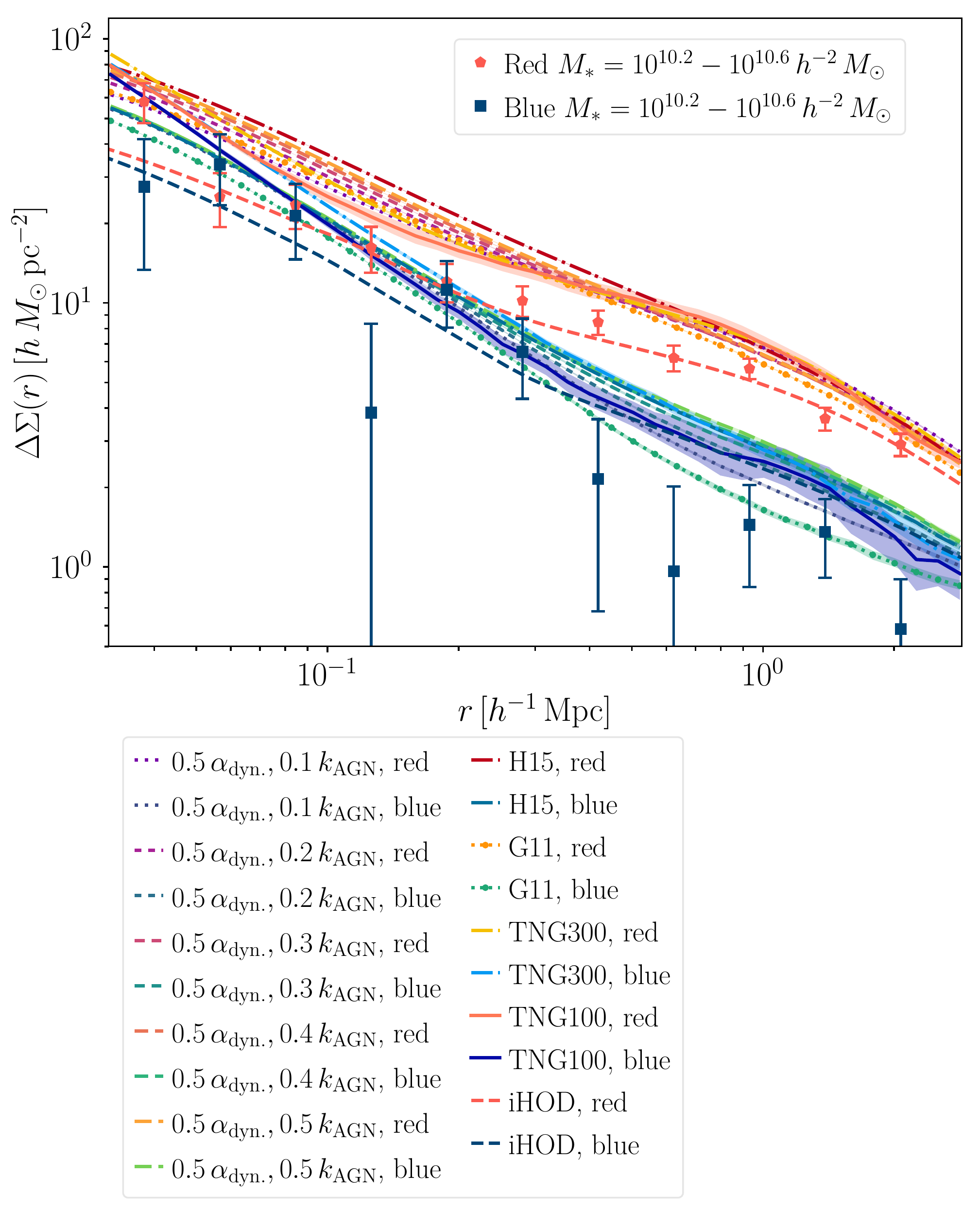}
\caption{Lensing predictions for red and blue galaxies for stellar masses $10.2 < \log_{10} M_\ast \left [ h^{-2}\,M_\odot \right ] < 10.6$ in SDSS using the \citet{2016MNRAS.457.4360Z} datasets and iHODs, with SAMs with different $k_\text{AGN}$ feedback strengths shown. Here, the iHODs agree with observations for the red lensing signal whereas all SAMs, TNG100 and TNG300 predict excessive signals. For the blue galaxies, the \citetalias{Guo:2010ap} and the $(0.5\alpha_\text{dyn}, 0.1\,k_\text{AGN})$ model produces the best results, especially for scales around $r \approx 1\,h^{-1}\,\text{Mpc}$.}
\label{fig:zmRedBlueGGLIntermediateMass}
\end{figure}

In this Section we show a few comparisons between the lensing signal
separated according to colour and corresponding observations from SDSS-DR7, with the
\citet{2016MNRAS.457.4360Z} selections imposed, and compare the SAMs, TNG300 and TNG100 to the quoted integrated, empirical HOD predictions from that publication. We also attempt to fit the total signal from all main SDSS-DR7
galaxies from \citet{2016MNRAS.457.3200M}. As implied by the small difference in the colour cuts and the volume-selection arguments brought
forth in Section~\ref{sec:simulationMocks}, there are only minor
differences between these observations and those of
\citet{2016MNRAS.457.4360Z}; and these have a simpler selection function. We have verified that these datasets produce consistent lensing profiles for the overlapping stellar mass range. These are the hardest datasets which we consider in this Paper; since the models both have to account for potential differences in the satellite fractions as well as the colour distribution, including dust modelling, of the observed samples.

For low mass galaxies with $9.4 < \log_{10} M_\ast \left [ h^{-2}\,M_\odot \right ] < 10.2$, the SAMs perform on par with the iHODs (and TNG100 and TNG300 for blue galaxies and in the centres below $r\sim 0.1- 0.3\,h^{-1}\,\text{Mpc}$ for red galaxies), although there is considerable scatter in the data. Especially the two fiducial SAMs, \citetalias{Guo:2010ap} and \citetalias{2015MNRAS.451.2663H} do well in this mass range for red galaxies (the constraints on the blue lensing signals are weaker due to scatter in the data; most models are in agreement), as well as $0.1\,k_\text{AGN}$ and $0.1\,\epsilon_\text{reheat}$. The predictions from TNG300 are elevated with respect to the data with at most a factor of two (for the lowest mass bin) and typically around 50\,\% for $0.3-0.4 \lesssim r [h^{-1}\,\text{Mpc}] < 2-3$, but better agreement is reached in the central region. This could possibly be amended with an improved colour assignment scheme, as the model underpredicts the red fractions w.r.t. SDSS (35\,\% and 50\,\% compared to 44\,\% and 59\,\% in the data, respectively), and the blue lensing signal is typically lower. This is in line with the underpredicted quenched fraction of satellite galaxies in low mass host haloes for TNG300 with respect to SDSS \citep{2020arXiv200800004D}. This problem persists in TNG100 where the model was run at target resolution; the model is for instance 75\,\% above the red lensing data ($\approx 56\,\%$ above the upper error bar) at $r \approx 0.6\,h^{-1}\,\text{Mpc}$ for the $9.8 < \log_{10} M_\ast \left [ h^{-2}\,M_\odot \right ] < 10.2$ mass bin whereas TNG300 is 42\,\% above ($\approx 27\,\%$ above the upper error bar). Weakening the AGN feedback has the net effect for models with $0.5\,\alpha_\text{dyn.}$
of increasing the amplitude of the central bump on scales $r \approx
700\,h^{-1}\,\text{kpc}$ for the SAM derivatives, which means that the red lensing signal can
be used to constrain this combination, although it is sensitive to the
colour assignment scheme and dust model.

\begin{figure}
\includegraphics[width=1.02\columnwidth]{\figrelpath 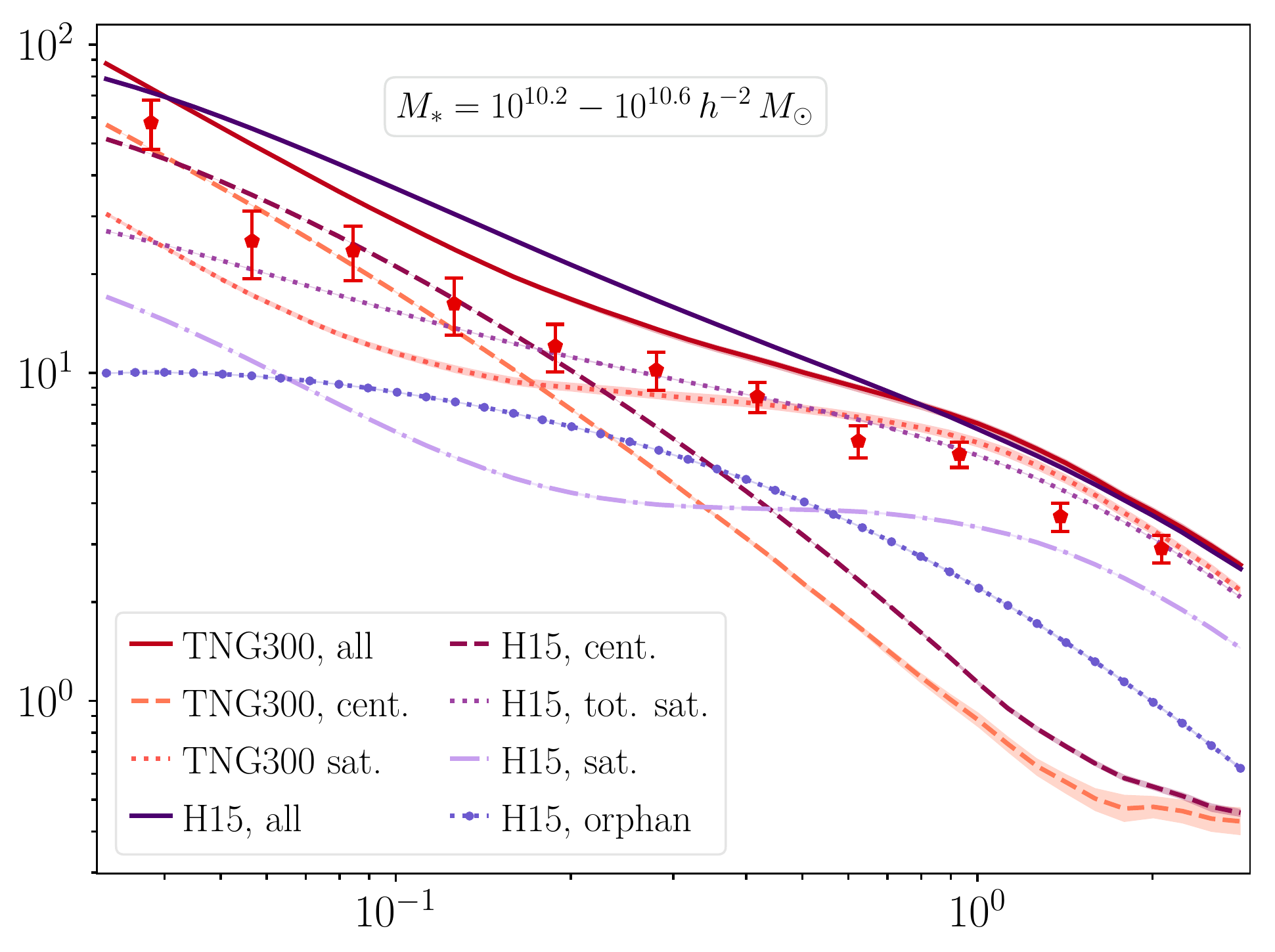}
\caption{Lensing predictions for red galaxies for stellar masses $10.2 < \log_{10} M_\ast \left [ h^{-2}\,M_\odot \right ] < 10.6$ compared to SDSS-DR7 using the \citet{2016MNRAS.457.4360Z} dataset, as in Fig.~\ref{fig:zmRedBlueGGLIntermediateMass} but with the different components shown for TNG300 ($f_\text{sat.} = 38\,\%$) and the \citetalias{2015MNRAS.451.2663H} SAM ($f_\text{sat.} = 40\,\%$). We can separate the satellite signal in the SAM into pure satellite (27\,\%) and orphan galaxy (13\,\%) contributions. In the outer region ($r \sim 1\,h^{-1}\,\text{Mpc}$), the data suggests a lower average host halo mass for satellite galaxies, whereas the inner data can point to lower central halo masses, satellites further out from the centre of the host haloes and/or lower host halo masses.}
\label{fig:zmRedGGLComponentSeparated}
\end{figure}

\begin{figure*}
\begin{centering}
\includegraphics[width=1.02\columnwidth]{\figrelpath 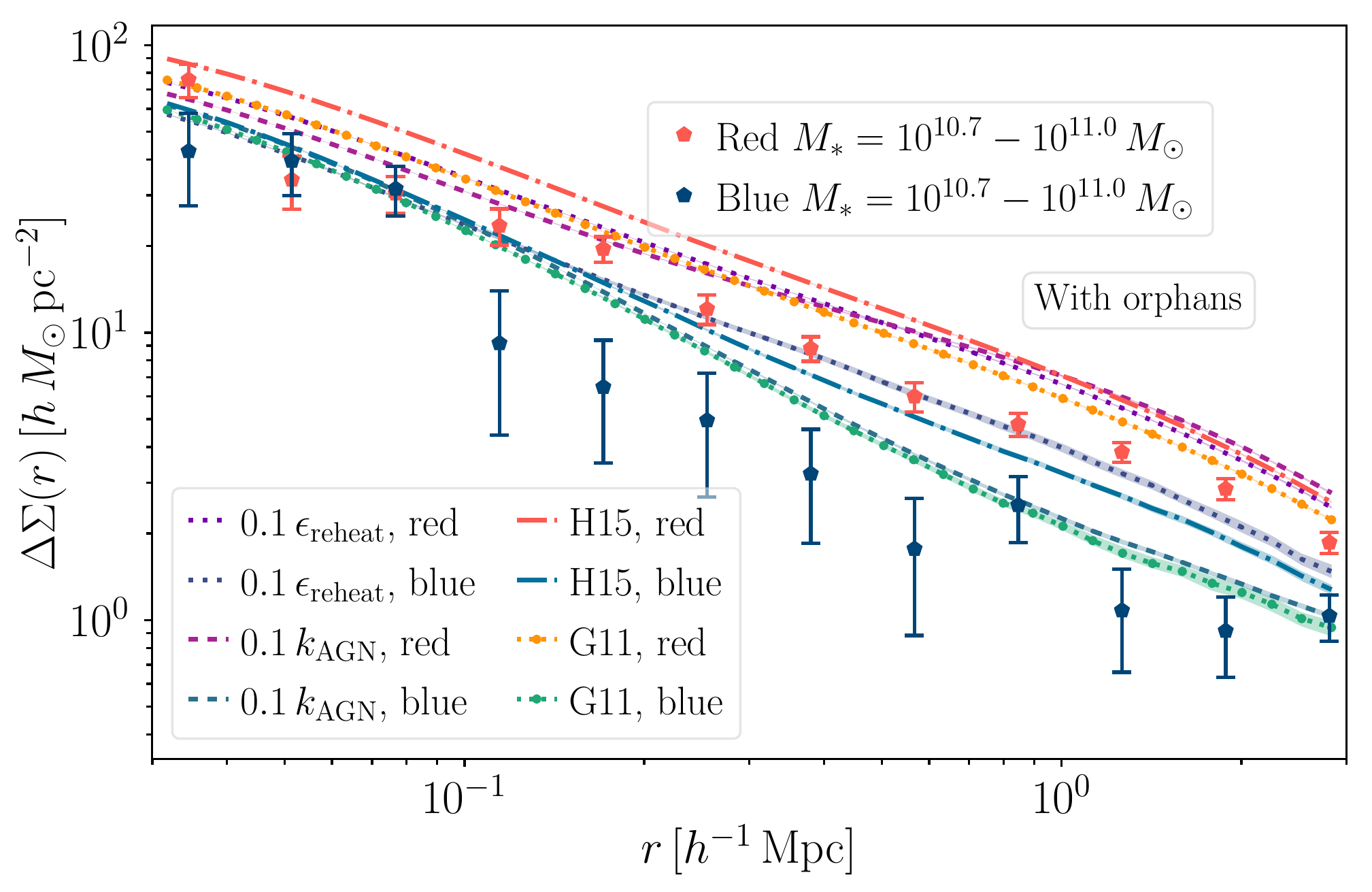}
\includegraphics[width=1.02\columnwidth]{\figrelpath 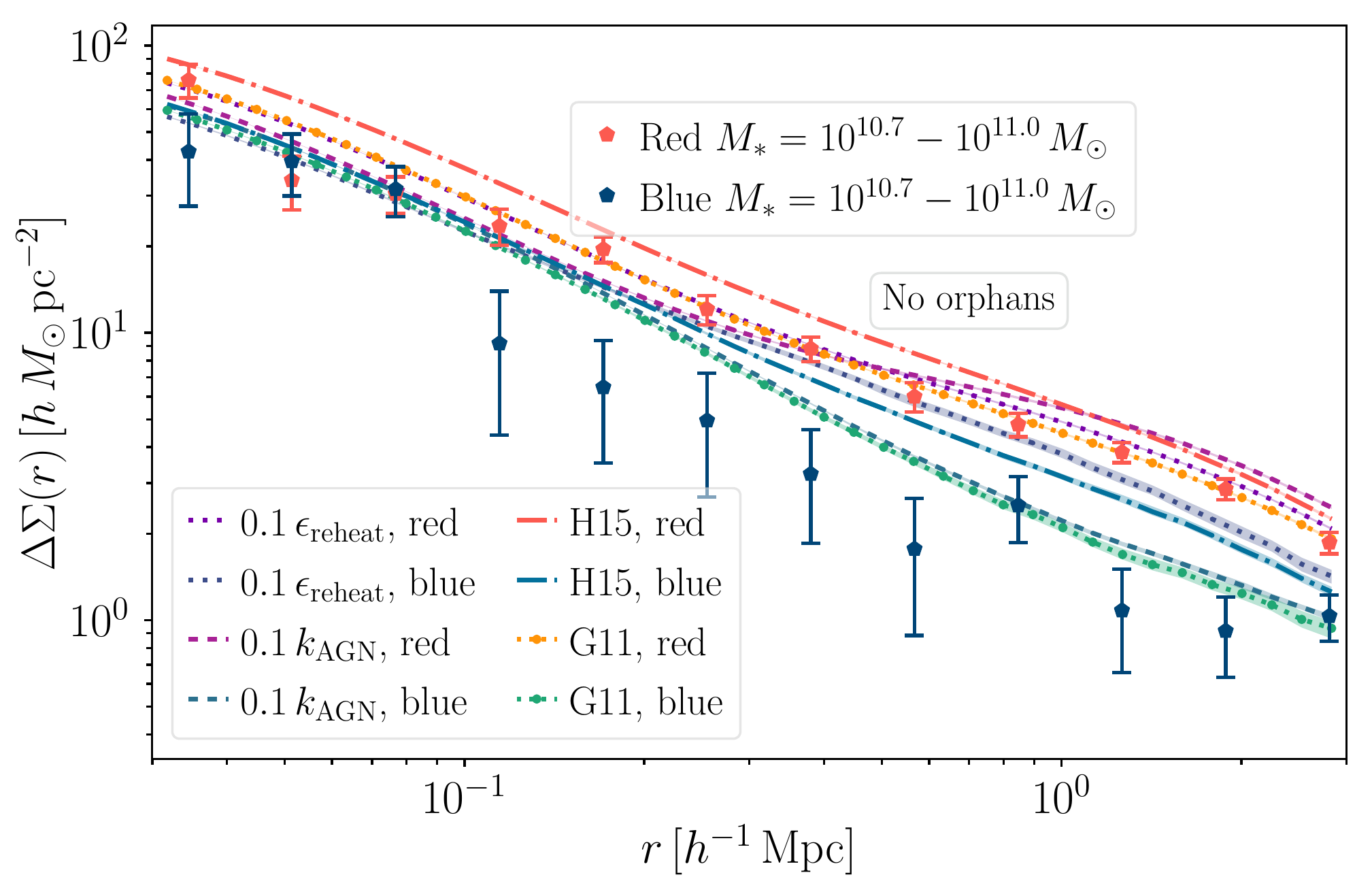}
\includegraphics[width=1.02\columnwidth]{\figrelpath 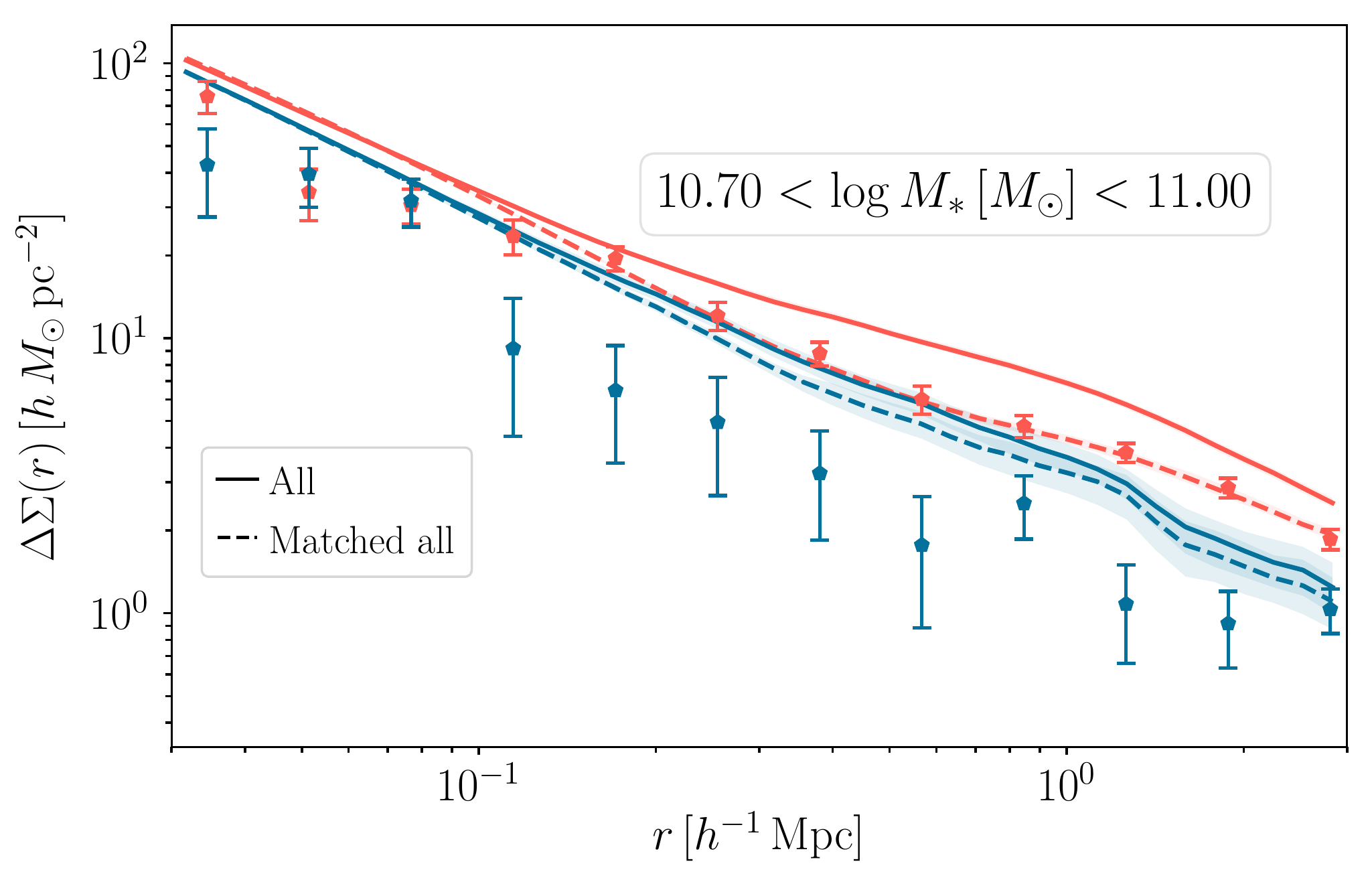}
\caption{Lensing predictions for all main SDSS red and blue galaxies with the same colour cut as for the LBGs with (\emph{left}) and without (\emph{right}) orphan galaxies (\emph{upper row}). If we consider the whole signal from SDSS \citep{2016MNRAS.457.3200M} there is little to no tension w.r.t. the \citet{2016MNRAS.457.4360Z} dataset for this mass range. Removing the orphan galaxies, which compose $\sim 10-15\,\%$ of the galaxies with a total satellite fraction of $\approx 35-45\,\%$, produces better agreement for the red lensing signal. In the lower row, we show lensing profiles from TNG300 for blue and red galaxies in SDSS for the same mass bin with the matched and total signal highlighted. If we restrict ourselves to matched subhaloes, effectively removing the structures corresponding to orphan galaxies in the SAM, the tension with respect to the red lensing signal drops. The red satellite fraction also drops from $f_\text{sat} = 0.34$ to $0.25$ with a central galaxy matching rate of 0.999 (red). The blue satellite lensing fraction also drops by a marginally smaller amount from 0.27 to 0.21 with a similar central matching rate 0.999, but the effect on the lensing signal is more modest.}
\label{fig:mSDSSRedBlueGGLIntermediateMass}
\end{centering}
\end{figure*}

Analysing the higher mass bins, we find that the SAMs and TNG300, as well as TNG100, systematically
overpredict the red lensing signal for $ 10^{10.2} \lesssim M_\ast \lesssim 10^{11.0} \,
M_\odot$, see example bins in Figs.~\ref{fig:zmRedBlueGGLIntermediateMass}, ~\ref{fig:zmRedGGLComponentSeparated} and~\ref{fig:mSDSSRedBlueGGLIntermediateMass}. For more
massive bins, the weak feedback models and 
TNG300 are once again in agreement with observations, reflecting the
results in Section~\ref{sec:mstarOnlyKiDSGAMA}. The results from TNG100 and TNG300 are similar across this stellar mass range, but the TNG100 profiles lie marginally below the data for $0.1 < r [h^{-1}\,\text{Mpc}] < 0.4$ for the red $11.0 < \log_{10} M_\ast [h^{-2}\,M_\odot] < 11.2$ mass bin; and marginally below the TNG300 profiles for the two most massive bins. Hence, we restrict parts of the analysis to TNG300 because of its greater volume. This red lensing signal tension exists for all SAMs, and none of the model variations listed in
Table~\ref{tab:samModelsParameters} produce acceptable solutions for
this intermediate mass range, with the \citetalias{Guo:2010ap} model performing the best. We also have issues with matching the blue lensing signals for $M_\ast > 10^{10.7} [M_\odot]$ in TNG300, but achieve acceptable results for some SAMs, such as the \citetalias{Guo:2010ap} model. The error bars are larger for the blue data than for the red, allowing for more model variation, and at the high mass end the analysis is obstructed by poor statistics. Hence, we primarily focus on the red signal. This red lensing signal mismatch can be caused by problems
matching the stellar masses in SDSS and enforcing the proper colour
separation, as we compare to Fig.~\ref{fig:vanUitert16ExtremeModels}
for the stellar mass-only sample from \citet{2016MNRAS.459.3251V} where both the
$0.1\,k_\text{AGN}$ and the $0.1\,\epsilon_\text{reheat}$ models are
able to match the lensing signal at the high mass end; and where all hydrodynamical simulations, except for Illustris, agree with the observations across the whole mass range (Fig.~\ref{fig:vanUitert16HydroComparison}). With respect to
the quoted satellite fractions for the samples listed in
\citet{2015MNRAS.454.1161Z}, the two fiducial SAMs are only a few
percent off\footnote{For the $10.2 < \log_{10} M_\ast \left [ h^{-2}\,
    M_\odot \right ] < 10.6$ bin, the quoted $f_\text{sat} = 0.37 \pm
  0.02$, and we measure $f_\text{sat} = 0.33$ and $f_\text{sat} = 0.34$
  for the \citetalias{Guo:2010ap} and \citetalias{2015MNRAS.451.2663H}
  models. The reported average host halo mass is $\expVal{\log_{10}
    M_\text{h}} = 12.15 (+ 0.03) (- 0.04) \left [ h^{-1}\, M_\odot
    \right ]$ and we find $12.16$ and $12.29$, respectively. In
  \citet{2016MNRAS.457.4360Z}, a red fraction $f_\text{red} = 0.71$
  for this mass bin is given, whereas we find $f_\text{red} = 0.87$
  and $f_\text{red} = 0.77$. Hence, we have more red galaxies, but for
  the \citetalias{2015MNRAS.451.2663H} model the difference should be
  negligible.}. In addition, the average host halo masses only differ
by about 0.1\,dex. These differences are too small to drive the large
biases we observe. Although we have no information on the satellite fractions from the observations separated according to colour, we can perform the signal decomposition in our models. This is shown in Fig.~\ref{fig:zmRedGGLComponentSeparated} for the red lensing signal compared to \citet{2016MNRAS.457.4360Z} data for TNG300 and the \citetalias{2015MNRAS.451.2663H} model. Both models suggest similar satellite fractions ($f_\text{sat} = 38\,\%$ for TNG300 and $f_\text{sat} = 40\,\%$, with an orphan fraction $f_\text{orphan} = 13\,\%$ for \citetalias{2015MNRAS.451.2663H}). In the outer region at $r \sim 1\,h^{-1}\,\text{Mpc}$, the satellite lensing signal constitute the bulk of the total signal and is similar between the two models (the predicted share is even larger for lower mass bins in this regime). If it is decreased a better agreement with the observations could be reached, assuming the model satellite fractions reflect the observations. More centrally at $r \sim 100\,h^{-1}\,\text{kpc}$, a combination of lowering the host halo masses for the satellites and lowering the central host halo masses could yield a better match.

We have computed estimates
for how much the dust extinction affects the signal amplitude for the two
fiducial SAMs for the \citet{2016MNRAS.457.4360Z} selection
function. At the high mass end for $M_\ast > 10^{11}\,h^{-2}\,M_\odot$
for the \citetalias{Guo:2010ap} model there are only small differences
for the red signal with and without dust whereas the dusty red signal
is suppressed for all masses for \citetalias{2015MNRAS.451.2663H} with
at most $\approx 15\,\%$ for the $10.6 < \log_{10} M_\ast \left [
  h^{-2}\,M_\odot \right ] < 11.0$ mass bin, closely followed by the
effects for the $10.2 < \log_{10} M_\ast \left [ h^{-2}\,M_\odot \right ] <
10.6$ mass bin. Not surprisingly, the dust correction thus works in the
opposite direction to reconcile the tension for the red lensing
signal. For the blue signal, the dust extinction boosts the
predictions by about a factor of 2 and 1.5 for the most massive bins
where there are many red galaxies and few blue, with smaller effects
for lower masses. For low mass systems below
$10^{10.2}\,h^{-2}\,M_\odot$ in the \citetalias{2015MNRAS.451.2663H}
model, there is a suppression for the central bump by about $\sim
15\,\%$ in the dust extinct signal. We attribute this effect to dusty
blue galaxies residing in less massive haloes, which are able to keep
more dust than their massive counterparts
\citep[e.g.][]{2013MNRAS.432.2298B, 2017MNRAS.471.3152P}, and thus a lower central signal.

We have also performed a crude comparison between the \citet{2016MNRAS.459.3251V} dataset and the \citet{2016MNRAS.457.4360Z} dataset using their quoted red fractions and they agree very well in the overlapping mass range, except possibly for the $9.4 < \log_{10} M_\ast \, [h^{-2}\,M_\odot] < 9.8$ mass bin where the \citet{2016MNRAS.457.4360Z} data points lie higher by approximately a factor of two for $0.3 < r\, [h^{-1}\,\text{Mpc}] < 1$, suggesting that the local lensing signal is fairly constant.

We have compared the stellar mass-only lensing predictions from TNG300 at $z = 0.3$ and $z = 0$ to see if we could gain some insights. The $z = 0$ signal is boosted with respect to the $z =0.3$ signal, but the boost is not universally radially. We compared the boosts at $r = 60\,h^{-1}\,\text{kpc}$ and at $r = 0.5\,h^{-1}\,\text{Mpc}$, and while the central boosts lie around $\sim 15-20\,\%$, the outer boost lies at around $30\,\%$ for $M_\ast < 10^{10.6}\,[h^{-2}\,M_\odot]$ (for the lowest mass bin, the $z = 0$ signal increases by $45\,\%$ in the outer range). For $M_\ast > 10^{10.6}\,[h^{-2}\,M_\odot]$, the boost is more universal across the whole radial range ($r < 2-3\,h^{-1}\,\text{Mpc}$) at $\sim 20\,\%$. As the outer boost is measured in the regime where the central host halo contribution term appears for the satellite lensing signal, the model might overestimate the evolution of this term which does not appear to evolve in the data. A redshift boost in general seems to be disfavoured by the data, except for the least massive bin, where TNG300 predicts a similar evolution as hinted at in our crude data comparison. If we decompose the signal into central and satellite contributions (and assume that the satellite fractions agree with the data sample), we see that the satellite signal evolves towards $z = 0$ and dominates the total signal in this radial range $r \sim 1\,h^{-1}\,\text{Mpc}$.

A connected interesting observation is shown in
Fig.~\ref{fig:mSDSSRedBlueGGLIntermediateMass} where we plot the predicted signals from
the SAMs and TNG300 with and without orphan galaxies and with and without unmatched
subhaloes, respectively, with respect to lensing observations from the all main
SDSS-DR7 sample \citep{2016MNRAS.457.3200M}. By removing the orphan galaxies in the SAMs, the
tension for the red galaxies is reduced and the
corresponding satellite fraction drops by about $10\,\%$. If we
just examine the orphan galaxy signal, see Fig.~\ref{fig:zmRedGGLComponentSeparated}, we find that it is similar to a
massive central term as the orphans reside close to the halo
centres. At lower masses the large abundance of low mass haloes
hosting central galaxies offsets the imprint of this signal and
gives agreement with observations. We find a corresponding effect
for TNG300 for the same observations when we remove all subhaloes
which lack a match in the gravity-only run and compare the lensing
signal to the full physics predictions. If we restrict ourselves to
matched substructures, much better agreement with data is
obtained. We note that the satellite fractions are comparable for the
red and blue signals, implying that the colour of a satellite galaxy
is not a good predictor for the likelihood of its host substructure to
still be present in the gravity-only run. Restricting the signal
to matched substructures has thus the effect of reducing the satellite
fraction by a similar amount for the blue and the red signal, although
the impact on the red lensing signal is more considerable as the
amplitude of the central host halo term drops for the satellite
signal. This can be caused by substructures merging and getting
disrupted more quickly in more massive host haloes, where galaxies on
average are redder and objects are excluded to a higher degree by the matching criterion. If
we look at more massive red galaxies, TNG300 is in agreement with
observations for $M_\ast > 10^{11}  M_\odot$, also
for scales around $r \approx 1\,h^{-1}\text{Mpc}$. For these masses the
signal is dominated by centrals, which are well-matched as we shall
see in the following Section.

Our tentative conclusion is that galaxy formation recipes, both SAMs and hydrodynamical simulations, which preferentially place quenched, red satellites galaxies in massive host haloes have to be redefined such that the lensing signals are matched. For TNG300, this seems to work fine at $z =0.3$, but the model suggests a redshift evolution which does not seem to appear in the data. This better matching could be achieved by, for instance, strengthening the gas stripping of satellites in group-scale haloes. Hence empirical models such as HODs still outperform physical modelling for this type of observational dataset.

\subsection{Lensing of locally brightest galaxies}\label{sec:lbgLensing}

By limiting our selection to LBGs, which by construction are mostly central galaxies, the predicted lensing signals drop
and are more compatible with the data for all models, both for the SAMs and TNG300. This has been established in previous studies \citep[see e.g.][]{2016MNRAS.456.2301W, 2016MNRAS.457.3200M}. Hydrodynamical simulations have not been compared to this type of data, and we are interested to see if the SAM variations which were preferred by the stellar mass-only lensing KiDS+GAMA observations in Section~\ref{sec:mstarOnlyKiDSGAMALG} also perform well against these datasets. Here we first study the predictions regardless of galaxy colour and then split the signals into red and blue, starting with the SAMs and then proceeding to the TNG300 results.

\subsubsection{LBG lensing for the SAMs}\label{sec:lbgLensingSAMs}

\begin{figure*}
\begin{centering}
\includegraphics[width=1.02\columnwidth]{\figrelpath 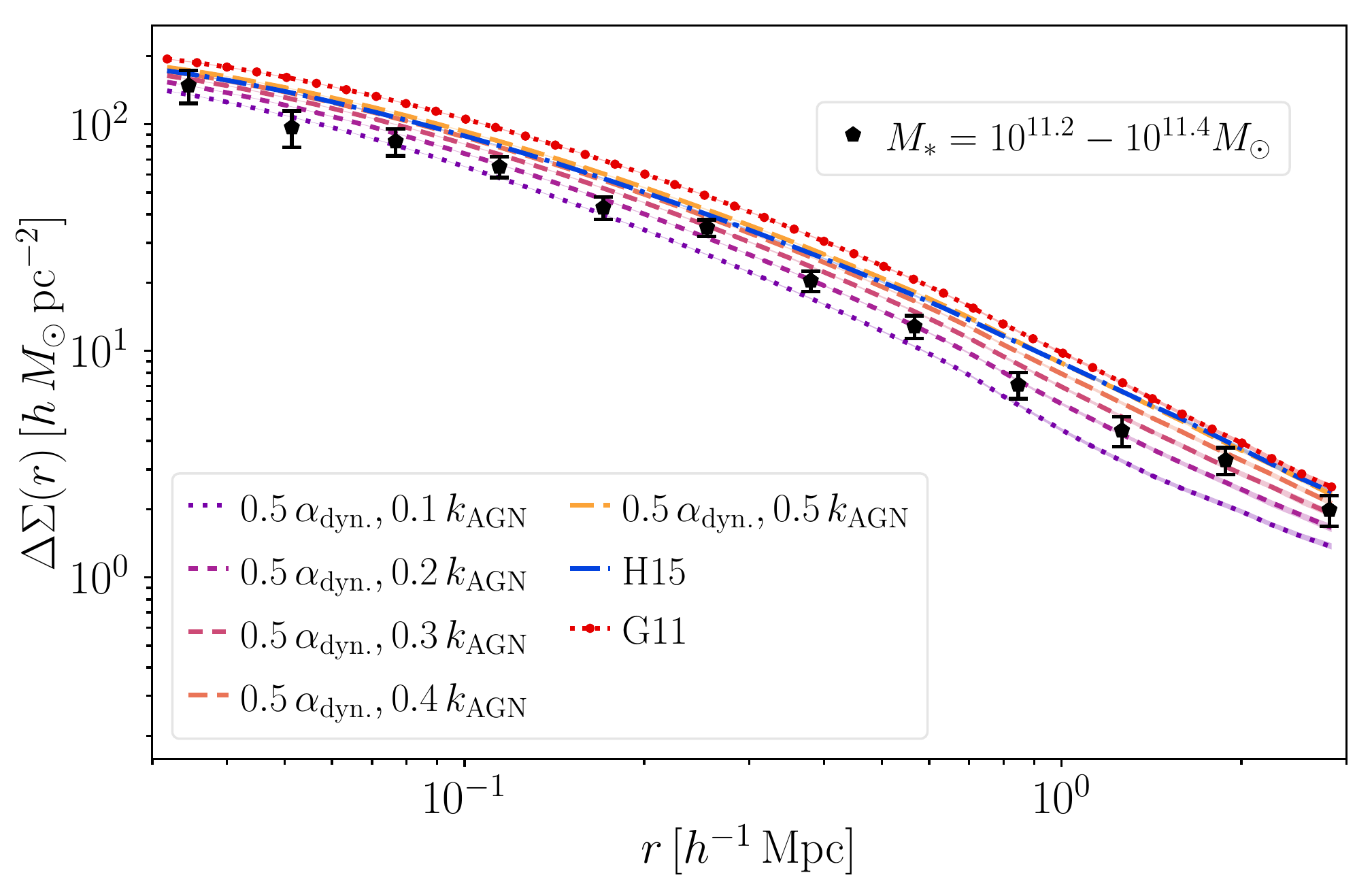}
\includegraphics[width=1.02\columnwidth]{\figrelpath 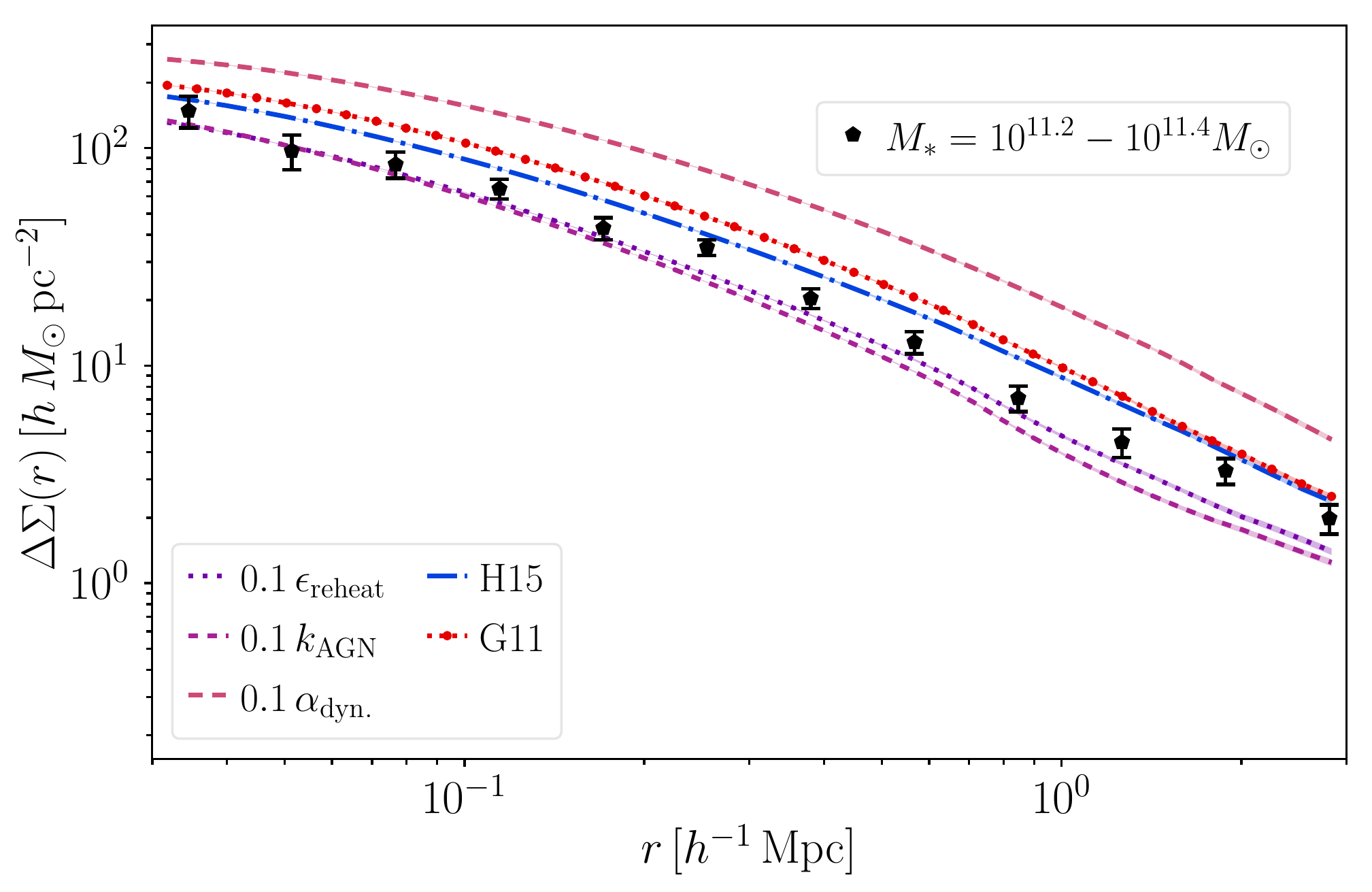}
\includegraphics[width=1.02\columnwidth]{\figrelpath 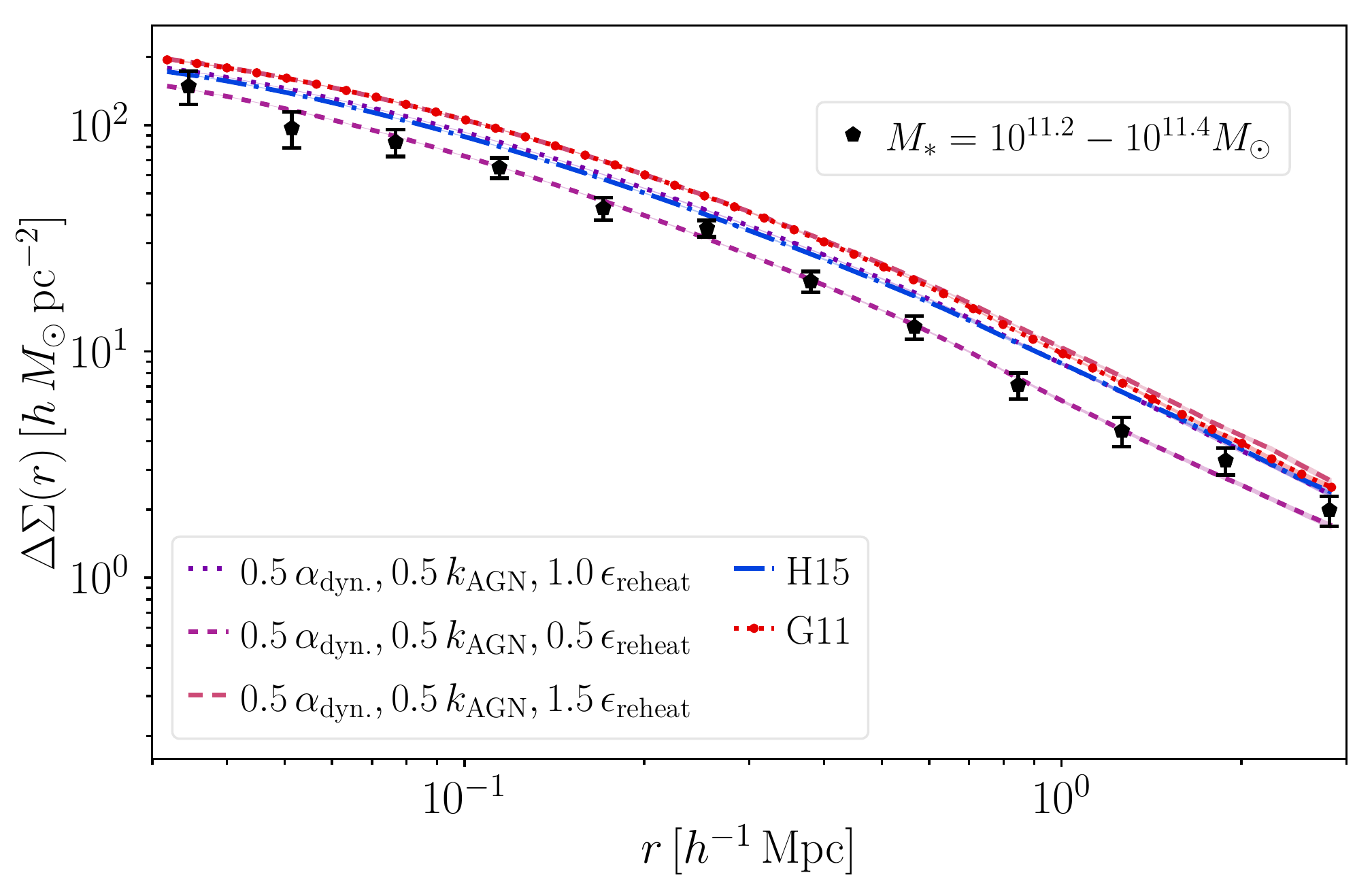}
\includegraphics[width=1.02\columnwidth]{\figrelpath 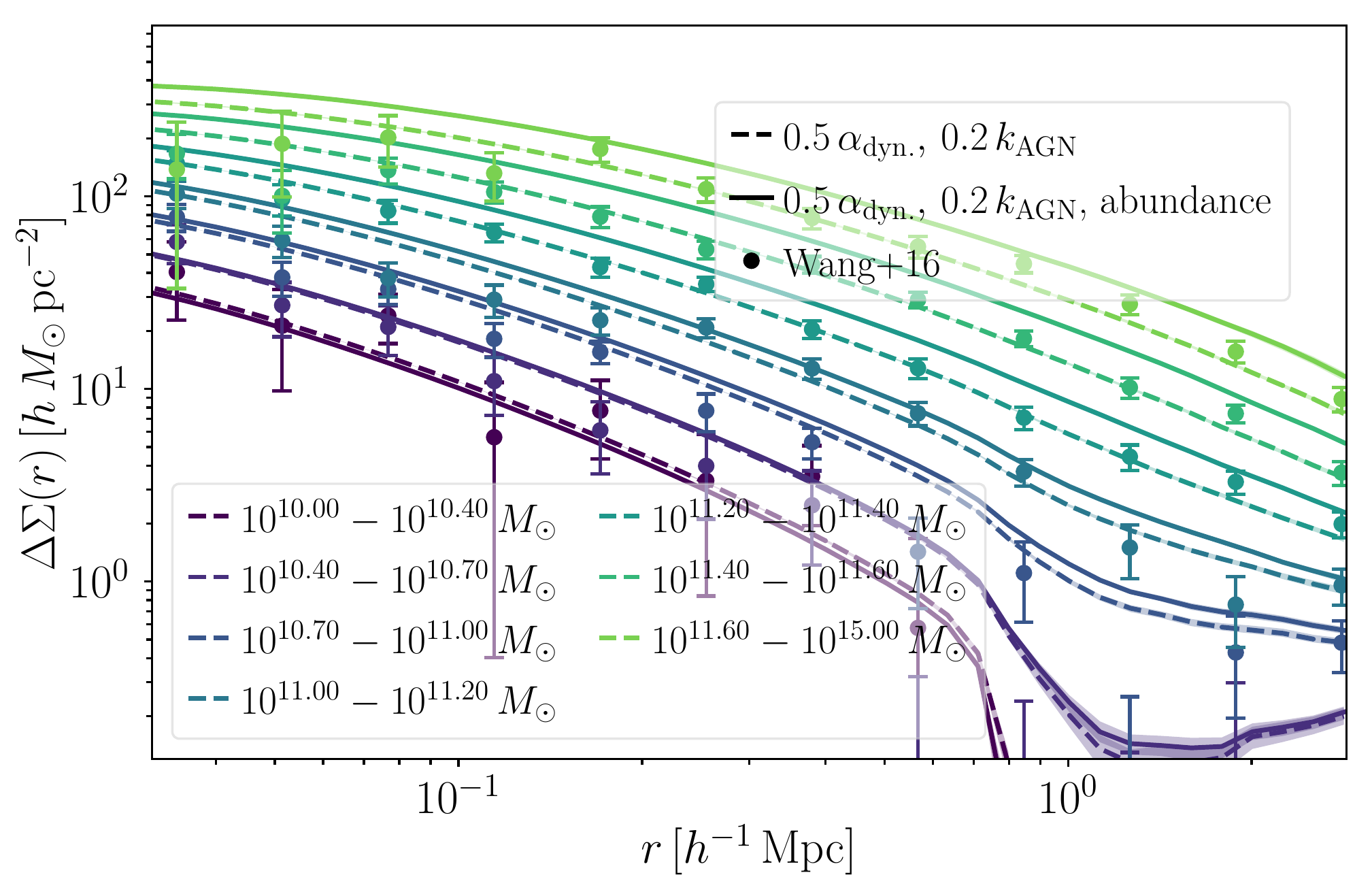}
\caption{Predicted GGL signals compared to observations from SDSS LBGs with data from \citet{2016MNRAS.456.2301W}. We show the effect of changing the $k_\text{AGN}$ strength (\emph{upper left}), where the effect is modest to none for intermediate masses and where it starts to have an effect on high mass systems. The predictions from the extreme models are ruled out by the LBG signal at the high mass end (\emph{upper right}); they decrease the signal more than what the observations allow. We are also able to produce reasonable agreements by reducing $\epsilon_\text{reheat}$ and $k_\text{AGN}$ at the same time (\emph{lower left}). Furthermore, we show the results for our best model $\left (0.5\,\alpha_{\mathrm{dyn.}}, \, 0.2\,k_{\mathrm{AGN}} \right )$ with the lowest figure-of-merit (\emph{lower right}) and we discern that the predictions are in excellent agreement with observations.}
\label{fig:lbgAllSDSS}
\end{centering}
\end{figure*}

{\renewcommand{\arraystretch}{1.4}
\tabcolsep=2pt\relax
\begin{table*}
	{\centering
	\begin{tabular*}{\textwidth}{l l l l l l l}
	\hline
	\footnotesize
		{\bf LBG (fiducial)}& {\bf All} & $\chi^2$ & {\bf Red} & $\chi^2$ & {\bf Blue} & $\chi^2$\\
		\hline
First & $\left (0.5\,\alpha_{\mathrm{dyn.}}, \, 0.2\,k_{\mathrm{AGN}} \right )$ & 1.59 & $\left (0.5\,\alpha_{\mathrm{dyn.}}, \, 0.5\,k_{\mathrm{AGN}}, \, 0.5 \,\epsilon_{\mathrm{reheat}} \right )$ & 1.12 & $\left (0.3\,\alpha_{\mathrm{dyn.}}, \, 0.1\,k_{\mathrm{AGN}} \right )$ & 1.67\\
		Second & $\left (0.1\,\alpha_{\mathrm{dyn.}}, \, 0.1\,k_{\mathrm{AGN}} \right )$ & 1.67 &$\left (0.5\,\alpha_{\mathrm{dyn.}}, \, 0.2\,k_{\mathrm{AGN}} \right )$ & 1.23 & $\left (0.5\,\alpha_{\mathrm{dyn.}}, \, 0.1\,k_{\mathrm{AGN}} \right )$ & 1.76 \\
		Third & $\left (0.4\,\alpha_{\mathrm{dyn.}}, \, 0.2\,k_{\mathrm{AGN}} \right )$ & 1.67 & $\left (0.4\,\alpha_{\mathrm{dyn.}}, \, 0.2\,k_{\mathrm{AGN}} \right )$ & 1.23 & $\left (0.4\,\alpha_{\mathrm{dyn.}}, \, 0.1\,k_{\mathrm{AGN}} \right )$ & 1.76 \\
		\hline
		{\bf LBG (abundance)}& {\bf All} & $\chi^2$ & {\bf Red} & $\chi^2$ & {\bf Blue} & $\chi^2$\\
		\hline
		First & \citetalias{2015MNRAS.451.2663H} & 2.89 & $0.1\,\epsilon_{\mathrm{reheat}}$ & 2.16 & $\left (0.5\,k_\text{AGN}, \, 0.5\, \alpha_\text{dyn}, \,1.5\, \epsilon_\text{reheat} \right )$ & 1.75\\
		Second &  $\left (0.5\, k_\text{AGN}, \, 0.5\, \alpha_\text{dyn}, \,0.5\, \epsilon_\text{reheat} \right )$ & 3.40 & \citetalias{2015MNRAS.451.2663H} & 2.36 & $\left (0.5\,k_\text{AGN}, \, 0.5\, \alpha_\text{dyn} \right )$ & 1.87\\
		Third & $0.1\,\epsilon_{\mathrm{reheat}}$ & 3.63 & $\left (0.5\,k_\text{AGN}, \, 0.5\, \alpha_\text{dyn}, \,0.5\, \epsilon_\text{reheat} \right )$ & 2.36  &  \citetalias{2015MNRAS.451.2663H} & 1.90\\
		\hline
	\end{tabular*}}
	\caption{The best fit SAM models for the LBG lensing predictions without and with abundance corrected masses. For the total LBG signal, the $ (0.5\,\alpha_{\mathrm{dyn.}}, \, 0.2\,k_{\mathrm{AGN}})$ model is the best and it also does reasonably well for the red signal. Performing the analysis with abundance corrections favours the \citetalias{2015MNRAS.451.2663H} and $\left (0.5\, k_\text{AGN}, \, 0.5\, \alpha_\text{dyn}, \,0.5\, \epsilon_\text{reheat} \right )$ models.}
	\label{tab:fomLensingLBGStellarMassOnly}
\end{table*}}

\begin{figure*}
\begin{centering}
\includegraphics[width=1.02\columnwidth]{\figrelpath 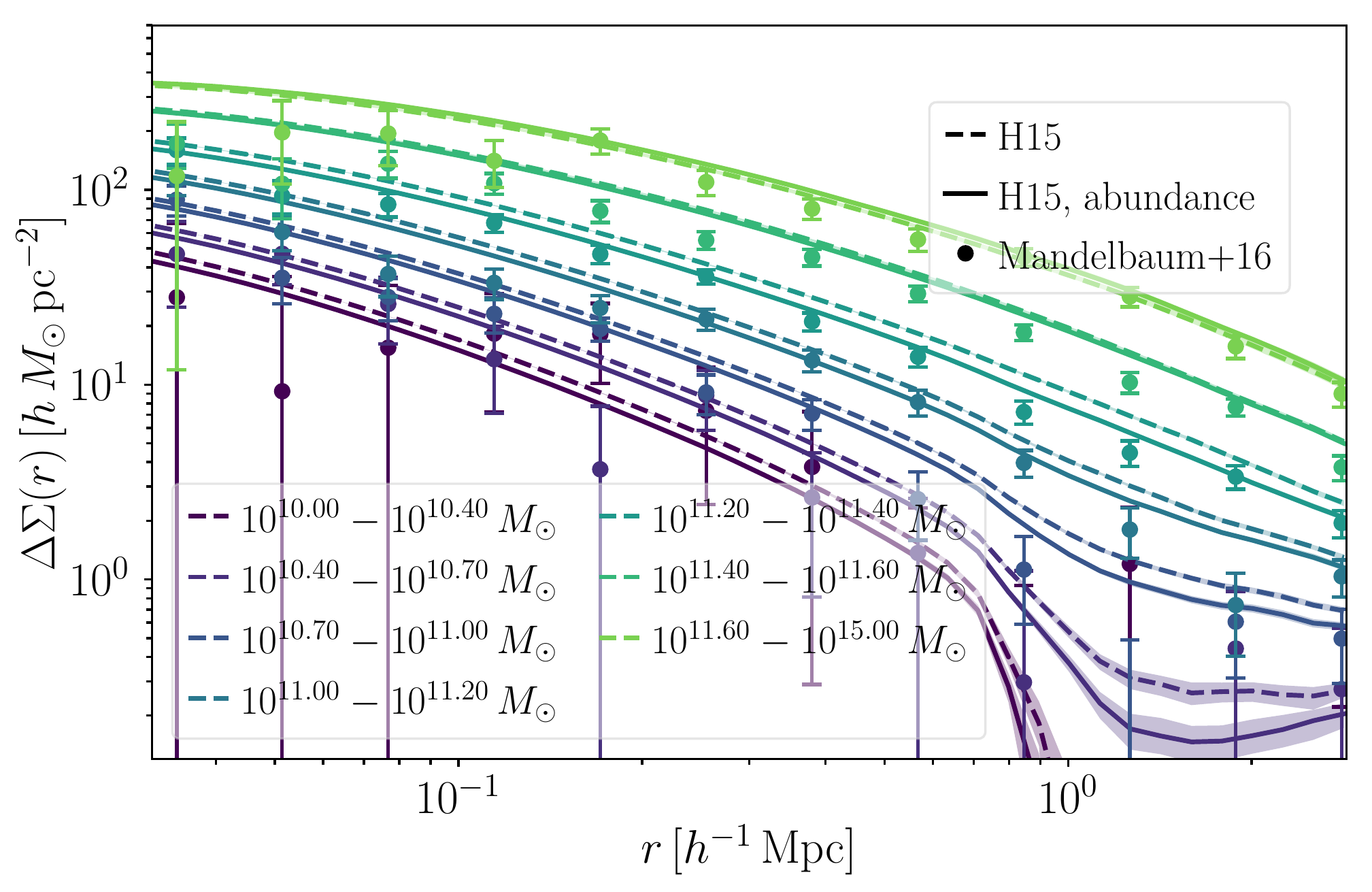}
\includegraphics[width=1.02\columnwidth]{\figrelpath 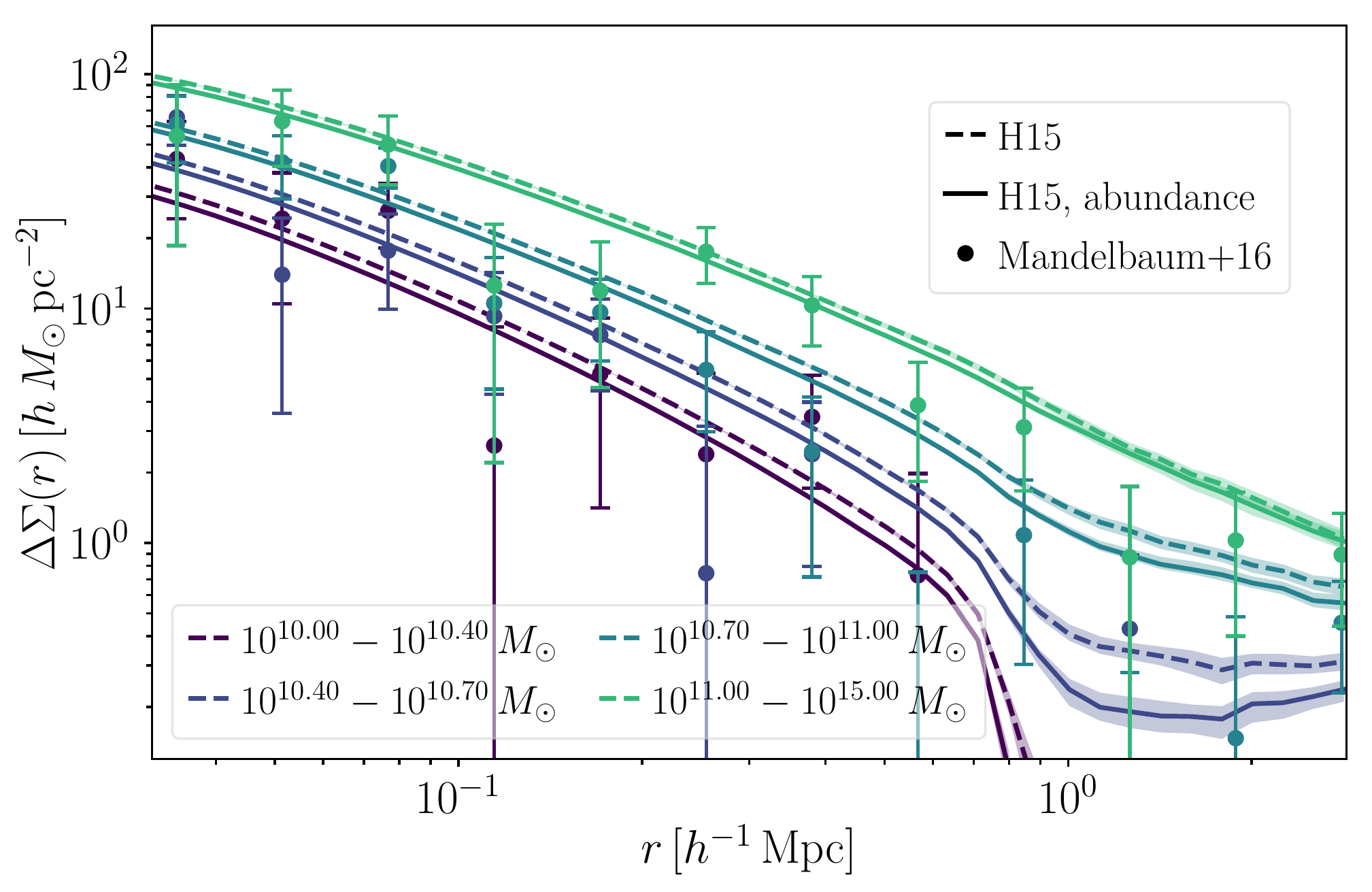}
\includegraphics[width=1.02\columnwidth]{\figrelpath 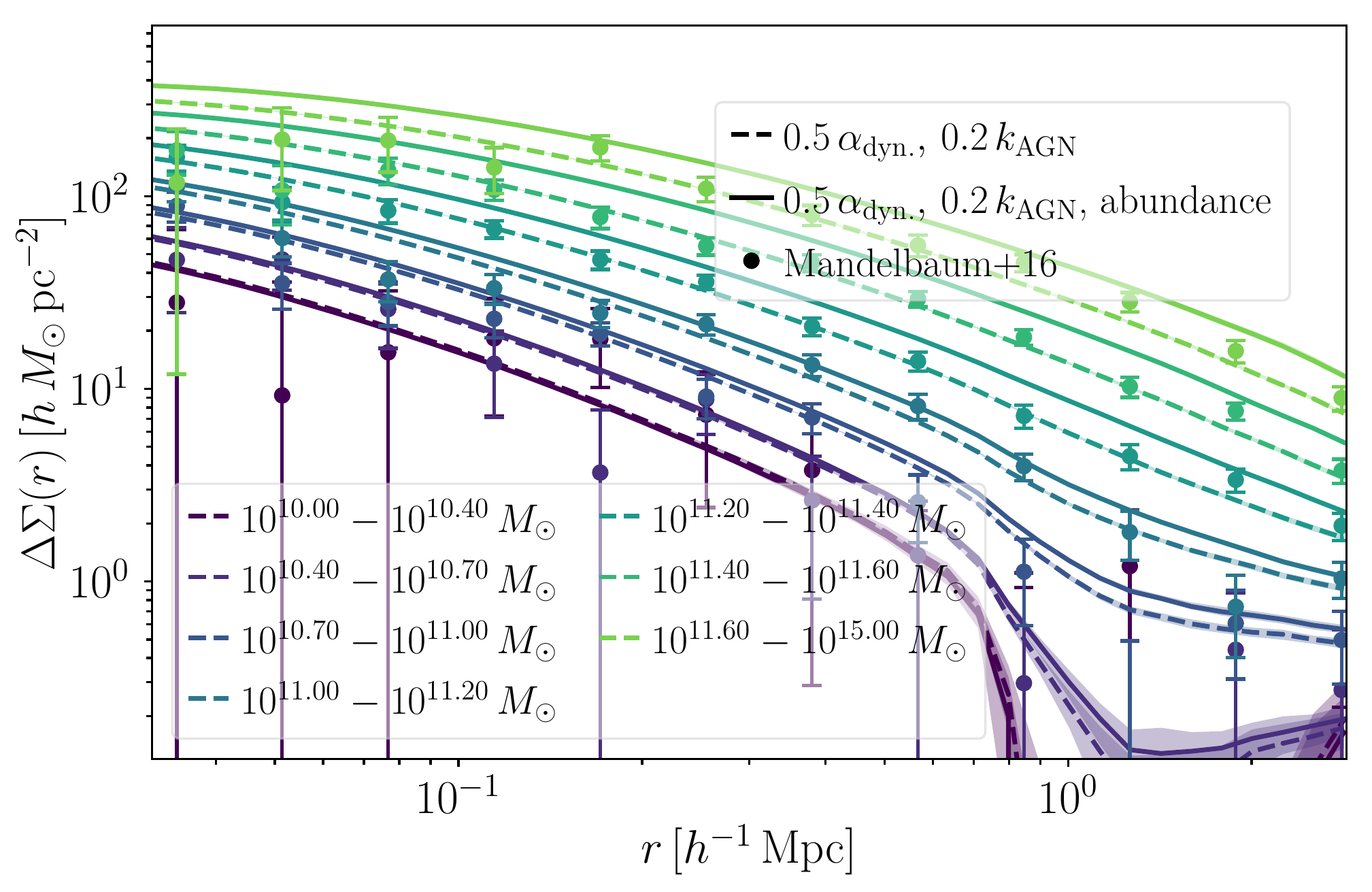}
\includegraphics[width=1.02\columnwidth]{\figrelpath 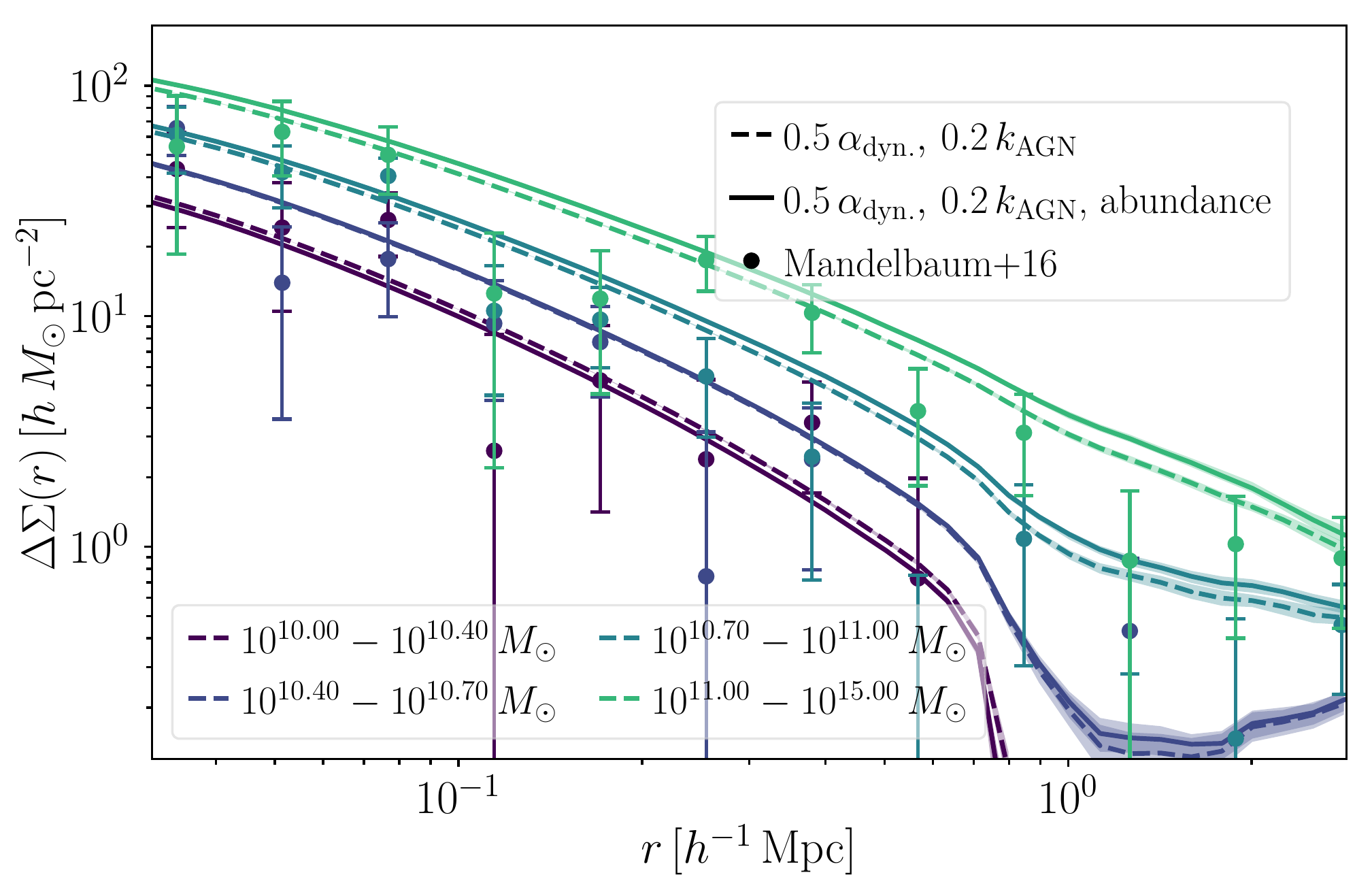}
\caption{Similar to Fig.~\ref{fig:lbgAllSDSS} but for LBGs separated according to colour and compared to the \citet{2016MNRAS.457.3200M} observations. In the upper row, we show how the \citetalias{2015MNRAS.451.2663H} model performs. The predicted signal for red galaxies ($\chi^2 = 4.00$) (\emph{upper left}) is excessive for masses $10^{11}\,M_\odot$, but the abundance correction mitigates the tension and the signal for blue galaxies ($\chi^2 = 2.43$) (\emph{upper right}) conforms with observations. In the bottom row we plot the predictions for the best fit model for all LBGs, $ (0.5\,\alpha_{\mathrm{dyn.}}, \, 0.2\,k_{\mathrm{AGN}})$, for red ($\chi^2 = 1.23$) and blue ($\chi^2 = 1.93$) LBGs, respectively, for comparison. The biggest improvement w.r.t. \citetalias{2015MNRAS.451.2663H} model is for the signal around massive red LBGs (\emph{lower left}).}
\label{fig:lbgColorSDSS}
\end{centering}
\end{figure*}

In Figs.~\ref{fig:lbgAllSDSS} and~\ref{fig:lbgColorSDSS}, we show the
SAM LBG lensing results from an assorted model collection with stellar mass and stellar mass + colour selection functions, respectively. In the first three figures in Fig.~\ref{fig:lbgAllSDSS}, we show how different model variations affect the signal for one mass bin with $10^{11.2} < M_\ast [M_\odot] < 10^{11.4}$ and in the last, the predictions from our best fit model across all mass bins. Number-wise, we have roughly $\sim 300\,000$ galaxies in the least
massive bin per axis for the SAMs run on the rescaled MR and $\sim 20\,000$ systems in TNG300 for the stellar mass only selection. We stress that the drop in the signal for the lowest mass bins around $r \approx 1\,h^{-1}\,\text{Mpc}$ is a consequence of the LBG selection function. This is less of a problem for the SAMs run on the rescaled MR simulation due to the improved statistical averages. We find that the predictions from the
\citetalias{2015MNRAS.451.2663H} model tend to agree better with
observations than the \citetalias{Guo:2010ap} curves, especially for
$M_\ast > 10^{11.2} M_\odot$, as seen in
Fig.~\ref{fig:lbgAllSDSS}. We are able to reproduce the results in \citet{2016MNRAS.456.2301W} by running
the \citetalias{2015MNRAS.451.2663H} model with the
\citetalias{Guo:2010ap} parameter inputs, which features a couple of improvements 
from the fiducial version published in \citetalias{Guo:2010ap}. This hybrid-model finishes on tenth place for the LBG lensing ($\chi^2 = 2.64$), whereas the \citetalias{Guo:2010ap} model ($\chi^2 = 7.84$) and the \citetalias{2015MNRAS.451.2663H} model ($\chi^2 = 5.01$) do considerably worse.

For intermediate stellar masses, fixing $\alpha_\text{dyn}$ and
varying $k_\text{AGN}$ has little to no effect on the profiles except for the transition
regime between the 1-halo and 2-halo terms at $r \sim 1\,h^{-1}\text{Mpc}$ where a weaker
$k_\text{AGN}$ yields a lower signal. Still, the variance of the
observations is quite large for these scales for low stellar
masses. If we move to higher stellar masses beyond the knee, the
different feedback prescriptions start to have an effect, see Fig.~\ref{fig:lbgAllSDSS}. We have excluded that this result is contaminated by the presence of
satellites and orphan galaxies in the sample, owing to the high central purity of
the signals ($\approx 85-95\,\%$ depending on the stellar mass bin and the examined model, lowest at $M_\ast = 10^{11}\,M_\odot$, and similarly for TNG300 with the lowest purity at $89\,\%$ at approximately the same mass).

For the stellar mass only selection, setting $k_\text{AGN} = 0.1 \,
k_\text{AGN}^\text{fid.}$ and $0.1\,\epsilon_\text{reheat}$ solves
the tension for group scale lenses, although the produced signals are
too low for $M_\ast > 10^{11}\, M_\odot$ systems -- see the upper right subfigure of
Fig.~\ref{fig:lbgAllSDSS}. For intermediate and high masses,
simultaneously reducing $k_\text{AGN}$ and $\epsilon_\text{reheat}$
improves the agreement as seen in the lower left figure, although there is still
tension for LBGs with $M_\ast < 10^{11}\, M_\odot$. Hence, this model
class is disfavoured by these lensing observations as we use all
stellar mass bins to construct our model ranking.

In Table~\ref{tab:fomLensingLBGStellarMassOnly}, we list the best
ranked models for the LBG sample with and without abundance matching
stellar mass corrections. Similarly as for the stellar mass-only sample,
the lensing data prefer a low AGN feedback efficiency, although here
the intermediate $ (0.5\,\alpha_{\mathrm{dyn.}}, \,
0.2\,k_{\mathrm{AGN}} )$ model is the best. We infer that this
shift is caused by the investigation of the signals from central-dominated samples, where the $ (0.5\,\alpha_{\mathrm{dyn.}}, \,
0.2\,k_{\mathrm{AGN}} )$ model produces fewer galaxies, but
they are also more isolated due to the shorter merger timescale. At
second place, we find the $ (0.1\,\alpha_{\mathrm{dyn.}}, \,
0.1\,k_{\mathrm{AGN}} )$ model, which also has more isolated
centrals due to the low $\alpha_{\mathrm{dyn.}}$. The signals for these two models are slightly elevated with respect to the data beyond $10^{10.79}\,h^{-2}\,M_\odot$ for the stellar mass
only selection. For the \citet{2016MNRAS.459.3251V} comparison these two models are thus only ranked
seven and six, respectively. The results after the abundance correction are
similar to the stellar mass-only lensing comparison with the fiducial
\citetalias{2015MNRAS.451.2663H} model with the best performance
followed by the low SN feedback efficiencies. We show the results for
the best model in the lower right figure in Fig.~\ref{fig:lbgAllSDSS}.

Separating the signal into red and blue according to
\eqref{eq:mandelbaum2016ColorCut} for the \citet{2016MNRAS.457.3200M}
dataset comparison, the $0.1\, k_\text{AGN}$ and $0.1\,
\epsilon_\text{reheat}$ solutions are again ruled out by the red signal
from systems with $M_\ast > 10^{11.2}\, M_\odot$ for $r \gtrsim 0.2\,h^{-1}\,\text{Mpc}$. We observe that
weakening the AGN feedback efficiency reduces the host halo bimodality. In addition, we note
that the \citetalias{2015MNRAS.451.2663H} model (upper row in Fig.~\ref{fig:lbgColorSDSS}) in general predicts a
stronger bimodality than the \citetalias{Guo:2010ap} model, and that
the former is not plagued by a tension with data for the blue LBG
lensing signal at the high mass end which was shown in \citet{2016MNRAS.457.3200M}. This
holds true for the red signal but to a smaller extent. Similarly as for the total signal, the $
(0.5\,\alpha_{\mathrm{dyn.}}, \, 0.2\,k_{\mathrm{AGN}} )$ model
does an excellent job where the results are shown in the lower row of Fig.~\ref{fig:lbgColorSDSS}. The fiducial \citetalias{2015MNRAS.451.2663H} model has $\chi^2 = 4.00$ (red LBGs) and $\chi^2 = 2.43$ (blue LBGs), which is worse than the \citetalias{Guo:2010ap} parameter values on top of this model which produces $\chi^2 = 2.25$ and $\chi^2 = 1.67$, respectively.

Also in Table~\ref{tab:fomLensingLBGStellarMassOnly} we list the best
fit models for red and blue LBGs with and without abundance corrected
masses. Compared to the stellar mass only selection, the $
(0.5\,\alpha_{\mathrm{dyn.}}, \, 0.5\,k_{\mathrm{AGN}}, \, 0.5
\,\epsilon_{\mathrm{reheat}})$ model now performs best,
followed by the $ (0.5\,\alpha_{\mathrm{dyn.}}, \,
0.2\,k_{\mathrm{AGN}})$ model. For the total LBG signal, this model
finishes in fourth place (with $\chi^2 = 1.81$), so there is reasonable concordance. If we switch to the
blue signal, we see a shift in preference towards models with short
$\alpha_{\mathrm{dyn.}}$ and weak $k_\text{AGN}$, with the biggest
gains on scales $r \sim 400\,h^{-1}\,\text{kpc}$ and outwards for
stellar masses $M_\ast > 10^{10.7}\,M_\odot$ w.r.t. the $
(0.5\,\alpha_{\mathrm{dyn.}}, \, 0.2\,k_{\mathrm{AGN}} )$
model, which is the sixth best with $\chi^2 = 1.93$. Still, the uncertainties in this
signal region are quite large, and there are only a few blue LBGs in
this mass range, meaning that we have more confidence in the red
signal. If we consider the model predictions without dust extinction, the
two colours do not agree on a single model, but we note that the
$( 0.5\,\alpha_{\mathrm{dyn.}}, \, 0.5\,k_{\mathrm{AGN}}, \, 0.5
\,\epsilon_{\mathrm{reheat}} )$ case fits best for red LBG lensing $\chi^2 = 2.17$ and also red clustering $\chi^2 = 10.63$, respectively. This model is still the best red LBG model if
we account for abundance corrections to the masses ($\chi^2 = 4.70)$.

\subsubsection{LBG lensing for TNG300}\label{sec:lbgLensingTNG}

\begin{figure}
\includegraphics[width=0.94\columnwidth]{\figrelpath 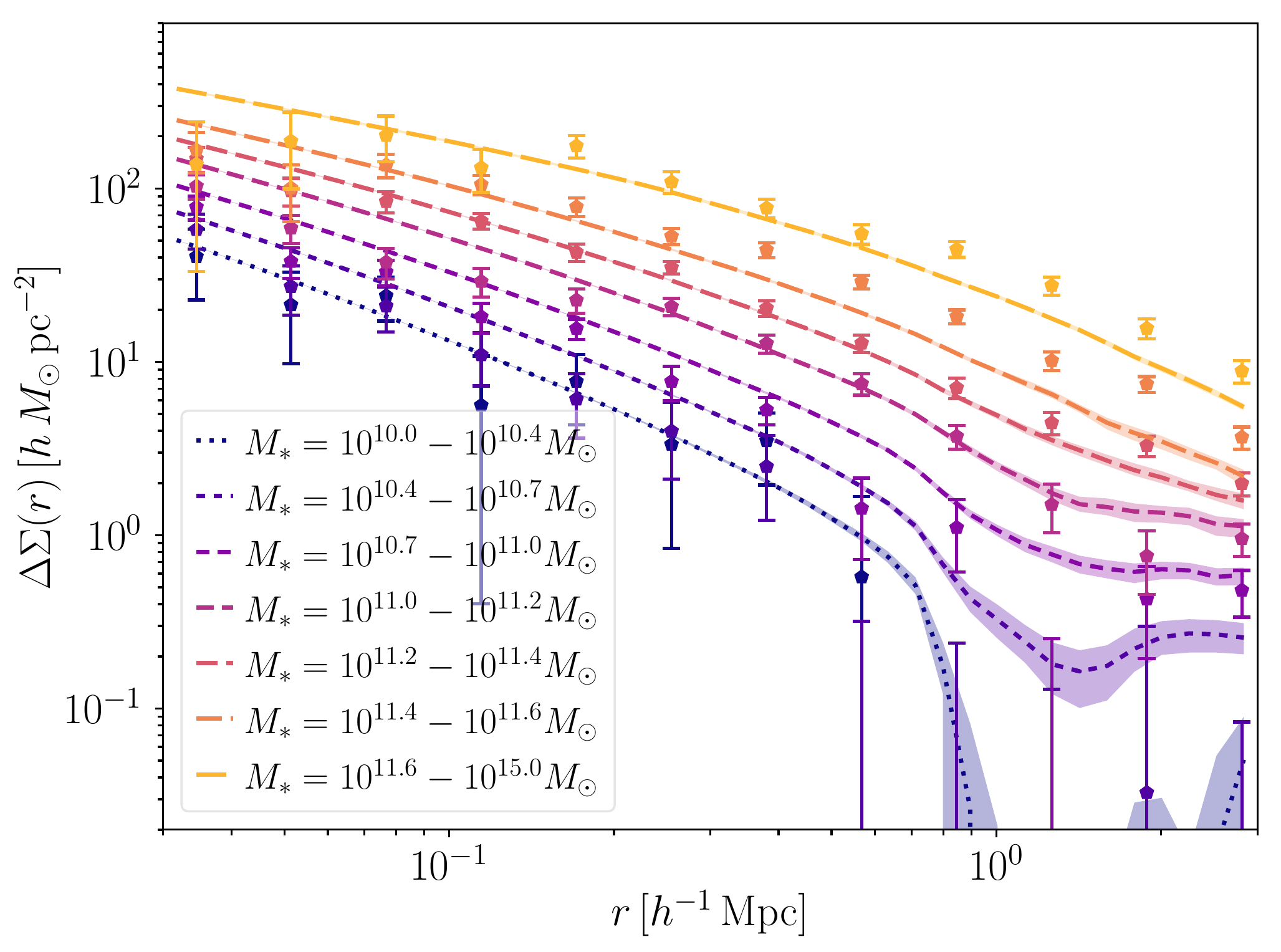}
\caption{LBG lensing signal from TNG300 compared to measurements from \citet{2016MNRAS.456.2301W}. The predicted signal lies above the data for the $10.7 < \log_{10} M_\ast \left [ M_\odot\right ] < 11.0$ mass bin on scales $r \sim 0.2 - 1 \,h^{-1}\,\text{Mpc}$ and bellow for the two most massive bins with $11.4 < \log_{10} M_\ast \left [ M_\odot\right ] < 15$ for $r > 0.6\, h^{-1}\,\text{Mpc}$ but overall the agreement is comparable to the SAMs. On scales $r < 0.7\,h^{-1}\,\text{Mpc}$, the median deviations for the model lies within $30\,\%$ of the data, apart from the $10.7 < \log_{10} M_\ast \left [ M_\odot\right ] < 11.0$ mass bin.}
\label{fig:tng300LBGMstarSelection}
\end{figure}

\begin{figure*}
\begin{centering}
\includegraphics[width=0.94\columnwidth]{\figrelpath 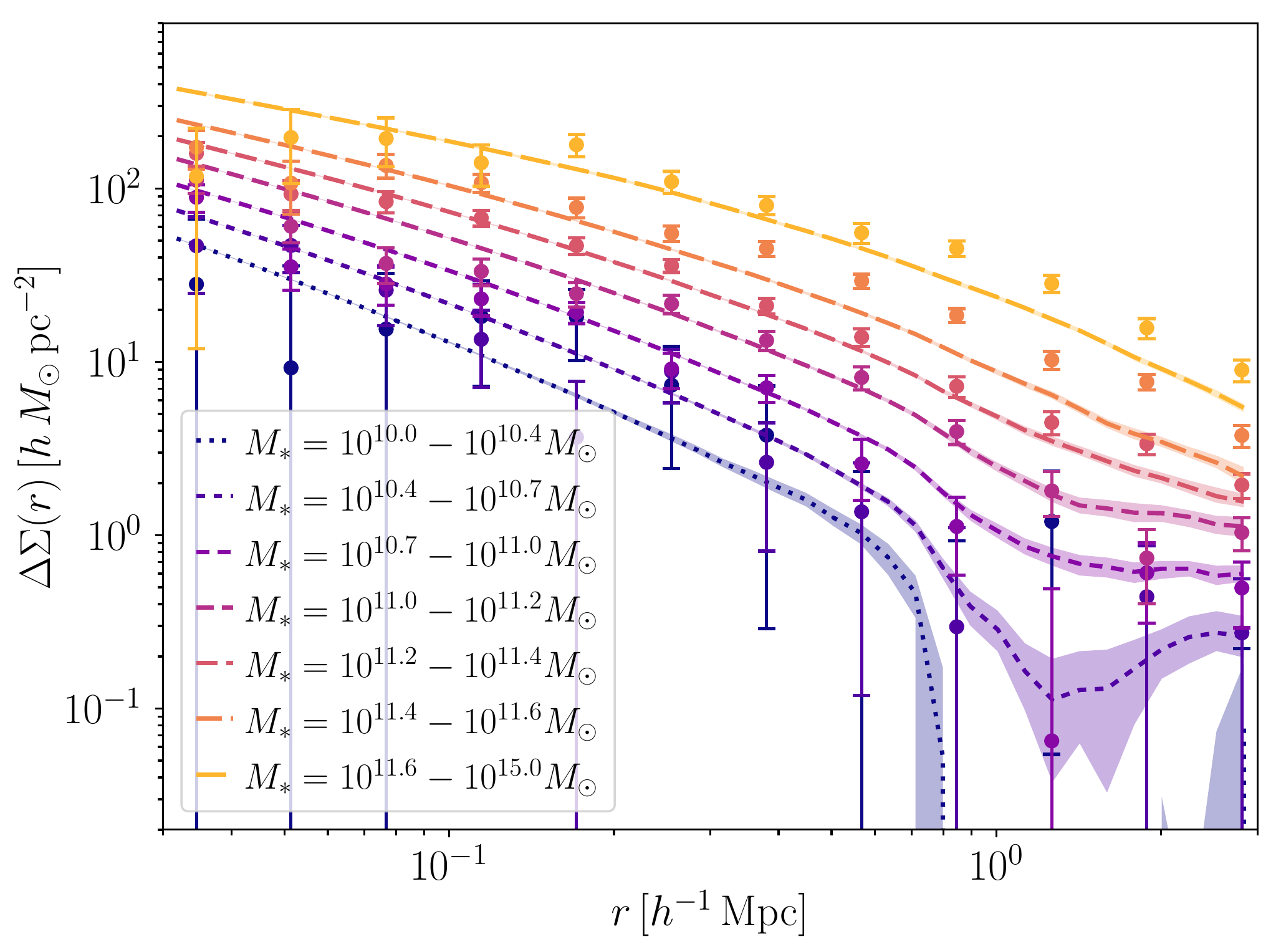}
\includegraphics[width=0.94\columnwidth]{\figrelpath 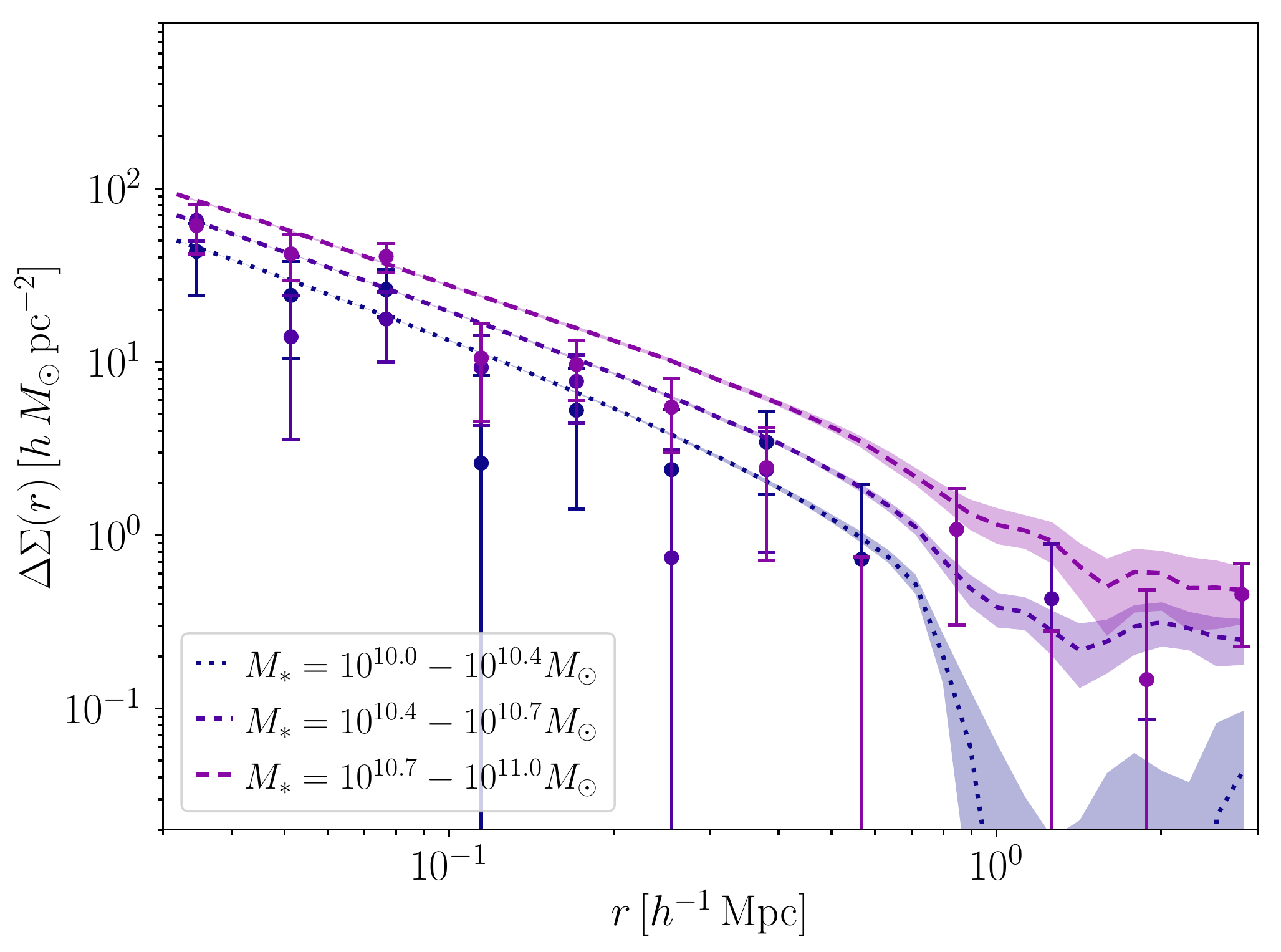}
\includegraphics[width=0.94\columnwidth]{\figrelpath 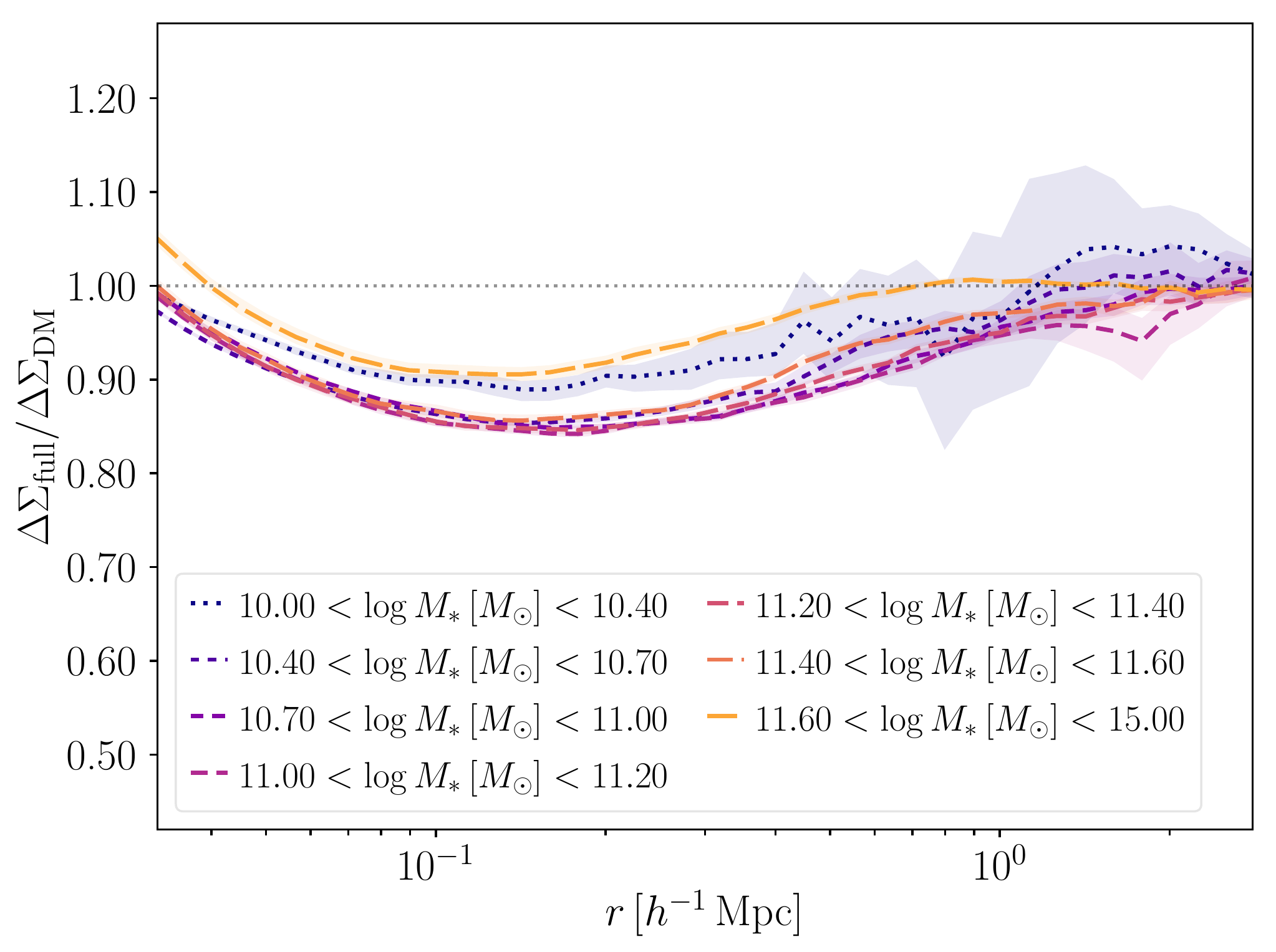}
\includegraphics[width=0.94\columnwidth]{\figrelpath 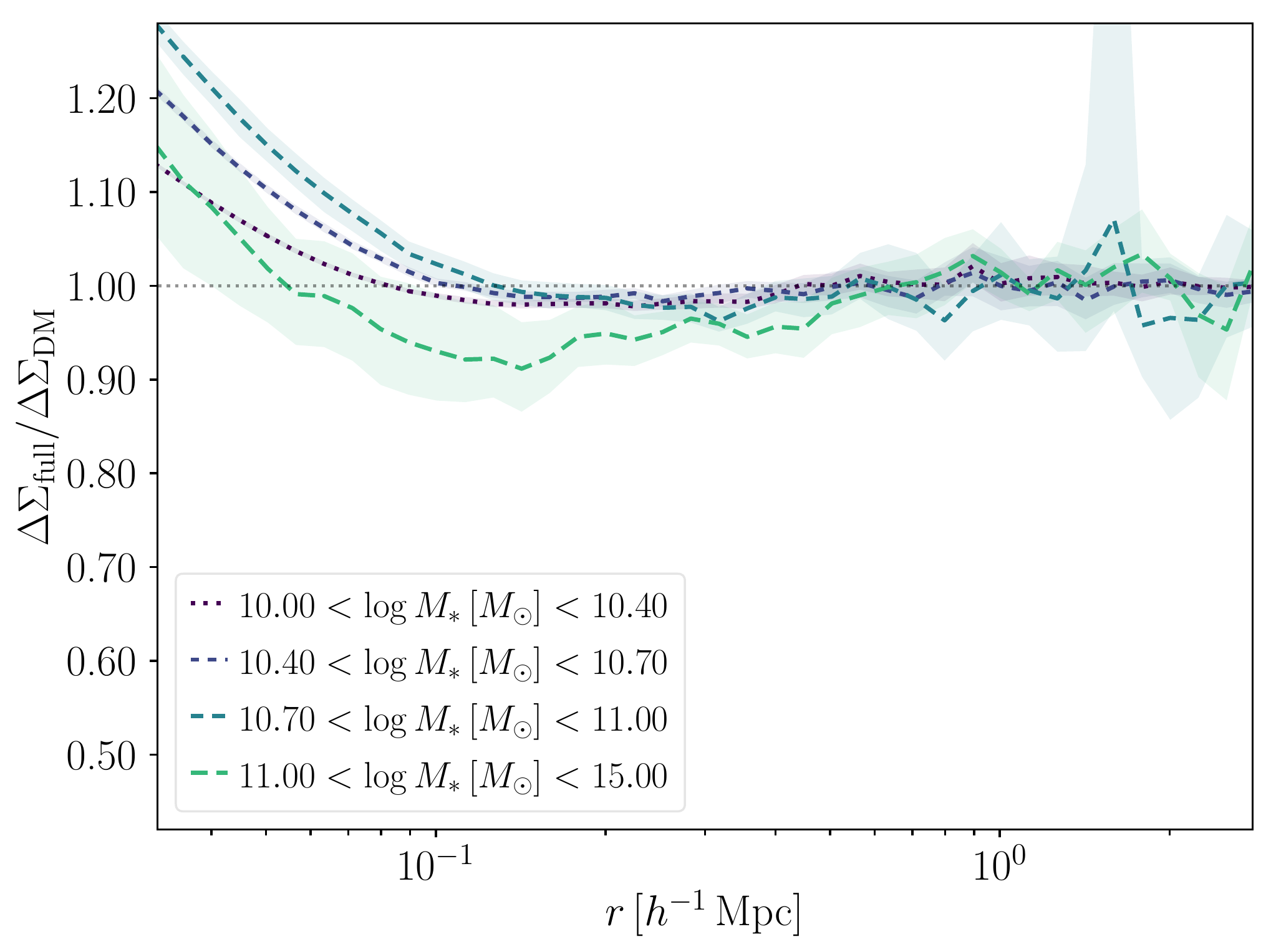}
\caption{Same as Fig.~\ref{fig:tng300LBGMstarSelection} but for the signal split into red (\emph{upper left}) and blue (\emph{upper right}) LBGs from TNG300 with bound stellar masses and dust extinction compared to measurements from \citet{2016MNRAS.457.3200M}. The simulation predictions are in good agreement with the observations, except for the most massive bins, where the model underestimate the signal for $r > 200\,h^{-1}\text{kpc}$. In the bottom row, we plot the baryonic deformations for matched central galaxies separated according to colour. The suppression is deeper for red galaxies, whereas blue galaxies have a more pronounced stellar term.}
\label{fig:tng300LBGMstarSelectionRedBlue}
\end{centering}
\end{figure*}

Moving on to the hydrodynamical predictions, we show in
Figs.~\ref{fig:tng300LBGMstarSelection} and
\ref{fig:tng300LBGMstarSelectionRedBlue} the LBG lensing predictions
for all, red and blue LBGs from TNG300. We have also conducted the analysis with corrected rTNG300 stellar masses and this mass choice
introduces a drop in the signal from the least massive bins and increases
the most massive signals from the fiducial bound mass predictions, but those predictions as well as those of TNG300 are in agreement with observations. We see that this
statement also applies for red and blue LBGs in
Fig.~\ref{fig:tng300LBGMstarSelectionRedBlue}, where we do not plot
the most massive blue signal due to poor statistics. If we compute the corresponding $\chi^2$-values, TNG300 yields $\chi^2 = 3.80$ for all LBGs, $\chi^2 = 3.06$ for red and $\chi^2 = 9.53$ for blue LBGs (which is reduced to $\chi^2 = 2.92$ if we omit the $11 < \log_{10} M_\ast [M_\odot] < 15$ mass bin where TNG300 predicts $\approx 110$ LBGs compared to a couple of thousands in the \citetalias{2015MNRAS.451.2663H} SAM). This means that TNG300 performs slightly better than the \citetalias{2015MNRAS.451.2663H} SAM for the total and red LBG signals, but worse for the blue LBG signal.

The effect of baryons
is mostly noticeable in the innermost bin due to the presence of the
stellar term, which produces an excessive signal for the two most
massive bins for the total and red LBG signal compared to the SAMs,
but otherwise the result conforms well with what we have previously
shown. With respect to our best fit LBG SAM $
(0.5\,\alpha_{\mathrm{dyn.}}, \, 0.2\,k_{\mathrm{AGN}} )$ which
slightly underpredicts the most massive red LBG lensing signal, it is
moderately increased for TNG300 as seen in
Fig.~\ref{fig:tng300LBGMstarSelection} and the upper left figure in
Fig.~\ref{fig:tng300LBGMstarSelectionRedBlue}. 

In the lower row of Fig.~\ref{fig:tng300LBGMstarSelectionRedBlue}, we restrict the comparisons to centrals and show the baryonic imprint for the two samples. We discover that the suppression for red centrals is larger than for blue centrals, which have more pronounced stellar terms. By fitting NFW profiles \citep{Navarro:1995iw, Navarro:1996gj} to the lensing signal \citep{2000ApJ...534...34W} for $0.1 < r \left [ h^{-1} \, \text{Mpc} \right ] < 1.0$ for centrals in the $10.7 < \log_{10} M_\ast \left [ M_\odot \right] < 11.0$ mass bin, we are able to translate this difference into a bias in the observed host halo bimodality. Using observationally motivated $1/r^2$ weights and assuming a lens redshift of $z = 0.11$, we find the best-fit parameter values in Table~\ref{tab:nfwMassBimodalityRedBlue}. Baryons cause a shift of almost $0.1\,$dex in the best fit host halo mass for the red sample, and while we still observe a host halo bimodality with red galaxies residing in more massive haloes by a factor of $\approx 1.33$, it is reduced by $\approx 14\, \%$ from the gravity-only run where the red-to-blue mass ratio is $\approx 1.55$. For the fitted masses for red galaxies, the suppression is $\sim 15\, \%$ for $10.4 < \log_{10} M_\ast \left [ M_\odot \right ] < 11.6$, after which the effect decreases. Baryonic effects also influence the measured concentrations $c$, with a shift of $\Delta c \sim 1$ for red systems with $10.7 < \log_{10} M_\ast \left [ M_\odot \right ] < 11.6$, whereas there are only small differences for blue galaxies. In Table~\ref{tab:nfwMassBimodalityRedBlue}, the host haloes for the blue galaxies are less concentrated in the gravity-only run, but they are more concentrated for all other mass bins, so we consider this a coincidence. The observed decrease in concentration in the full physics run conforms to previous findings in the literature for 3D density profiles \citep[e.g.][]{2010MNRAS.405.2161D, 2017MNRAS.471..227M}. For this mass range, red central galaxies in the baryonic run reside in less concentrated host haloes in the full physics run than blue galaxies, but these correspond to more concentrated haloes in the gravity-only run. It is well-known that concentration correlates with formation time \citep[e.g.][]{Navarro:1996gj, 2007MNRAS.377L...5G}, with older haloes on average being more concentrated which would host older galaxies which on average are redder. However, in the full physics run, feedback processes, whose effects appear to be irreversible \citep[e.g.][]{2017MNRAS.466.3876Z}, have had more time to change the appearance of these older haloes, thus lowering their concentrations with respect to the subhaloes hosting younger blue systems. 

Here, we have applied the dust model from \citet{2018MNRAS.475..624N} to the colours, but we have also checked that these results hold without dust, as well as apply to rTNG300. We have performed the same analysis in TNG100 for matched centrals and note comparable results, but with more scatter, with the largest suppression for red galaxies at $\approx 10-14\,\%$ for $10.4 < \log_{10} M_\ast [M_\odot] < 11.6$. We note the same trend with ($\approx 2-5\,\%$) shallower deformations in TNG100 than in TNG300 for red galaxies, whereas any difference for blue galaxies is hard to notice except for $11 < \log_{10} M_\ast [M_\odot] < 15$.

{\renewcommand{\arraystretch}{1.5}
\begin{table}
	\centering
	\begin{tabular}{l c c}
	\hline
		{\bf Fitted parameter} & {\bf Full physics} & {\bf Gravity-only}\\
		\hline
$\log_{10} M_{200\text{c}} \left [ h^{-1}\,M_\odot\right ]$ (red) & 12.42 & 12.50\\
		$\log_{10} M_{200\text{c}} \left [ h^{-1}\,M_\odot\right ]$ (blue) & 12.30 & 12.31 \\
		Concentration (red) & 5.67 & 6.70 \\
		Concentration (blue) & 6.02 & 5.74 \\
		\hline
	\end{tabular}
	\caption{Fitted NFW parameters for central red and blue galaxies with $10.7 < \log_{10} M_\ast \left [ M_\odot \right] < 11.0$ in TNG300 compared to their matches in the dark matter-only run TNG300-DMO.}
	\label{tab:nfwMassBimodalityRedBlue}
\end{table}}

We thus conclude that the TNG suite is equally good as SAMs at producing predictions for LBGs. That there is only a significant disagreement between the two modelling frameworks for the innermost radial bin is very promising for the interpretation of observational LBG data with SAMs; i.e. that baryonic effects can mostly be ignored when determining host halo masses from lensing, as these small scales can be excluded or down-weighted.

\subsection{Clustering}\label{sec:clustering}

\begin{figure*}
\begin{centering}
\includegraphics[width=0.9\columnwidth]{\figrelpath 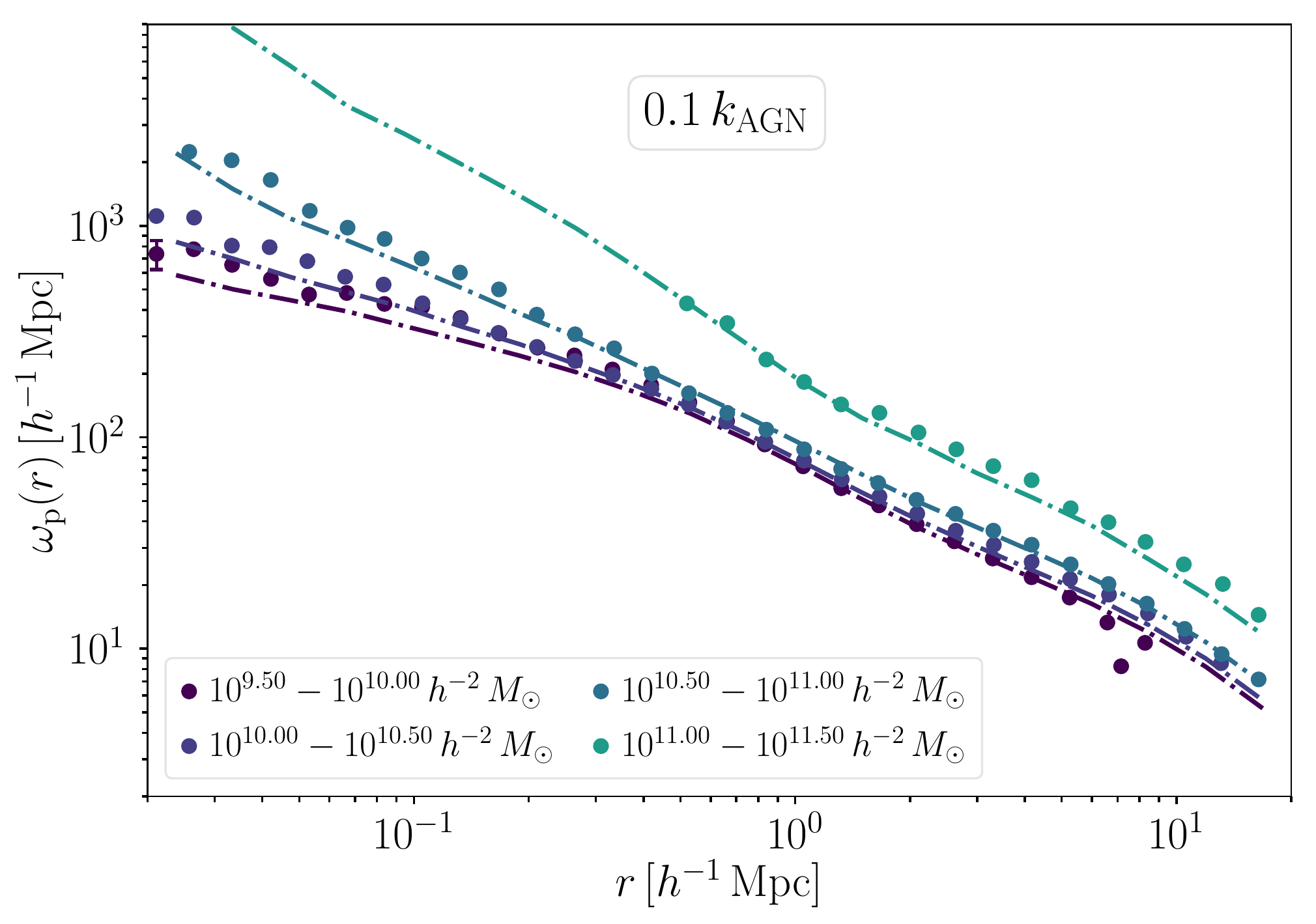}
\includegraphics[width=0.9\columnwidth]{\figrelpath 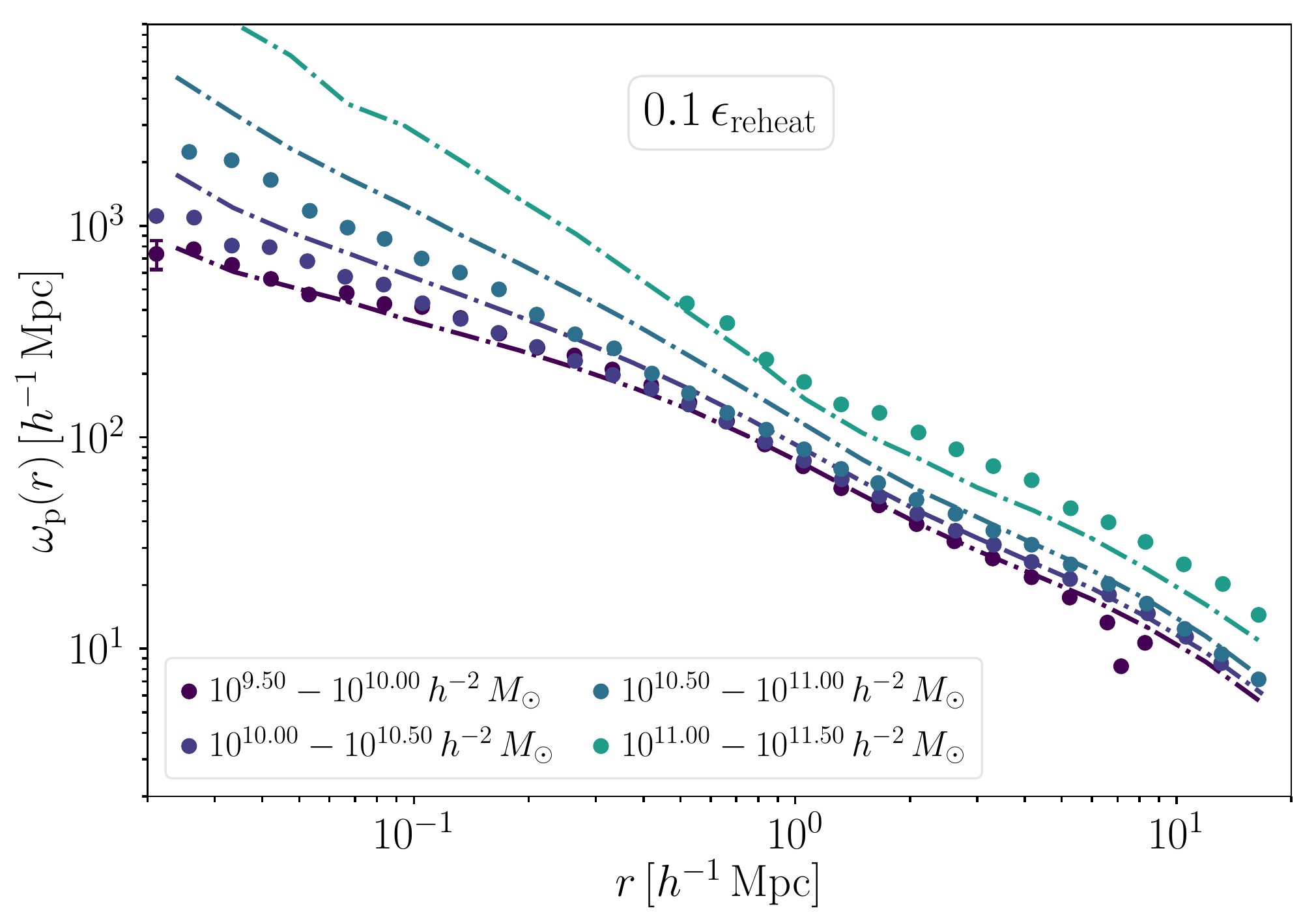}
\caption{Clustering predictions for SAM galaxies in the best fit $0.1\,k_\text{AGN}$ model ($\chi^2 = 0.72$, \emph{left}) and for galaxies in the $0.1\,\epsilon_\text{reheat}$ model ($\chi^2 = 3.83$, \emph{right}) w.r.t. SDSS observations from \citetalias{Guo:2010ap}. By comparing the two, we realise that the data favours weaker AGN feedback as opposed to weaker SN feedback.}
\label{fig:clustering0p1kAGN}
\end{centering}
\end{figure*}

{\renewcommand{\arraystretch}{1.4}
\begin{table*}
	\centering
	\begin{tabular}{l l l l l l l}
	\hline
		{\bf Clustering (fiducial)}& {\bf All} & $\chi^2$ & {\bf Red}& $\chi^2$  & {\bf Blue} & $\chi^2$\\
		\hline
First &  $0.1\, k_\text{AGN}$ & 0.72 & $\left (0.3\, \alpha_\text{dyn}, \, 0.2\,k_\text{AGN} \right )$  & 4.05 & $\left (0.3\, \alpha_\text{dyn}, \, 0.1\,k_\text{AGN} \right )$ & 2.61 \\
		Second & $\left (0.5\, \alpha_\text{dyn}, \, 0.2\,k_\text{AGN} \right )$ & 0.94 & $\left (0.4\, \alpha_\text{dyn}, \, 0.2\,k_\text{AGN} \right )$ & 4.36 & $\left (0.4\, \alpha_\text{dyn}, \, 0.1\,k_\text{AGN} \right )$ & 2.94\\
		Third & $\left (0.5\, \alpha_\text{dyn}, \, 0.1\,k_\text{AGN} \right )$  & 0.98 &  $\left (0.5\, \alpha_\text{dyn}, \, 0.3\,k_\text{AGN} \right )$ & 4.89  & $\left (0.5\, \alpha_\text{dyn}, \, 0.1\,k_\text{AGN} \right )$ & 3.05\\
		\hline
	\end{tabular}
	\caption{The best fit models according to galaxy clustering. Our best LBG lensing model $\left (0.5\, \alpha_\text{dyn}, \, 0.2\,k_\text{AGN} \right )$ is a runner up and finishes in fourth place for the red clustering ($\chi^2 = 4.94$) and fifth for blue ($\chi^2 = 3.95$). The $0.1\, k_\text{AGN}$ model which is best for the total clustering signal does considerably worse for the red clustering ($\chi^2 = 15.74$) than the blue ($\chi^2 = 3.22$). The clustering performance of the fiducial models \citetalias{2015MNRAS.451.2663H} ($\chi^2 = 3.78$),  \citetalias{Guo:2010ap} ($\chi^2 = 5.66$) and the \citetalias{Guo:2010ap} parameter values on the \citetalias{2015MNRAS.451.2663H} model ($\chi^2 = 2.08$) vary.}
	\label{tab:fomClusteringStellarMassOnly}
\end{table*}}

In this Section, we compare the clustering predictions from the different \textsc{L-Galaxies} variations and TNG300, with and without resolution correction to the stellar masses. Here, we do not change the stellar masses of the SAMs to match abundances
but only focus on the baseline model predictions.

For the stellar mass-only clustering, we determine the best fit models
through \eqref{eq:fom} by the mean values for all four clustering bins
with the results given in Table~\ref{tab:fomClusteringStellarMassOnly}. We find that both 2-pt
statistics point towards a consistent picture with the lowest, best
fit values reached for the weak AGN feedback models. In
Table~\ref{tab:fomClusteringStellarMassOnly} we see that the best
agreement is reached for the $0.1\,k_\text{AGN}$ model, plotted in
Fig.~\ref{fig:clustering0p1kAGN}, which also gave the best stellar
mass-only lensing predictions in Table~\ref{tab:fomLensingStellarMassOnly} and that the $ (0.5\,
\alpha_\text{dyn}, \, 0.2\,k_\text{AGN} )$ is number two,
which was the best for LBG stellar mass-only lensing in
Table~\ref{tab:fomLensingLBGStellarMassOnly}. The latter model
predictions are very similar to the former and thus we refrain from
showing them. In the case of lensing, it
was not apparent at the high mass end whether the weak AGN feedback
models or the weak SN feedback models were to be preferred, but if we
compare the results in Fig.~\ref{fig:clustering0p1kAGN}, we recognise
that the $0.1\,\epsilon_\text{reheat}$ model is disfavoured ($\chi^2 = 3.83$) by the
massive clustering signals. Moderately weaker $\epsilon_\text{reheat}$ values, such as the $ (0.5\,
\alpha_\text{dyn}, \, 0.5\,k_\text{AGN}, \, 0.5\,\epsilon_\text{reheat})$ model, are also marginally less preferred by the data ($\chi^2 = 1.23$).

We compare the projected red and blue clustering signal to SDSS DR7
data from \citet{2016MNRAS.457.4360Z}. In Table~\ref{tab:fomClusteringStellarMassOnly} we list the
best models for red and blue clustering. If we compare the two, both prefer
weaker AGN feedback. Our best model for the LBG lensing finishes in
fourth place for the red clustering although the top three is
dominated by its close siblings in parameter space. It is interesting
to note that the top blue clustering models in
Table~\ref{tab:fomClusteringStellarMassOnly} closely resemble the top
models in Table~\ref{tab:fomLensingLBGStellarMassOnly} for blue LBG
lenses. For low mass systems differences between the \citetalias{2015MNRAS.451.2663H} and
\citetalias{Guo:2010ap} models are substantial due to the
overproduction of red galaxies in the \citetalias{Guo:2010ap} model
which are more clustered. For these systems, the AGN feedback
strength also has a significant effect on the amplitude of the 1-halo term
for red galaxies. The TNG300 results are similar to the lowest feedback
model predictions, although the signal drops towards the centre.

\begin{figure}
\includegraphics[width=1.02\columnwidth]{\figrelpath 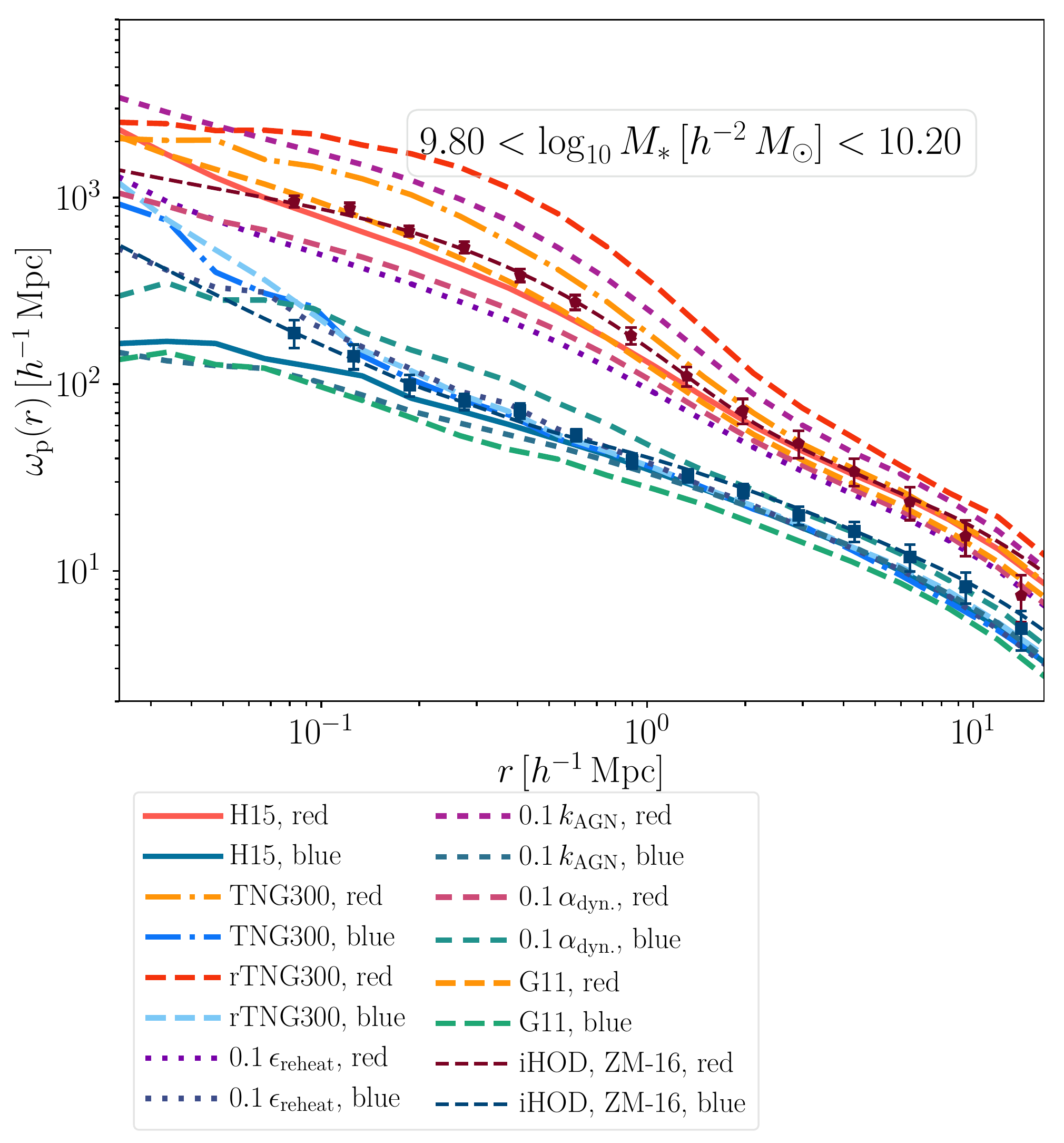}
\caption{Predicted red and blue clustering for galaxies with stellar masses $9.8 < \log_{10} M_\ast \left [ h^{-2}\, M_\odot \right ] < 10.2$ compared to the \citet{2016MNRAS.457.4360Z} observations. Here we compare the clustering predictions for TNG300 with and without resolution corrections with dust extinction, as well as the predictions of the most extreme SAMs. Contrary to the mild impact on the lensing signal, the resolution correction significantly alters the clustering predictions for the model.}
\label{fig:clusteringHenriques17ExtremeVaryingBin2RedBlue}
\end{figure}

In Fig.~\ref{fig:clusteringHenriques17ExtremeVaryingBin2RedBlue}, we
also illustrate how the \citet{2018MNRAS.475..648P} resolution
correction together with dust extinction affects the predicted
clustering signal for red and blue galaxies in TNG300 vs. rTNG300. As
previously reported in \citet{2018MNRAS.475..676S}, there was a
tension for the predicted clustering signal for red galaxies with $9.5
< \log_{10} M_\ast \left [ h^{-2}\, M_\odot \right ] < 10.0$ without dust
extinction w.r.t. SDSS observations. If we use dust corrected colours
the signal decreases and the tension is mitigated as can be seen in
Fig.~\ref{fig:clusteringHenriques17ExtremeVaryingBin2RedBlue}. The changes in the
amplitude due to dust extinction are strongest in this low mass range
since there are only a few red galaxies present and there is a rapid
transition between red and blue. Still, if we apply the additional
resolution correction, the tension is re-introduced as the red
sequence is artificially shifted into the blue, leaving the most
clustered galaxies. As hinted by the different satellite fractions at the knee of the SMF, the resolution correction introduces a
similarly large tension for galaxies with $10.5 < \log_{10} M_\ast \left [
  h^{-2}\,M_\odot \right ] < 11.0$ w.r.t. data, as the one highlighted
in \citet{2018MNRAS.475..676S}. Note that we are using 30 pkpc masses, so the result can differ slightly from the bound mass results. Further work has to be undertaken
to clarify which corrections are necessary for which observable and to
quantify the magnitude of the induced biases.

Fig.~\ref{fig:clusteringHenriques17ExtremeVaryingBin2RedBlue} suggests that the red clustering can be used to constrain the feedback efficiencies, but the results are sensitive to the dust model. We quantify this variation by comparing the clustering signals for the \citetalias{Guo:2010ap} and \citetalias{2015MNRAS.451.2663H} model with and without dust. For stellar masses $<10^
{11}\,h^{-2}\,M_\odot$, there is a clear smooth suppression of the
signal for red galaxies when dust is included compared to no dust, as more dusty star-forming blue galaxies
which are on average less clustered are counted as red. This primarily
affects the 1-halo term and the effect can amount to $30-40\,\%$
whereas the effect for the 2-halo term is $\sim 10-20\,\%$ depending
on the galaxy formation model. This effect is greater for the
\citetalias{Guo:2010ap} model due to its many low mass red galaxies,
and it is greater for lower masses since most galaxies in that range
are blue. For blue galaxies the situation is less clear; we observe a
mild suppression for the two lowest mass bins for the
\citetalias{2015MNRAS.451.2663H} model, but the result at higher
masses contains a lot of scatter.

We also show results for our extreme SAMs in
Fig.~\ref{fig:clusteringHenriques17ExtremeVaryingBin2RedBlue}, where
we spot a clear tension between the $0.1\, \epsilon_\text{reheat}$
model and the observations. This is the reason the low SN feedback
models do not feature among our best. The
$0.1\, k_\text{AGN}$ model is not as extreme as the resolution
corrected TNG300 result in Fig.~\ref{fig:clusteringHenriques17ExtremeVaryingBin2RedBlue} for red
galaxies, which holds true for more massive systems. As for the $0.1\,
\alpha_\text{dyn}$ model, it produces an excessive blue clustering
signal, although it is in agreement with data for the least massive
bin, and a too low red signal. The iHOD predictions are also plotted for comparison, and we see that they agree very well with the predictions from the SAMs and TNG300. 

As shown in previous studies \citep[e.g.][]{2017MNRAS.469.2626H, 2018MNRAS.475..676S}, SAMs and hydrodynamical simulations in
cosmological volumes are able to produce very accurate clustering
predictions and it is nice to see the concordance between the iHODs
and these two other frameworks.

\subsection{Galaxy group lensing}\label{sec:groupCriteria}

\begin{table*}
	\centering
	\begin{tabular}{l c c  c c c c c c c c c}
	\hline
		$\log_{10} M_\ast [M_\odot]$ & $\bar{d}_\text{H15}$ & $\bar{d}_\text{E}$ & $M_{200\text{c}}^\text{cen., H15}$ & $M_{200\text{c}}^\text{sat., H15}$ & $M_{200\text{c}}^\text{cen., E}$ & $M_{200\text{c}}^\text{sat., E}$ & $N_\text{gal}^\text{H15}$ & $N_\text{gal}^\text{E}$  & $M_\ast^\text{lim, H15}$ & $M_\ast^\text{lim, E}$ & $f_\text{sat}$ \\
		\hline
		$10.3 - 10.6$ &  0.686 & 0.590 & 13.19 & 13.61 & 12.29 & 13.78 & 95\,467 & 354 & 9.98 & 9.46 & 0.98 \\
		$10.6 - 10.9$ & 0.728 & 0.725 & 13.45 & 13.74 & 12.75 & 13.92 & 60\,289 & 150 & 10.22 & 9.91 & 0.95 \\
		$10.9 - 11.2$ & 0.763  & 0.902 & 13.64 & 13.83 & 12.96 & 13.97 & 26\,387 & 68 & 10.26 & 9.96 & 0.81 \\
		$11.2 - 11.5$ & 0.859 & 1.151 & 13.89 & 14.08 & 13.22 & 14.02 & 6\,698 & 22 &  10.36 & 10.33 & 0.50 \\
		$11.5 - 11.8$ & 0.976 & 1.877 & 14.05 & 14.30 & 13.52 & 14.07 & 1\,908 & 29 & 9.86 & - & 0.21 \\
		\hline
	\end{tabular}
	\caption{In this Table, we compare the properties of the \citet{2017MNRAS.471.2856V} comparison simulation samples (H15 = \textsc{L-Galaxies-15}, E = EAGLE) with all mean halo masses $M_{200\text{c}}$ (host FOF groups) in units of $\log_{10} h^{-1}\,M_\odot$ and all stellar masses and $M_\ast^\text{lim}$ in units of $\log_{10} M_\odot$. The satellite fractions $f_\text{sat}$ match the ones in the \textsc{GAMA} group catalogue. $\bar{d}$ is the average 3D distance between the satellite galaxies and their centrals and $N_\text{gal}$ the total number of galaxies. We note that $M_\ast^\text{lim}$ is higher for the \citetalias{2015MNRAS.451.2663H} model than in EAGLE for the three lowest stellar mass bins, which yields higher average $M_{200\text{c}}^\text{cen.}$ values, which in turn produce higher lensing signals.}
	\label{tab:velliscigEAGLELGComparison}
\end{table*}

\begin{table*}
	\centering
	\begin{tabular}{l c c c c c}
	\hline
		$\log_{10} M_\ast [M_\odot]$ & $\bar{d}$ & $M_{200\text{c}}^\text{cen.}$ & $M_{200\text{c}}^\text{sat.}$ & $N_\text{gal}$ & $\log_{10}M_\ast^\text{lim} [ M_\odot]$\\
		\hline
		{\bf \citetalias{2015MNRAS.451.2663H} model on the gravity-only TNG100-DMO}\\
		\hline
		$10.3 - 10.6$ & 0.732  & 13.12 & 13.80 & 400 & 9.87\\
		$10.6 - 10.9$ & 0.757 & 13.23 & 13.83 & 324 & 10.13\\
		$10.9 - 11.2$ & 0.937 & 13.60 & 13.92 & 157 & 10.39 \\
		$11.2 - 11.5$ & 0.766 & 13.93 & 14.08 & 24 & 10.75\\
		$11.5 - 11.8$ & 0.818 & 13.80 & 14.31 & 6 & -\\
		\hline
		{\bf $0.5\, \alpha_\text{dyn}$ and $0.2 \, k_\text{AGN}$ model on TNG100-DMO}\\
		\hline
		$10.3 - 10.6$ &  0.751 & 13.08 & 13.80 & 345 & 9.83\\
		$10.6 - 10.9$ & 0.759 & 12.97 & 13.87 & 238 & 10.24\\
		$10.9 - 11.2$ & 0.808 & 13.16 & 13.79 & 275 & 9.49 \\
		$11.2 - 11.5$ & 0.964 & 13.52 & 13.87 & 108 & 9.95\\
		$11.5 - 11.8$ & 0.925 & 13.48 & 14.13 & 46 & -\\
		\hline
	\end{tabular}
	\caption{Average host halo masses, 3D distances between the satellite galaxies and the central galaxy in each FOF group in units of $h^{-1} \, \text{Mpc}$ and number counts for the \citetalias{2015MNRAS.451.2663H} model applied to the gravity-only TNG100-DMO. If we compare these values with those quoted in Table~\ref{tab:velliscigEAGLELGComparison}, they are consistent with one another. For the average satellite distances, the \citetalias{2015MNRAS.451.2663H} model predicts more coherent values across the whole mass range w.r.t. EAGLE for the rescaled MR, but the distribution is different in the TNG100-DMO. We also list the statistics for our best fit $0.5\, \alpha_\text{dyn}$ and $0.2 \, k_\text{AGN}$ model on TNG100-DMO in the lower part of the table. Compared to the values for the fiducial \citetalias{2015MNRAS.451.2663H} model, the average central host halo masses are reduced for the four most massive bins by a maximum of $\sim 0.4\,$dex and we have more galaxies in the three most massive bins. There is a small effect on the average host halo masses for the satellites for the three most massive bins where it drops with 0.1-0.2\,dex.}
	\label{tab:velliscigLGComparisonsFiducialBestNew}
\end{table*}

\begin{figure*}
\begin{centering}
\includegraphics[width=1.6\columnwidth]{\figrelpath 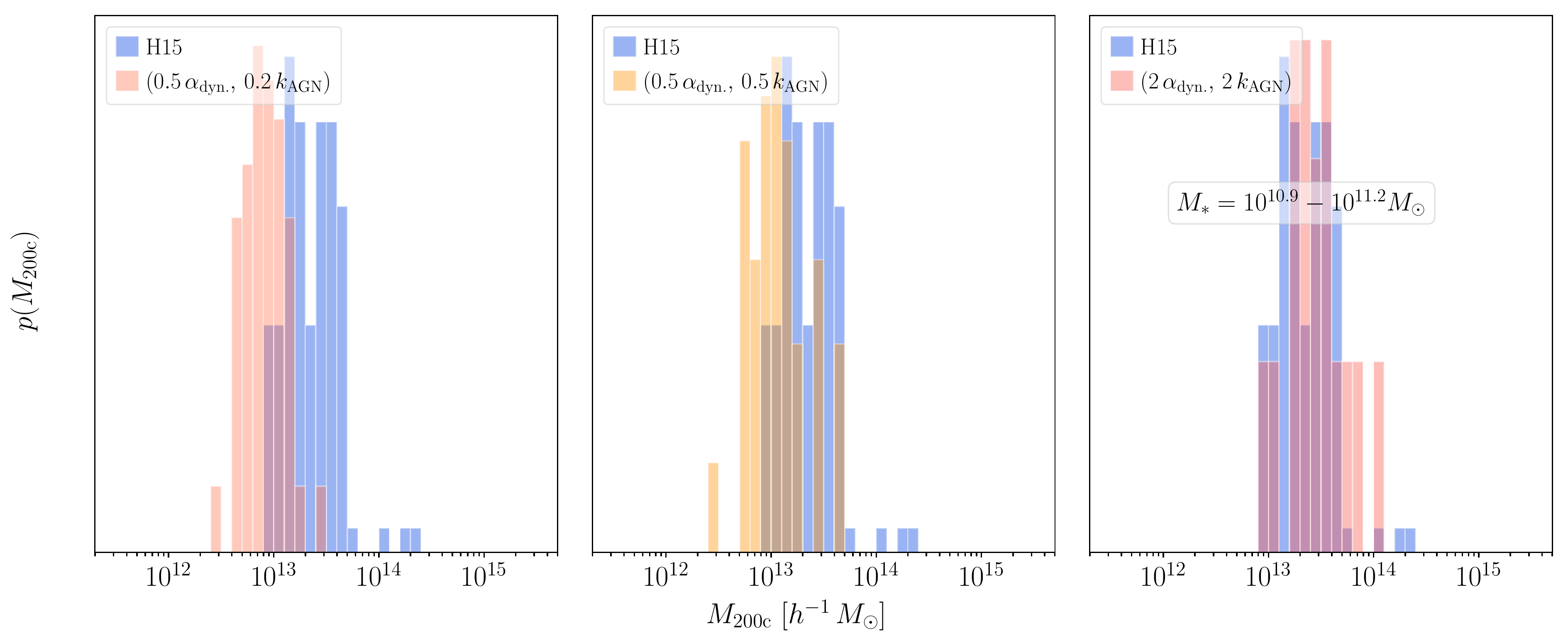}
\caption{Host halo masses for central galaxies in the $10.9 < \log_{10} M_\ast \left [ \, M_\odot \right ] < 11.2$ mass bin for the \citet{2017MNRAS.471.2856V} selection for $ ( 0.5 \, \alpha_\text{dyn}, \,  0.2\, k_\text{AGN})$ (\emph{left}), $( 0.5 \, \alpha_\text{dyn}, \, 0.5\, k_\text{AGN} )$ (\emph{middle}), the fiducial \citetalias{2015MNRAS.451.2663H} (\emph{in all figures}) model and $ ( 2 \, \alpha_\text{dyn}, \, 2\, k_\text{AGN} )$ (\emph{right}) run on TNG100-DMO. Reducing the dynamical friction parameter as well as the AGN feedback efficiency brings about better agreement with the observational constraints, as already indicated by the LBG lensing, whereas we obtain the opposite effect by increasing these two values. Moreover, there is a difference in the host halo mass distribution between $( 0.5 \, \alpha_\text{dyn}, \, 0.5\, k_\text{AGN} )$ and \citetalias{2015MNRAS.451.2663H}, not evident from the SMFs (Fig.~\ref{fig:smfDifferentkAGN}) or the stellar mass-only lensing (for $ 10^{10.79}< M_\ast [h^{-2}\,M_\odot] < 10^{11.19}$), underscoring that there are differences between these models.}
\label{fig:hostHaloDistThirdMassBinVelliscig17}
\end{centering}
\end{figure*}

In this comparison, we are testing our new SAMs with altered parameter combinations from the previous constrained measurements from the stellar mass-only and LBG lensing and clustering samples against observations from an independent survey, in this case KiDS+GAMA galaxy group lenses. This serves to demonstrate that the new parameter combinations can match independent datasets, which will inform on the validity to use local redshift two-point statistics to tune a SAM and then use it to provide accurate predictions for upcoming higher redshift observations from surveys such as \emph{Euclid} and LSST. We also present the corresponding results from the TNG suite, as well as Illustris, and discuss differences with respect to the original study \citep{2017MNRAS.471.2856V}, which compared data to the EAGLE simulation.

\subsubsection{L-Galaxies vs. model variations}\label{sec:lgalaxiesGroupLensing}

Here we focus on a few SAMs from Table~\ref{tab:samModelsParameters}, and especially our best fit LBG model with
$(0.5\, \alpha_\text{dyn}, \, 0.2\,k_\text{AGN} )$. Firstly we compare various statistical properties from the fiducial \citetalias{2015MNRAS.451.2663H} run on the rescaled MR simulation with respect to EAGLE in Table~\ref{tab:velliscigEAGLELGComparison}. Then we reduce the volume and run the SAMs on the gravity-only TNG100-DMO to obtain similar statistics as for the EAGLE simulation, see Table~\ref{tab:velliscigLGComparisonsFiducialBestNew}, which was used in the original comparison, and especially to test claims on volume effects on the large-scale signal around $r \sim 1 \,h^{-1}\,\text{Mpc}$. To show that we obtain the opposite shift of the host halo mass distribution, we also include a modified model with $(2\, \alpha_\text{dyn}, \, 2\,k_\text{AGN})$ in the comparison. We also list the corresponding host
halo masses and other properties for the rescaled MR to show that they
are consistent across simulation volumes.

\begin{figure*}
\begin{centering}
\includegraphics[width=0.8\columnwidth]{\figrelpath 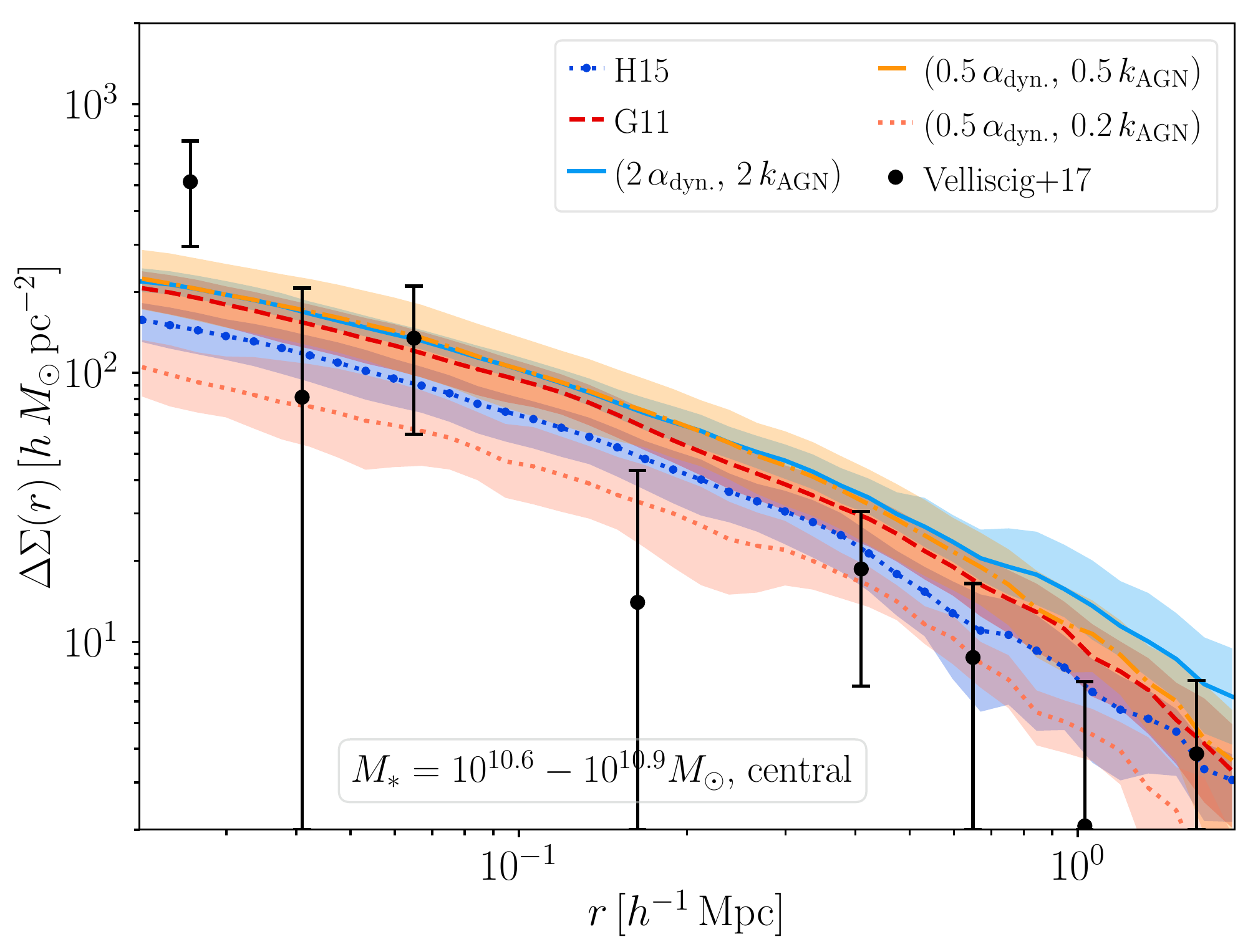}
\includegraphics[width=0.8\columnwidth]{\figrelpath 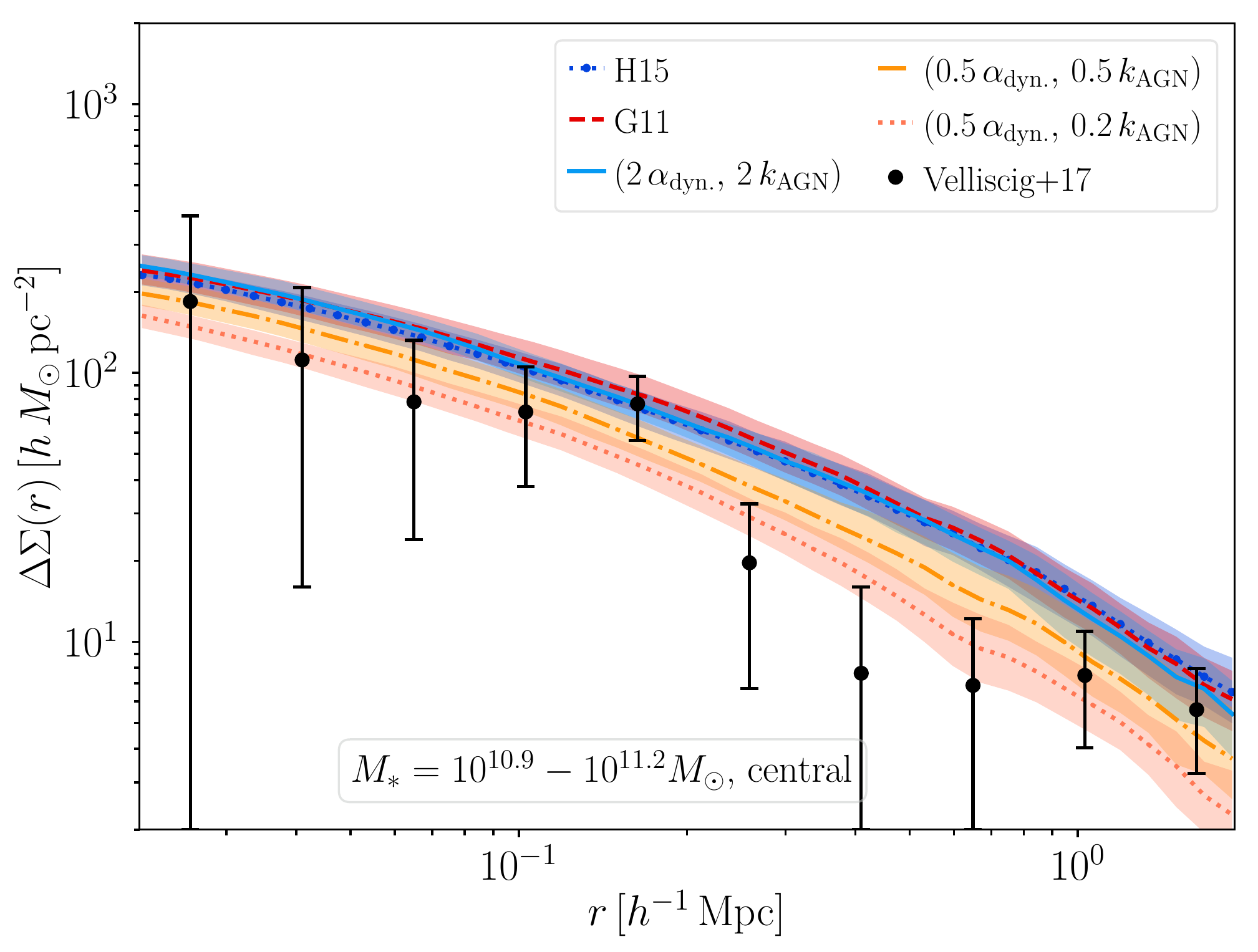}
\includegraphics[width=0.8\columnwidth]{\figrelpath 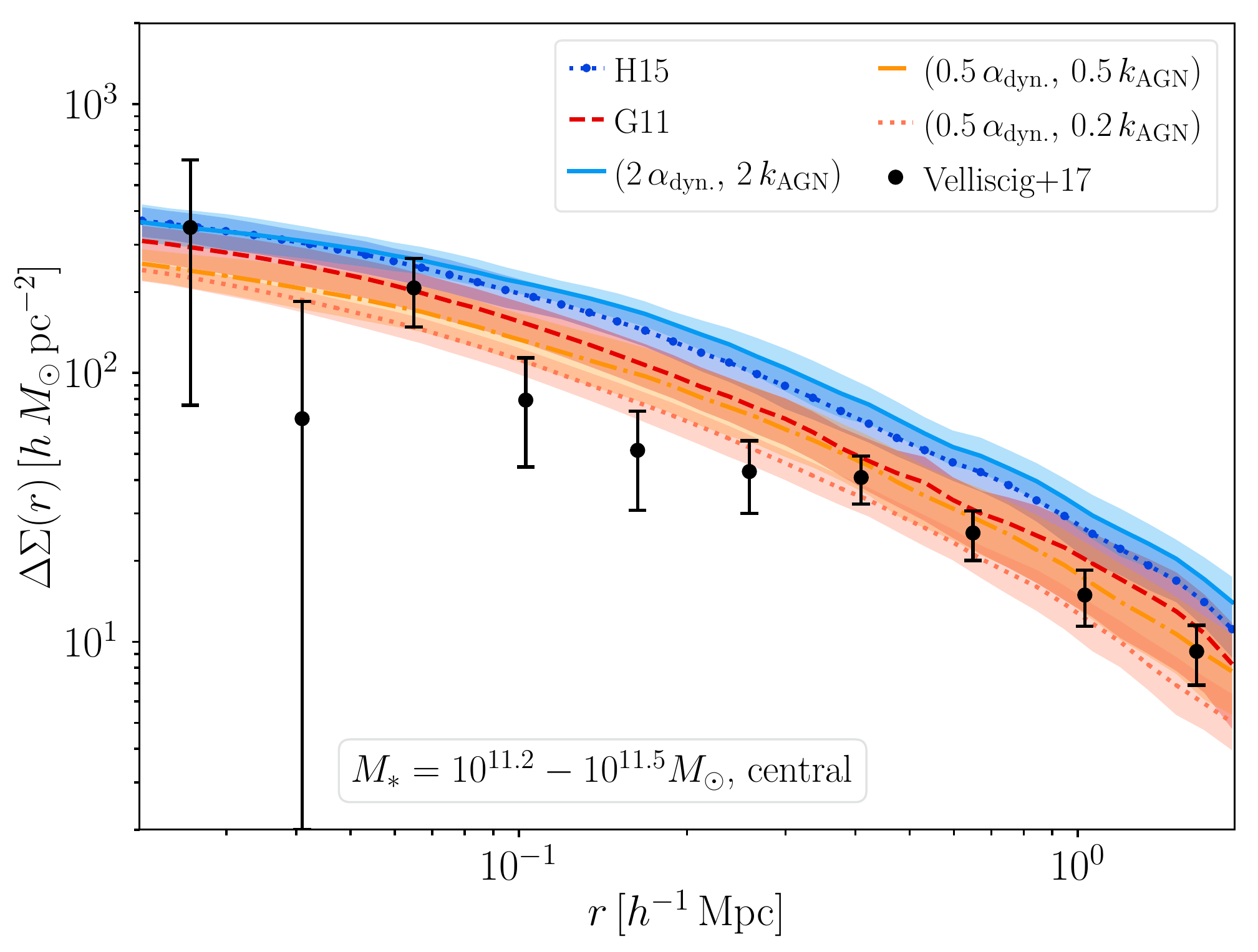}
\includegraphics[width=0.8\columnwidth]{\figrelpath 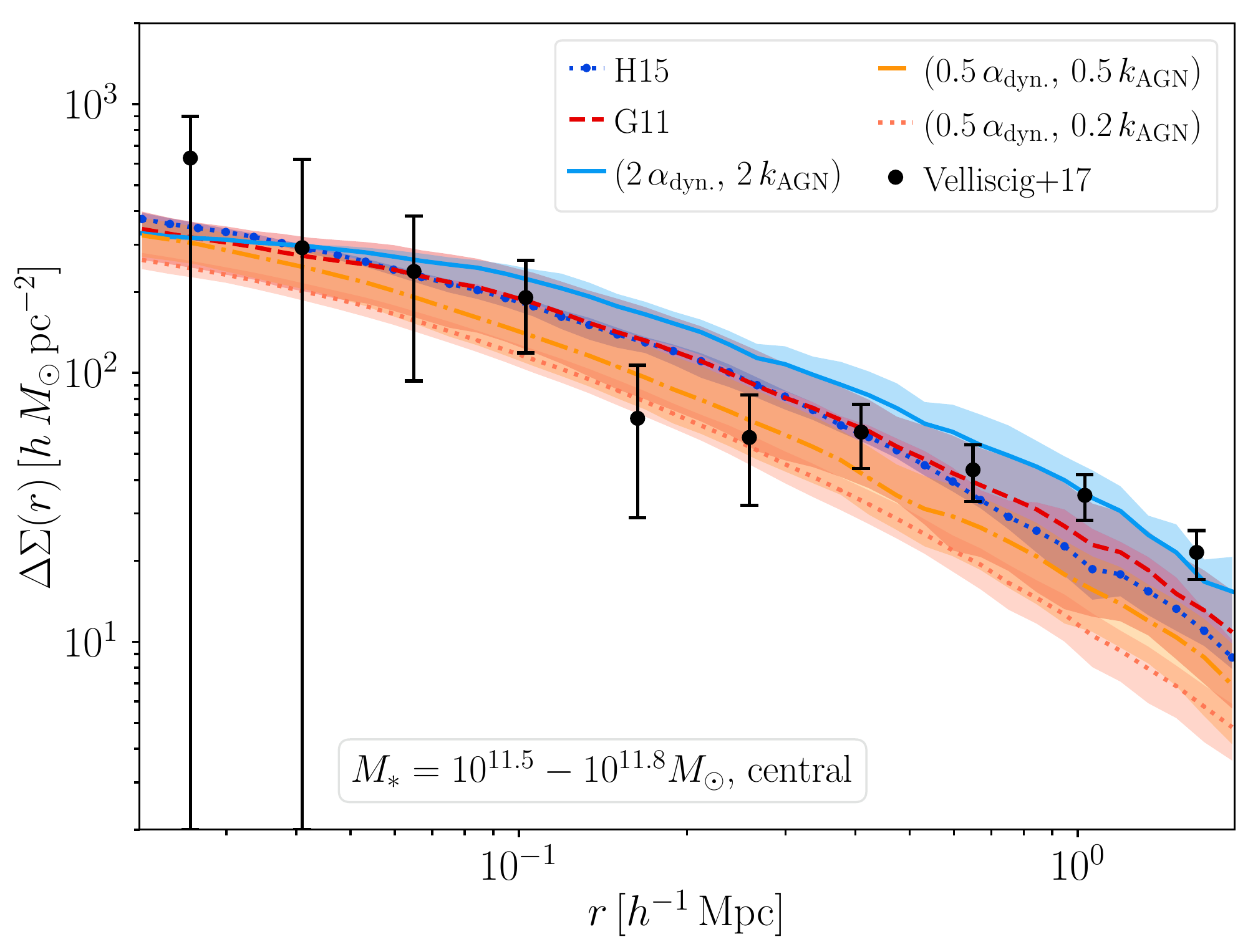}

\caption{GGL signals for central galaxies w.r.t. data from \citet{2017MNRAS.471.2856V}. We see that the best fit $ (0.5\, \alpha_\text{dyn}, \, 0.2\,k_\text{AGN} )$ model produces accurate predictions across the whole mass range although the signal around $1\,h^{-1}\,\text{Mpc}$ in the lower right figure for the most massive bin is somewhat low. For this bin the best agreement is reached for the $ (2\, \alpha_\text{dyn}, \, 2,k_\text{AGN} )$ model in this regime, but we attribute the tension to small number statistics.}
\label{fig:centralSignalVelliscig17}
\end{centering}
\end{figure*}

\begin{figure*}
\begin{centering}
\includegraphics[width=0.80\columnwidth]{\figrelpath 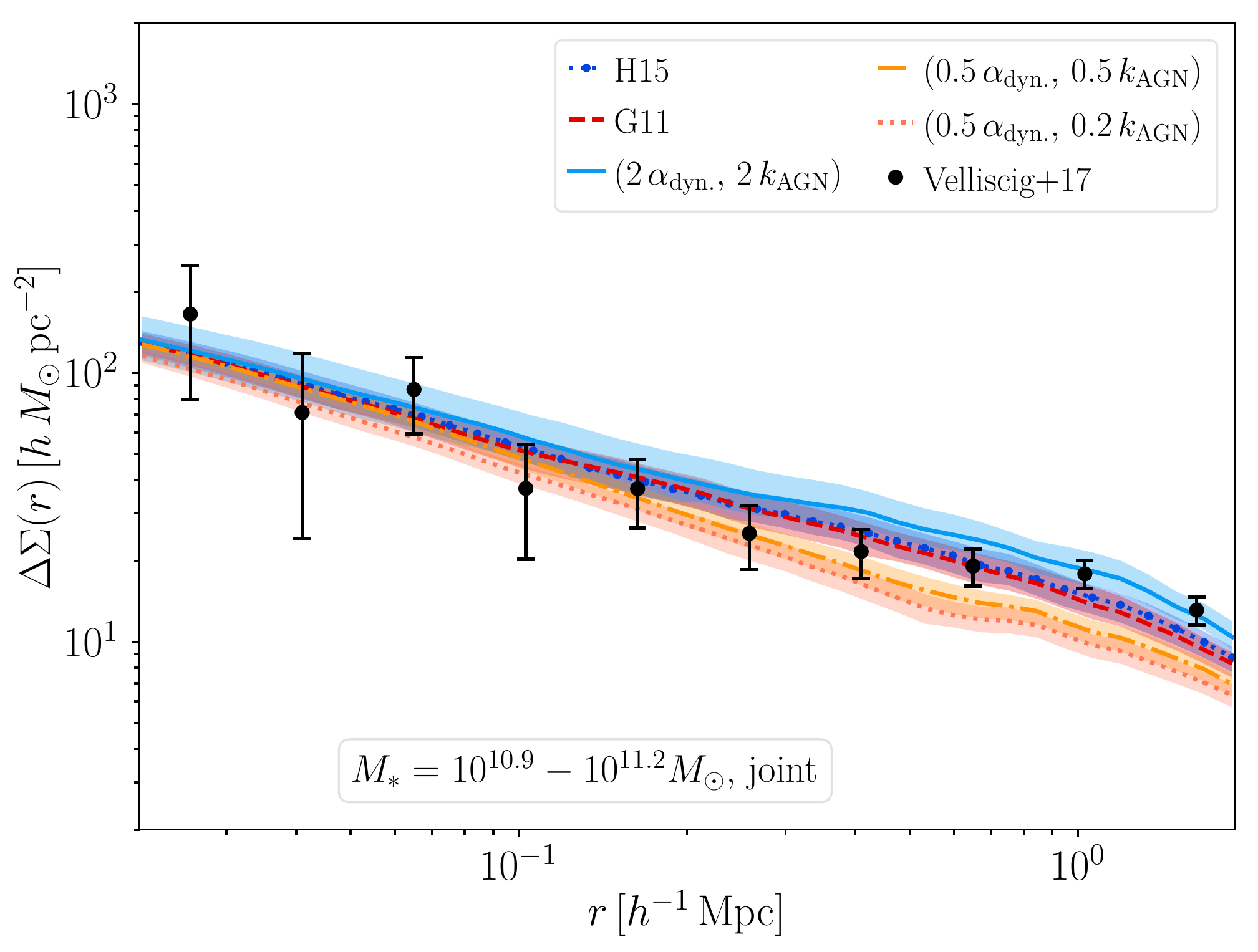}
\includegraphics[width=0.80\columnwidth]{\figrelpath 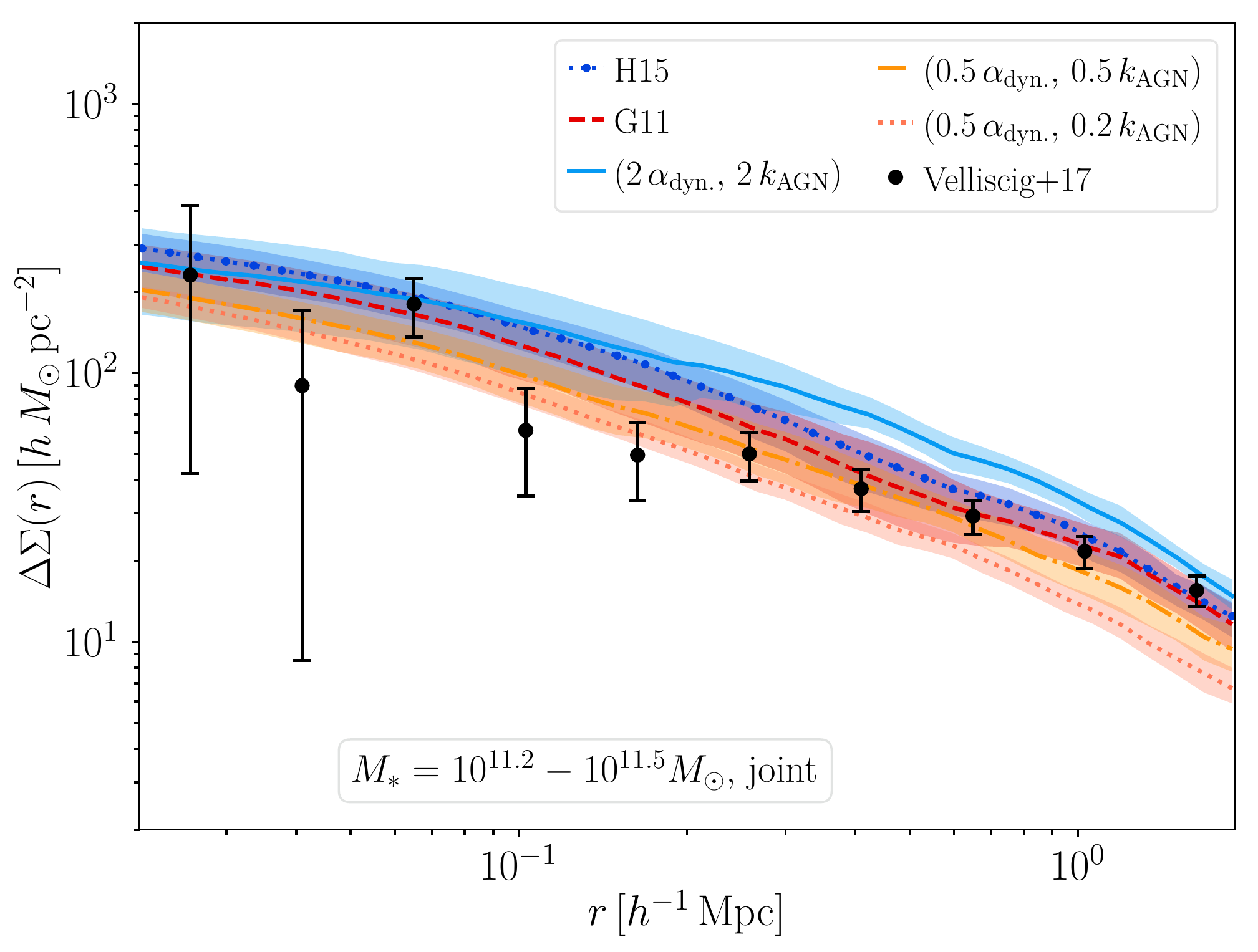}
\caption{Joint GGL signals for central and satellite galaxies w.r.t. data from \citet{2017MNRAS.471.2856V}. Here we see that matched satellite fractions could ensure an agreement with observations, in relation to the \citet{2016MNRAS.459.3251V} comparison, where the fractions did not necessarily agree. Originating from their problems to match the large scale ($r \sim 1\,h^{-1}\,\text{Mpc}$) satellite lensing signal, the $ (0.5\, \alpha_\text{dyn}, \, 0.5\,k_\text{AGN} )$ and $ (0.5\, \alpha_\text{dyn}, \, 0.2\,k_\text{AGN} )$ models do not conform well with the joint signal data points for the three most massive bins in this radial range.}
\label{fig:jointSignalVelliscig17}
\end{centering}
\end{figure*}

Compared to the quoted values in \citet{2017MNRAS.471.2856V} listed in
Table~\ref{tab:velliscigEAGLELGComparison} for the hydrodynamical
EAGLE simulation \citep{2015MNRAS.446..521S, 2015MNRAS.450.1937C}, we
see that the limiting stellar masses $M_\ast^\text{lim}$ are around 0.3-0.5\,dex higher in the SAM. This translates to higher host halo masses in
Table~\ref{tab:velliscigLGComparisonsFiducialBestNew} in order to satisfy the observational matched satellite fraction
$f_\text{sat}$ criteria. By comparing the values in
Table~\ref{tab:velliscigEAGLELGComparison}
and~\ref{tab:velliscigLGComparisonsFiducialBestNew}, we observe that they are
consistent with one another, although the rescaled MR has better
statistics than TNG100-DMO, which means that volume should have a negligible effect on the central galaxy lensing signal. In general, a more massive stellar mass bin requires a
higher $M_\ast^\text{lim}$ for the group membership criteria, although
this is not necessarily true for all derivative
\citetalias{2015MNRAS.451.2663H} models, which we see in Table~\ref{tab:velliscigLGComparisonsFiducialBestNew}. By changing the merger criteria for the satellites, we are able to reduce $M_\ast^\text{lim}$ for the $10.9 < \log_{10} M_\ast [M_\odot] < 11.5$ mass bins; but it is raised for the $10.6 < \log_{10} M_\ast [M_\odot] < 10.9$ bin. We shall see that this does not affect the agreement with the observations in this mass bin; and the reduction in the other two bins help to bring about better lensing predictions.

One may
question how sensitive the host halo distributions and the lensing signals are to the explicit value of
$M_\ast^\text{lim}$. It is true that the satellite fraction $f_\text{sat}$ only evolves
slowly with an increased $M_\ast^\text{lim}$ for the lowest group mass
bins due to the large number of satellites, which allows for a larger range of
viable $M^\text{lim}_\ast$. By computing the average host halo masses as well as the altered satellite fractions for $M^\text{lim}_\ast$ in the range $M^\text{lim,\,fid}_\ast \pm 50\,\%$ for the \citetalias{2015MNRAS.451.2663H} model run on the rescaled MR simulation, we discern that the average host halo mass for
centrals is robust to moderate variations of $M^\text{lim}_\ast$ with
only $\pm \,0.1\, \text{dex}$ changes which induce modest relative changes to $f_\text{sat}$ compared to $f_\text{sat, \,GAMA}$ (at most $\sim 15-20\,\%$ for the second most massive bin).

We determine that the host halo masses for the lowest mass bin differ
by about 1\,dex between EAGLE and the SAMs, the
predicted lensing signal for central galaxies from the SAMs are still consistent with the observations due to the large error bars; and the models yield similar signals with a slight shift between the \citetalias{2015MNRAS.451.2663H} and \citetalias{Guo:2010ap} models. For the central signal in
Fig.~\ref{fig:centralSignalVelliscig17}, all models are in agreement
with data below $10^{10.9}\,M_\odot$. If we start to
modify the SAMs to achieve better agreement for the higher stellar mass bins, changing
$\alpha_\text{dyn}$ and $k_\text{AGN}$, can affect the
signal and average host halo masses, with differences starting to show up from the $\log_{10} M_\ast [M_\odot] > 10.6$ bin onward. If we reduce $\alpha_\text{dyn}$
and $k_\text{AGN}$ we are able to obtain more consistent values with
the central galaxy signals for $10.6 < \log_{10} M_\ast \left [ M_\odot
  \right ] < 11.5$. This is especially true for stellar masses
exceeding $10^{10.9}\, M_\odot$ where we illustrate the shift in the
host halo mass distribution in
Fig.~\ref{fig:hostHaloDistThirdMassBinVelliscig17} compared to that of
the reference \citetalias{2015MNRAS.451.2663H} model. Increasing
$\alpha_\text{dyn}$ and $k_\text{AGN}$ produces a similar SMF, but
causes a shift in the host halo mass distribution away from the
observations. The best fit LBG model $(0.5\,
\alpha_\text{dyn}, \, 0.2\,k_\text{AGN} )$ gives the lowest average host halo
mass and the best central lensing signal for $10.6 < \log_{10}
M_\ast \left [ M_\odot \right ] < 11.5$. When it comes to the
satellite signal, however, this model does not do quite as well as the
fiducial models for $10.9 < \log_{10} M_\ast \left [ M_\odot \right ] <
11.5$. This small error then propagates
into the joint signal as seen in
Fig.~\ref{fig:jointSignalVelliscig17}. Hence, while it is the best
model for LBG lensing and for most of the central galaxy signals in
\citet{2017MNRAS.471.2856V}, it still needs refinements to adhere to the satellite lensing signal. To conclude, we see that our best
constraint model $ (0.5\, \alpha_\text{dyn}, \,
0.2\,k_\text{AGN} )$ conforms reasonably well with the new
dataset for centrals, validating our approach.

If we consider the two fiducial models
\citetalias{2015MNRAS.451.2663H} and \citetalias{Guo:2010ap}, they
give similar predictions, especially for the satellite lensing signals. For the central
galaxies, the \citetalias{Guo:2010ap} model predicts a lower signal
for the $10.3 < \log_{10} M_\ast \left [ M_\odot \right ] < 10.6$ bin and is
greater by an almost equal amount for $10.6 < \log_{10} M_\ast \left [
  M_\odot \right ] < 10.9$ and the two are equal for $10.9 < \log_{10}
M_\ast \left [ M_\odot \right ] < 11.2$ and $11.5 < \log_{10} M_\ast \left
[ M_\odot \right ] < 11.8$. Thus this dataset cannot be used to
discriminate them against each other.

Compared to the stellar mass-only selection, we are able to agree with
the joint and satellite lensing measurements in \citet{2017MNRAS.471.2856V} with all SAM models in the
comparison, highlighting the need for isolation and group membership
information from future surveys. We also obtain better agreement for
the satellite lensing signal than EAGLE, which had a suppressed
amplitude of the central bump, for $11.2 < \log_{10} M_\ast [M_\odot] < 11.8$. They argued that this was caused by the
small simulation volume, but as we see for the SAMs run on TNG100-DMO this
does not necessarily have to be an issue, although we have very few
galaxies in our most massive bins; see the good agreement around $r \sim 1-2 \, h^{-1}\,\text{Mpc}$ in Fig.~\ref{fig:jointSignalVelliscig17}. If we compare the lensing signal
for the \citetalias{2015MNRAS.451.2663H} model run on the rescaled MR simulation,
we find that the signal amplitude is slightly higher due to the
presence of more massive haloes, specifically for the $10.6 < \log_{10}
M_\ast \left [ M_\odot \right ] < 10.9$, $10.9 < \log_{10} M_\ast \left [
  M_\odot \right ] < 11.2$ and $11.5 < \log_{10} M_\ast \left [ M_\odot
  \right ] < 11.8$ bins (the rescaled MR actually gives a slightly
lower lensing signal for the $11.2 < \log_{10} M_\ast \left [ M_\odot
  \right ] < 11.5$ bin). For the satellite lensing signal the central
bumps are less prominent due to better statistics and larger spread in the
average distances between the satellites and their centrals, which
introduces a smoothing between the central subhalo lensing signal and its
host central. We have to wait for larger observational datasets to see
if this feature is also present in such surveys, and not intrinsic to how we construct mock catalogues in different simulation volumes.

To recap, we have shown that the modified SAM with $ (0.5\, \alpha_\text{dyn}, \,
0.2\,k_\text{AGN} )$ is able to fit the central galaxy group signals across all covered stellar mass bins in the sample, and that it does moderately well for joint central and satellite galaxy group lensing. This bodes well for the possibility to forecast the signal for upcoming high-$z$ lensing surveys.

\subsubsection{Hydrodynamical predictions and baryonic impact}\label{sec:groupCriteriaTNG}

\begin{figure*}
\begin{centering}
\includegraphics[width=0.65\columnwidth]{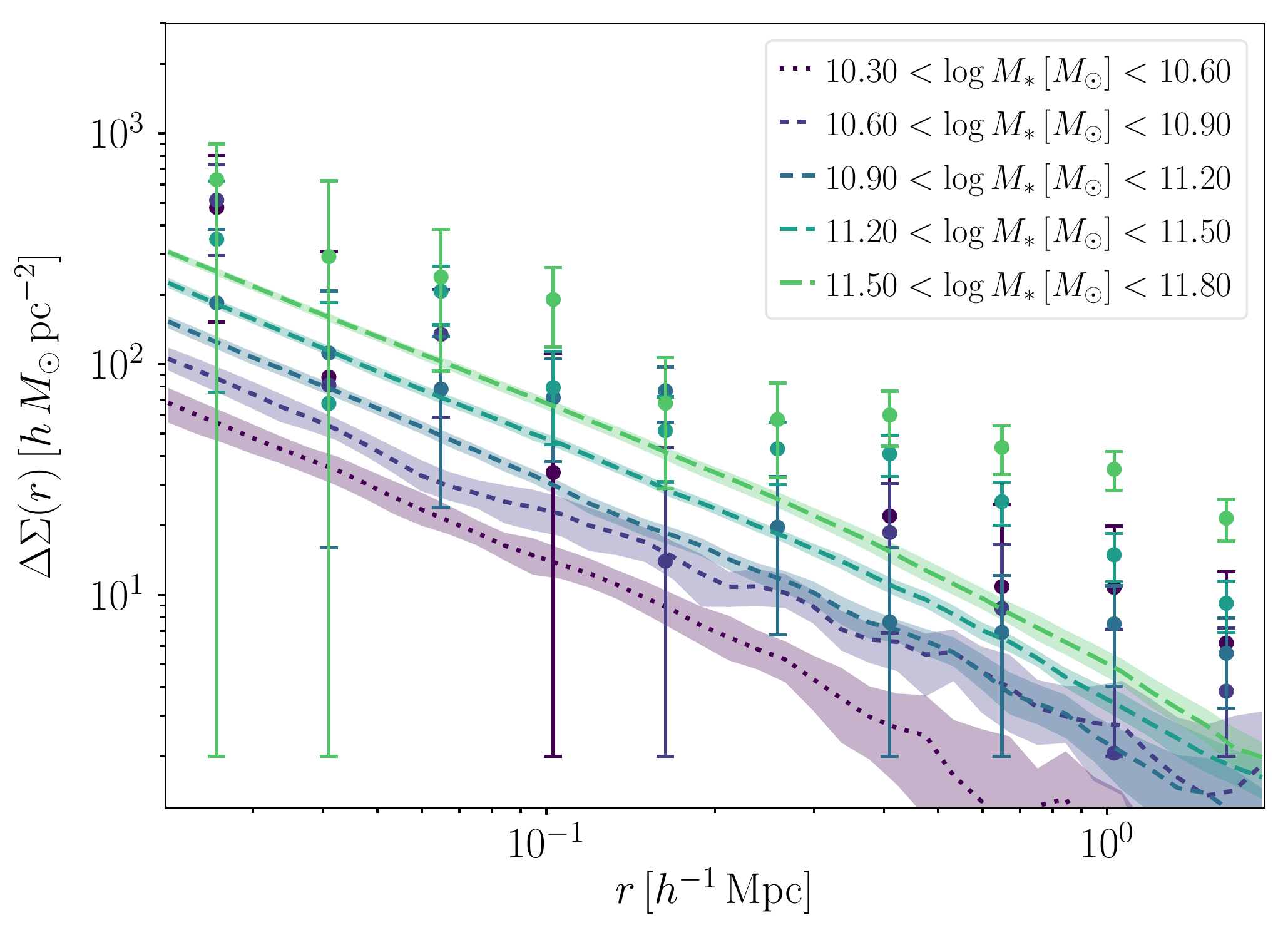}
\includegraphics[width=0.65\columnwidth]{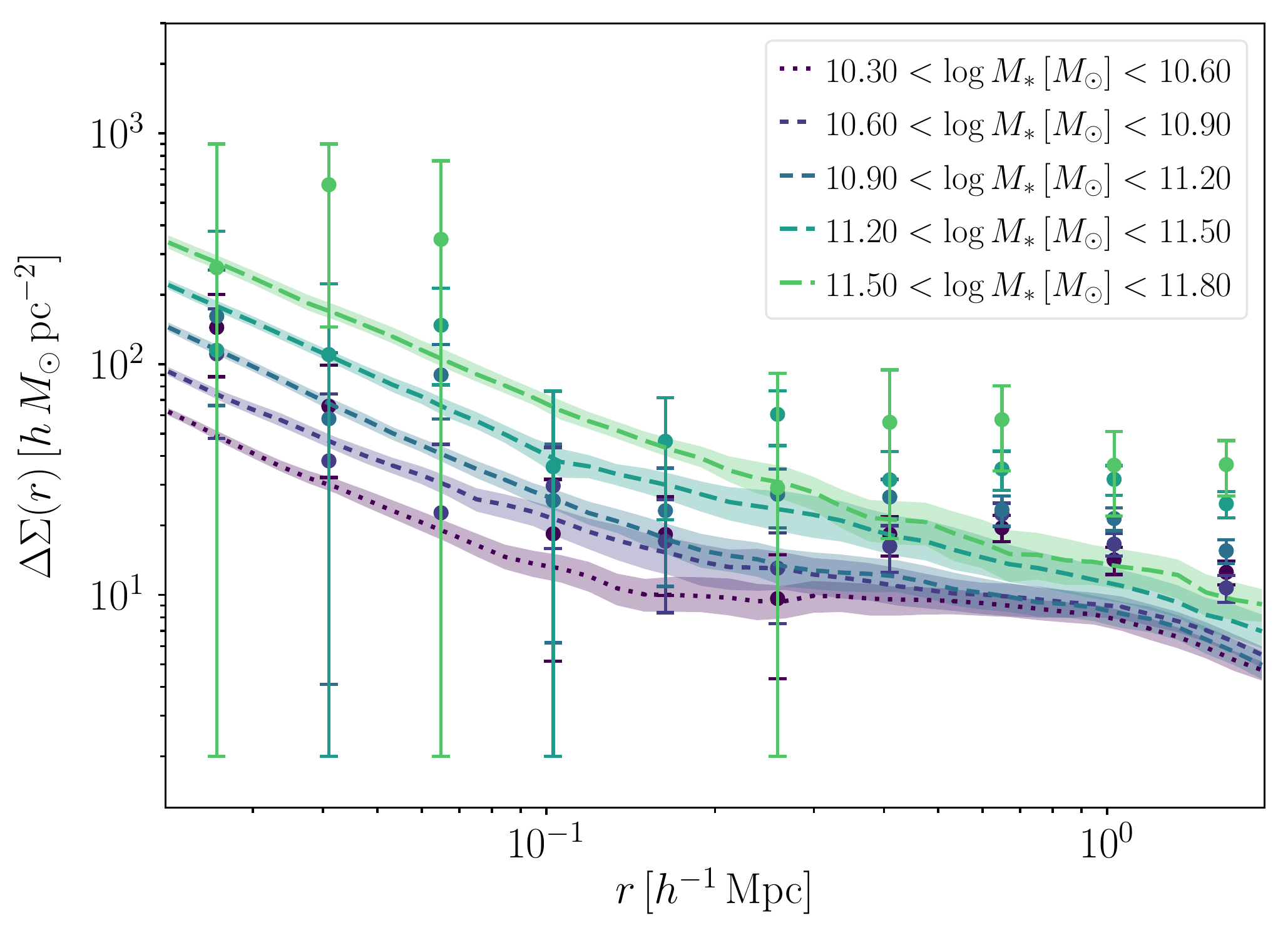}
\includegraphics[width=0.65\columnwidth]{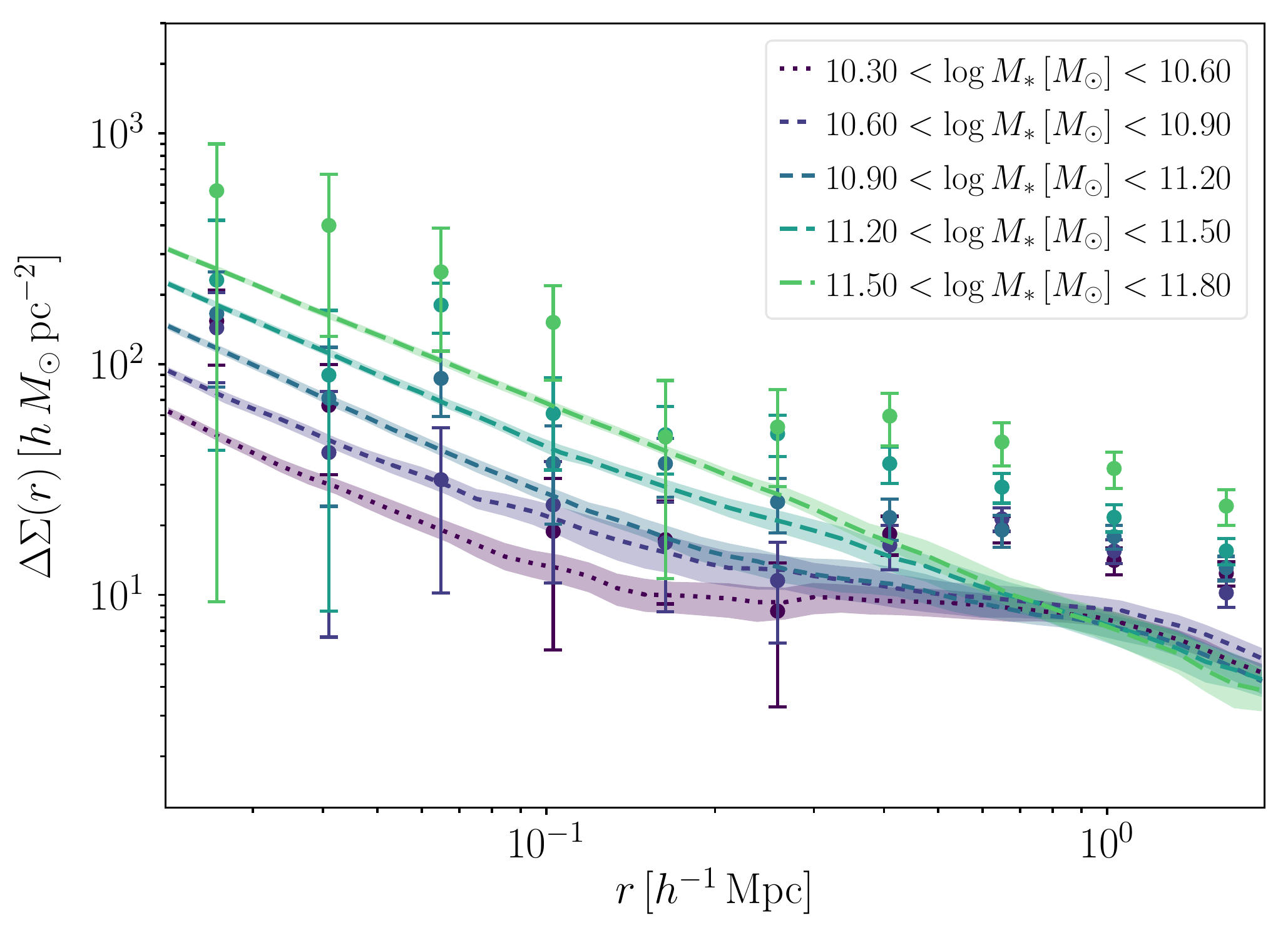}
\includegraphics[width=0.65\columnwidth]{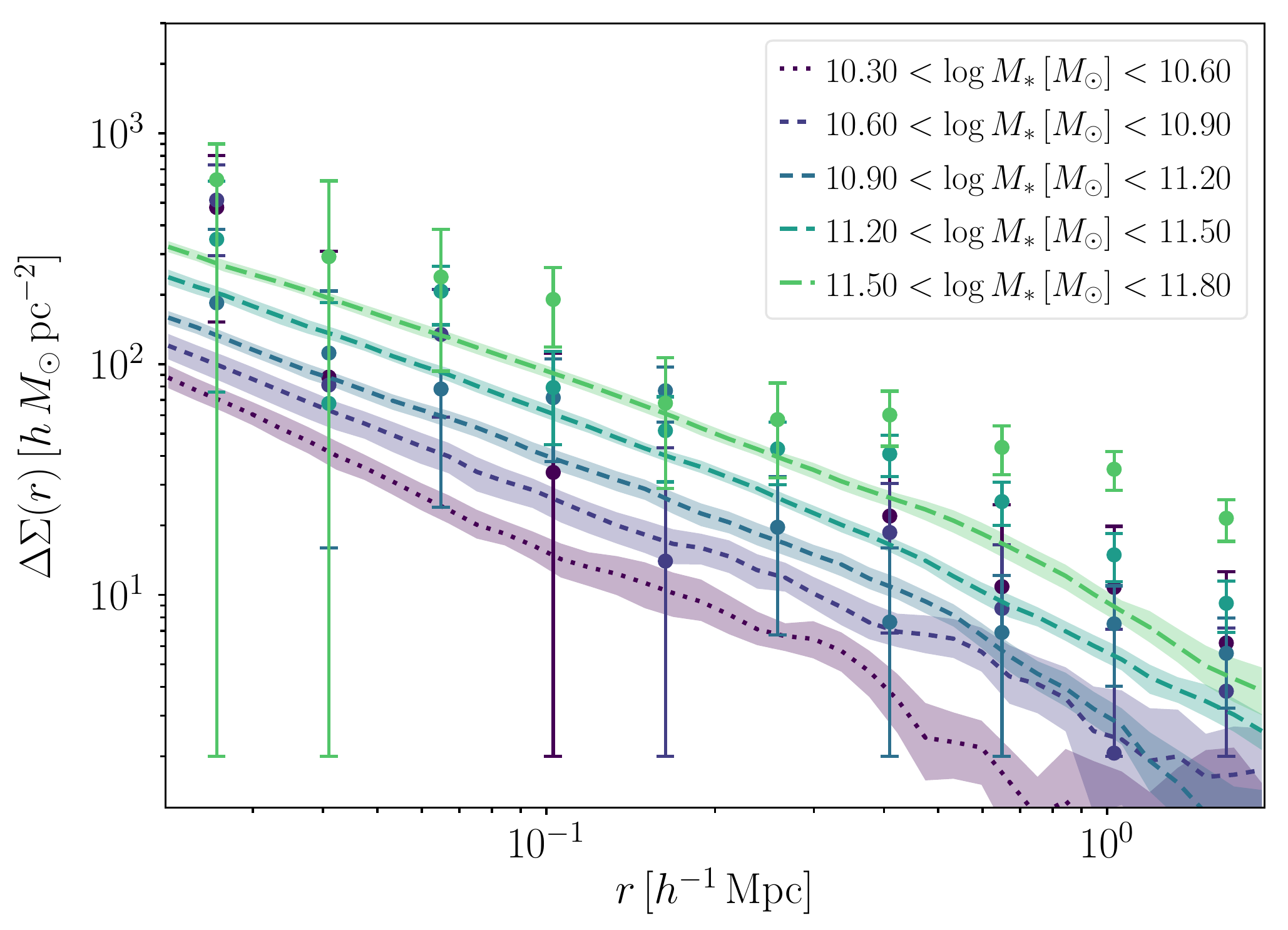}
\includegraphics[width=0.65\columnwidth]{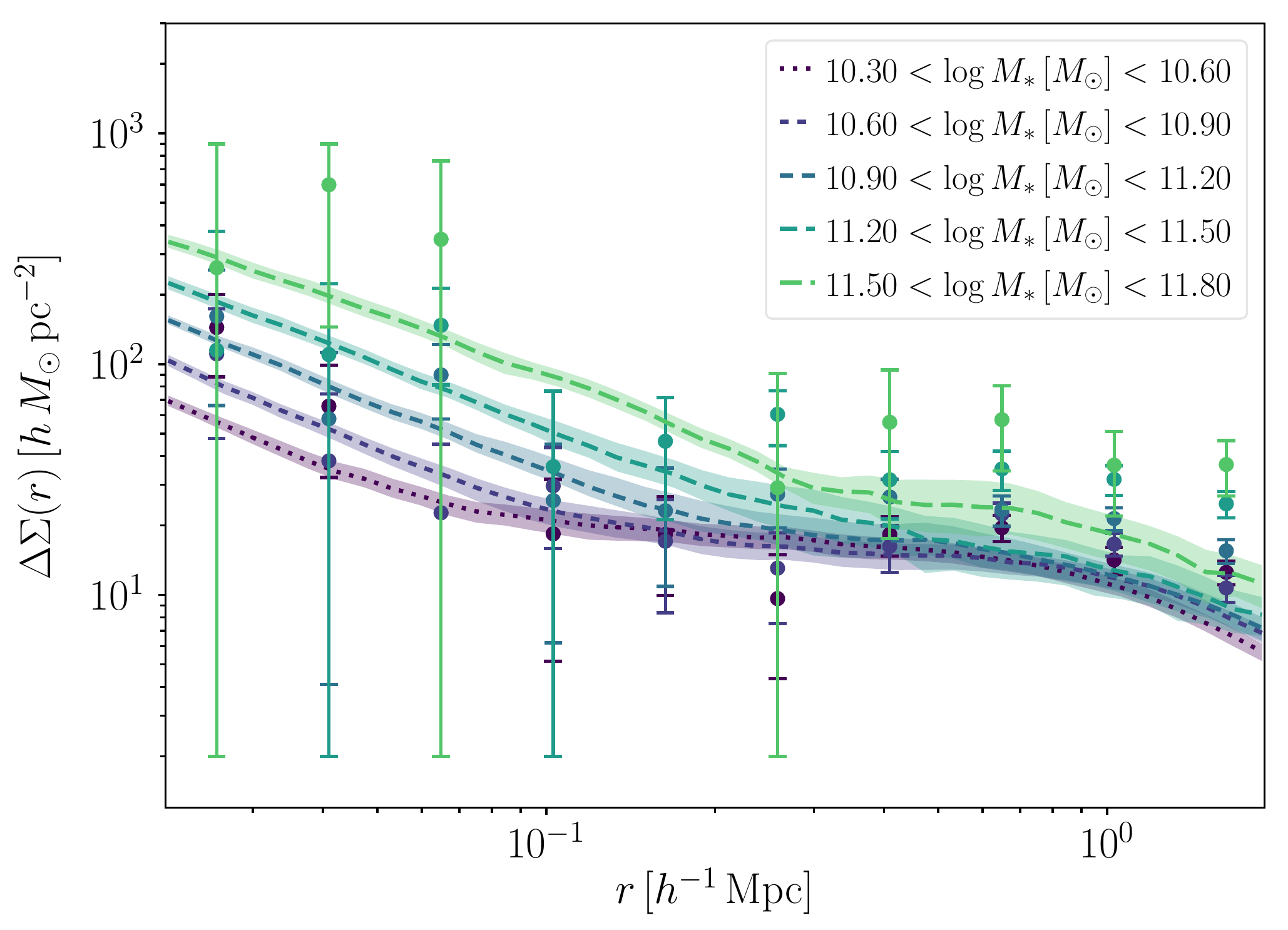}
\includegraphics[width=0.65\columnwidth]{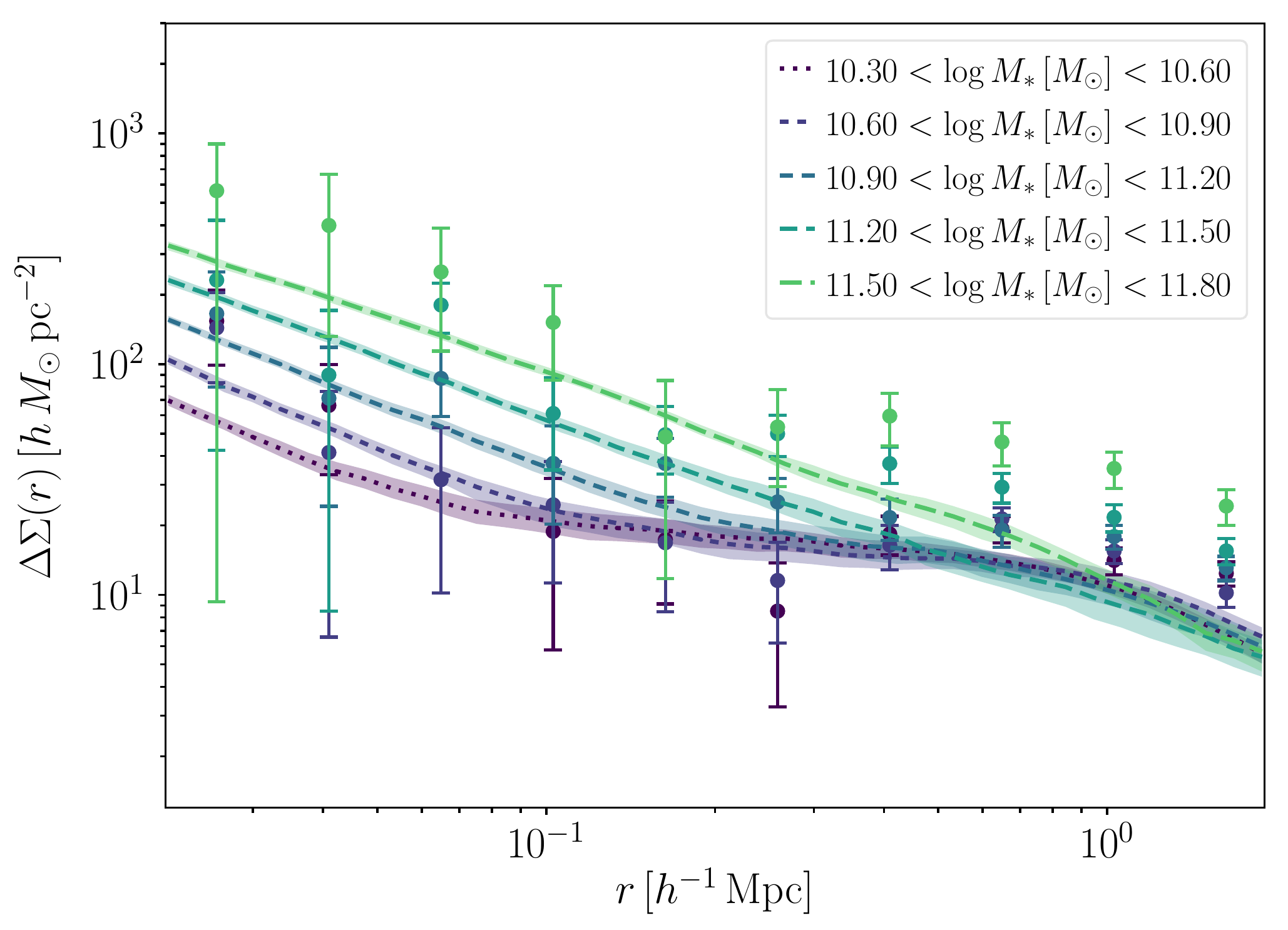}
\includegraphics[width=0.65\columnwidth]{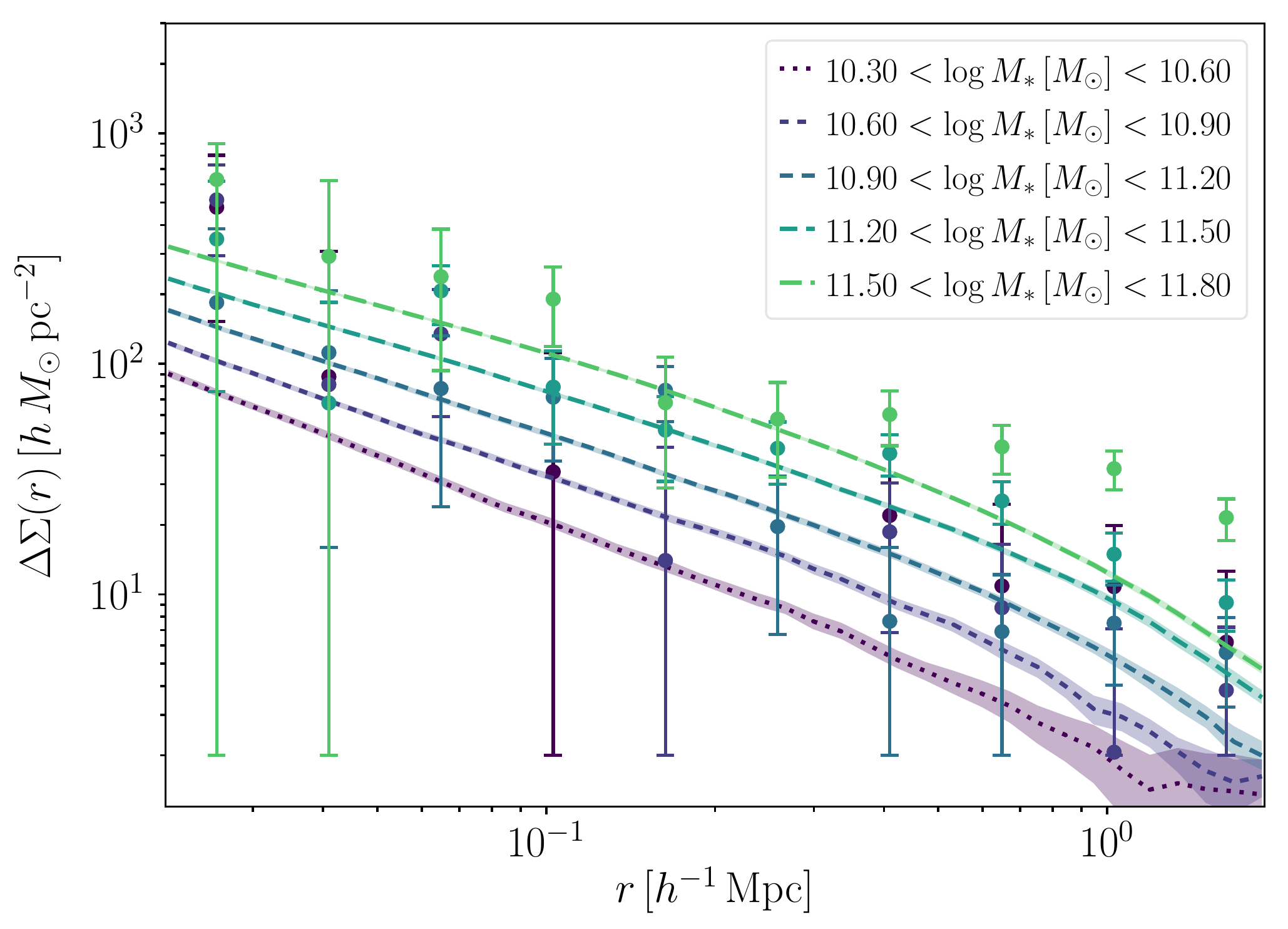}
\includegraphics[width=0.65\columnwidth]{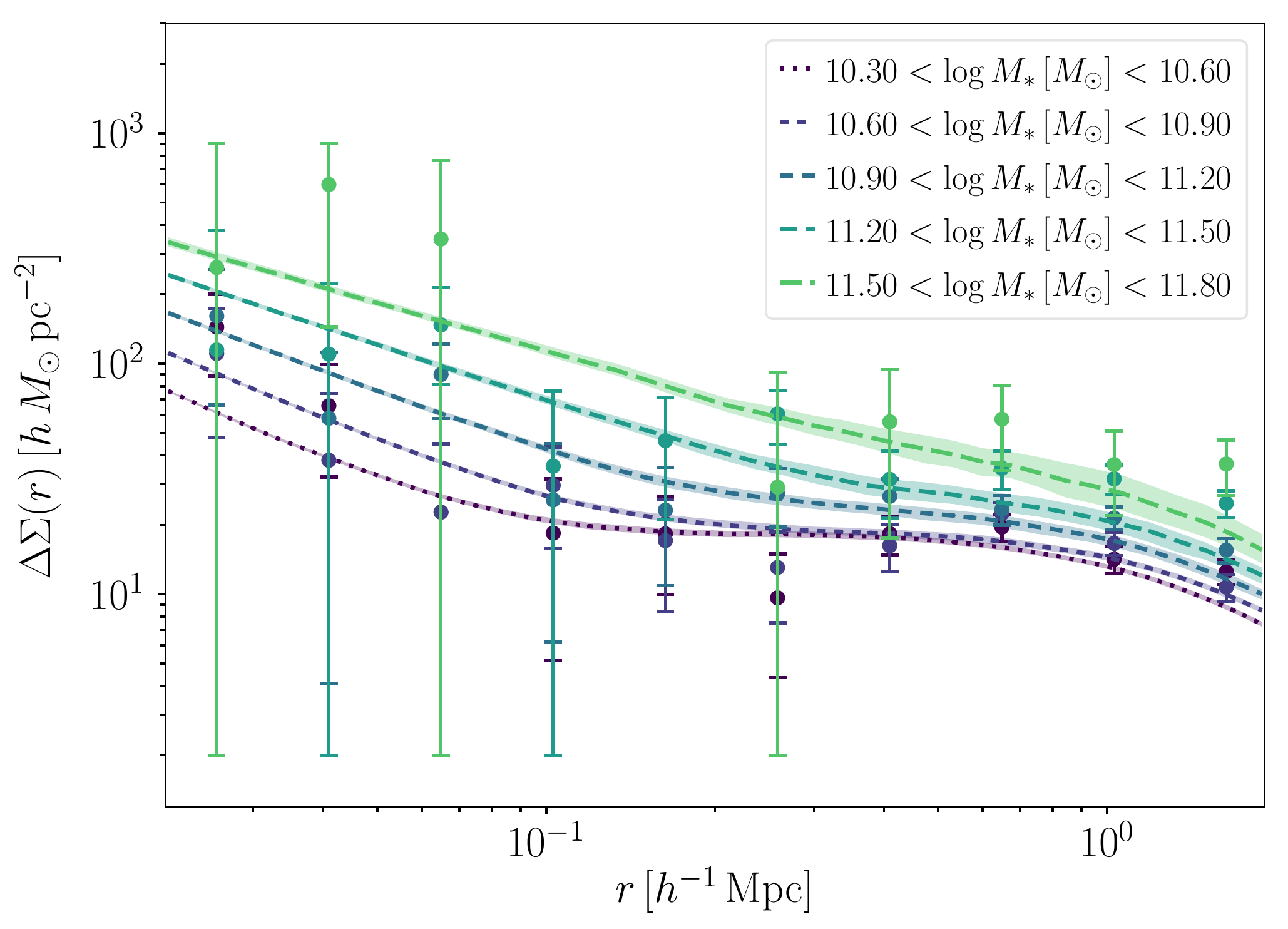}
\includegraphics[width=0.65\columnwidth]{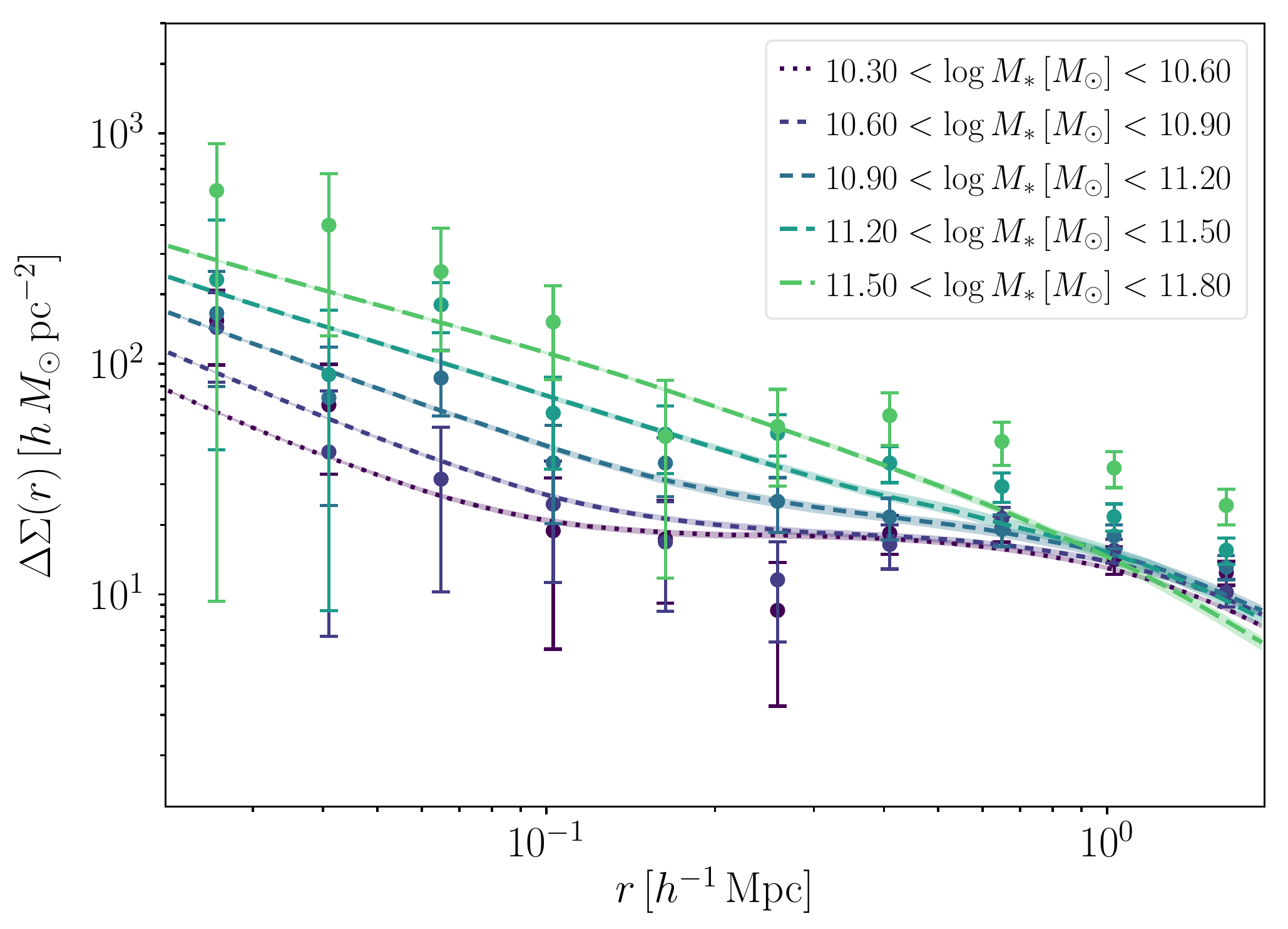}
\caption{Predicted lensing signals from Illustris (\emph{first row}), TNG100 (\emph{middle row}) and TNG300 (\emph{last row}) compared to observations for galaxy group members from \citet{2017MNRAS.471.2856V} for centrals (\emph{first column}), satellites (\emph{middle column}) and the joint combined signal (\emph{last column}). Illustris generally has problems to match the observations, particularly for centrals (\emph{upper left figure}), whereas TNG100 and TNG300 agree better except for in the outer region for the two most massive bins, similarly as EAGLE. Compared with the SAM signals in Figure~\ref{fig:jointSignalVelliscig17}, the hydrodynamical simulations do not succeed at predicting the joint lensing signal around $r \sim 1\,h^{-1}\,\text{Mpc}$, although the larger volume of the TNG300 simulation brings about a better agreement for the satellite signals.}
\label{fig:vellisig17AllIllustris}
\end{centering}
\end{figure*}

\begin{table*}
	\centering
	\begin{tabular}{l c c c c c  c c c c c}
	\hline
		$\log_{10} M_\ast [M_\odot]$ & $\bar{d}_\text{I}^\text{all}$ & $\bar{d}_\text{I-DMO}$ & $\bar{d}_\text{I}^\text{matched}$ & $\bar{d}_\text{I-DMO}^\text{matched}$ & $M_{200\text{c}}^\text{cen., I}$ & $M_{200\text{c}}^\text{sat., I}$ & $M_{200\text{c}}^\text{cen., I-DMO}$ & $M_{200\text{c}}^\text{sat., I-DMO}$ & $N_\text{gal}$ & $M_\ast^\text{lim}$\\
		\hline
		{\bf Illustris} & & & & & & & & & & \\
		\hline
		$10.3 - 10.6$ & 0.657 & 0.725 & 0.752 & 0.725 & 12.06  & 13.52 & 12.02 & 13.59 & 419 (292)  & 9.34  \\
		$10.6 - 10.9$  & 0.698 & 0.855 & 0.894 & 0.860 & 12.41 & 13.56 & 12.53 & 13.66 & 265 (174) & 9.64\\
		$10.9 - 11.2$ & 0.686 & 0.767 & 0.842 & 0.767 & 12.53 & 13.51 & 12.63 & 13.60 & 256 (184) & 9.39\\
		$11.2 - 11.5$ & 0.785 & 1.017 & 0.981 & 1.024 & 12.79 & 13.68 & 12.90 & 13.74 & 159 (131) & 9.51\\
		$11.5 - 11.8$ & 1.283 & 1.208 & 1.370 & 1.203 & 12.99  & 13.81 & 13.14 & 13.84 & 125 (119) & -\\
		\hline
	{\bf TNG100} & & & & & & & & & & \\
	\hline
		$10.3 - 10.6$ & 0.498  & 0.684 & 0.627 & 0.685 & 12.12 & 13.65 & 12.11 & 13.62 & 619 (377)  & 8.94 \\
		$10.6 - 10.9$ & 0.636 & 0.783 & 0.750 & 0.783 & 12.56 & 13.72 & 12.51 & 13.71 & 407 (290) & 9.44 \\
		$10.9 - 11.2$  & 0.668 &  0.864 & 0.811 & 0.864 & 12.78 & 13.75 &12.83 & 13.74 & 216 (160) & 9.74 \\
		$11.2 - 11.5$ & 0.934 & 1.207 & 1.060 & 1.116 & 13.02 & 13.88 & 13.06 & 13.91 & 104 (90) & 9.90 \\
		$11.5 - 11.8$ & 1.103 & 1.256 & 1.234 & 1.232 & 13.34 & 13.90 & 13.36 & 13.83 & 80 (77) & -\\
		\hline
		{\bf TNG300} & & & & & & & & & & \\
	\hline
		$10.3 - 10.6$ &  0.588 & 0.791 & 0.733 & 0.792 & 12.35 & 13.85 & 12.26 & 13.84 & 9427 (5795)  & 9.12 \\		$10.6 - 10.9$ & 0.675 & 0.886 & 0.833 & 0.886 & 12.70 & 13.92 & 12.71 & 13.90 & 5652 (3662) & 9.64 \\
		$10.9 - 11.2$  & 0.806 & 1.059 & 0.995 & 1.057 & 12.98 & 14.03 & 13.02 & 14.00 & 1981 (1354) & 10.00 \\
		$11.2 - 11.5$ & 0.990 & 1.250  & 1.164 & 1.242 & 13.28 & 14.10 & 13.32 & 14.06 & 978 (831) & 10.16 \\
		$11.5 - 11.8$ & 1.397 & 2.020 & 1.565 & 1.989  & 13.48 & 14.23 & 13.50 & 14.19 & 872 (842) & -\\
		\hline
	\end{tabular}
	\caption{Tabulated values for the \citet{2017MNRAS.471.2856V} dataset comparison for Illustris, TNG100 and TNG300 using bound stellar masses. The $f_\text{sat}$ values for the most massive bin are 0.26, 0.25 and 0.13 for Illustris, TNG100 and TNG300, respectively. We compare the galaxy groups to their dark matter counterparts, to the extent it is possible to find matches. The matching rate for centrals lies between $99-100\,\%$, but it is harder with satellite structures, meaning that we attain lower satellite fractions in the gravity-only groups. These fractions drop by $\approx 2-4\,\%$ for the $10.3 < \log_{10} M_\ast [M_\odot] < 10.9$ mass bins and by $\approx 10\,\%$ for $10.9 < \log_{10} M_\ast [M_\odot] < 11.2$, and $3-6\,\%$ for the most massive bin in the gravity-only runs. We list the average 3D distance between the satellites and the central galaxy for the full physics runs in column $\bar{d}_\text{I}^\text{all}$, for the gravity-only in $\bar{d}_\text{I-DMO}$, and for the matched satellite structures in the full physics and gravity-only runs in columns $\bar{d}_\text{I}^\text{matched}$ and $\bar{d}_\text{I-DMO}^\text{matched}$, respectively. The total number of galaxies in the full physics (gravity-only) runs are quoted in the $N_\text{gal}$ column.}
	\label{tab:velliscigEAGLEIllustrisTNGComparison}
\end{table*}

In Fig.~\ref{fig:vellisig17AllIllustris}, we show the corresponding profile measurements for Illustris, TNG100 and TNG300 for the central, satellite and the joint signal, respectively. Corresponding host halo masses and other statistical properties are quoted in Table~\ref{tab:velliscigEAGLEIllustrisTNGComparison}. The TNG suite is well-apt at predicting the group lensing signals, both for centrals, satellites and joint samples, yielding similar predictions as EAGLE. However, these simulations have difficulties in matching the joint lensing signal at scales $r \sim 1\,h^{-1}\,\text{Mpc}$. This was also previously found for EAGLE \citep{2017MNRAS.471.2856V}. TNG300 yields the best results in this regard, where the signal is boosted by $\sim 60\,\%$ with respect to TNG100 for the joint lens sample for the $11.2 < \log_{10} M_\ast [M_\odot] < 11.5$ mass bin at $r \sim 1 \,h^{-1}\,\text{Mpc}$. However, several of the SAMs run on TNG100-DMO manage better, suggesting that this cannot be fully attributed to the simulation volume. Still, since the TNG300 predictions match the lensing signal well for satellite galaxies (in the lower middle row) whereas the central signals (left plot in the lowest row) for the two most massive bins are low in the outer region, one could correct the deficiency by adding more centrals in more massive haloes, which are prevalent in larger boxes. If we plot the same quantities with rTNG300, this problem is alleviated and there is only a mismatch for the most massive bin. Hence, some caution should be taken when interpreting these results. Illustris (in the first row) does not yield sufficiently high central lensing signals, although the model performs reasonably well in the inner radial regions for satellite galaxies.

Regarding the tabulated properties, comparing Tables~\ref{tab:velliscigEAGLELGComparison} and \ref{tab:velliscigEAGLEIllustrisTNGComparison}, we find that the satellites in EAGLE are on average positioned marginally further out than those in TNG300 and the average central host halo masses in these two simulations agree fairly well. This also applies to the average satellite host halo masses, apart from the two most massive bins where the TNG300 values are more massive with $0.1\,$dex. The smaller volume TNG100 has smaller average host halo masses than TNG300 with shifts of $\sim 0.1-0.2\,$dex. These differences are larger for the average satellite host halo masses where they lie around $0.2-0.3\,$dex. This produces a better agreement for the satellite lensing signal (compare the signal from $r \sim 300\,h^{-1}\,\text{kpc}$ outwards in the middle column in the middle and last row in Fig.~\ref{fig:vellisig17AllIllustris}). Illustris has lower average central host and satellite halo masses than the other hydrodynamical simulations, and on average predicts more massive stellar masses for its galaxies, which yields a discrepancy in the number counts in the $N_\text{gal}$ column at the high mass end. Concerning the cutoff stellar masses, $M^\text{lim}_\ast$, Illustris, TNG100 and TNG300, in increasing order, prefer lower values than EAGLE; meaning lower values still than the SAMs. Since EAGLE, TNG100 and TNG300 are able to match the observations, the cutoff stellar masses should probably lie within their quoted range. Thus, here, the more consistent treatment of the satellite galaxies in galaxy groups in hydrodynamical simulations than in the SAMs affect the quality of the predictions.

Since we have had access to the gravity-only companion simulations, we can gauge how the statistics differ between the full physics and gravity-only runs. We find that the average host halo masses for the centrals and the satellite agree well for the TNG100 and TNG300 simulations. For Illustris, the central hosts are on average 0.1\,dex more massive and this also applies to the satellite hosts, apart from the two most massive bins. Although we cannot match all satellite galaxies between the runs, we can generally say for all simulations here that the satellites in the gravity-only runs are on average situated further away from the host halo centres. Closer objects have probably already merged with the central galaxies in the gravity-only runs, which yields lower satellite fractions in the gravity-only runs for the groups. If we look at the substructures which are bijectively matched, they lie on average further out than all satellite galaxies in the full physics run, but closer to the centres than their gravity-only counterparts, which supports the hypothesis. Neglecting the lower matching rates for the satellite structures, which puts the validity of the comparison into question, the position shift results in a shift of the central bump in the 2D projections. This introduces significant scatter in the baryonic deformations on scales $100\,h^{-1}\,\text{kpc} < r < 1\,h^{-1}\,\text{Mpc}$, which is hard to model. In addition, if one limits the analysis to the matched satellite signal (for $M_\ast < 10^{11.2}\,M_\odot$), it resembles the total satellite signal in the centre, but there are large reductions of the matched signal compared to the fiducial signal; and it has a steeper slope between the satellite term and the central bump term, which is smoothed for the full signal. In this region, the matched satellite signal is reduced by approximately a factor of two\footnote{This implies a lower host halo mass for the matched satellites than in the full signal.}, which together with its altered shape makes it a poor proxy. This is less of a problem for the two most massive bins, as the central bump decreases in prominence, but there are still considerable differences on scales $r > 300\,h^{-1}\,\text{kpc}$. These statements hold for all simulations in Table~\ref{tab:velliscigEAGLEIllustrisTNGComparison}; but the situation is slightly better in the EAGLE simulation for the $10.3 < \log_{10} M_\ast [M_\odot] < 11.2$ mass bins where these differences are the most prominent. The drop of the matched satellite signal with respect to the total satellite signal for the $10.3 < \log_{10} M_\ast [M_\odot] < 10.6$ mass bin at $r = 300\,h^{-1}\,\text{kpc}$ is $15\,\%$ compared to $\sim 50\,\%$ for the other models. We attribute this difference to the slightly different subhalo matching techniques. Further studies could clarify if varying the matching criterion could produce better total signal proxies to determine baryonic effects for satellite galaxies. These matching issues impede the ability to gauge the impact of baryons on the lensing signal, particularly for samples with high $f_\text{sat}$ fractions. For our analysis, this problem particularly affects the joint signal analysis for the bins with $M_\ast < 10^{11.2}\,M_\odot$, where the central fraction starts to exceed 50\,\%. This suggests that these effects are something future analyses can marginalise over for high-confidence central-dominated lens samples.

We already observed in Fig~\ref{fig:mSDSSRedBlueGGLIntermediateMass} that the matched and total red lensing signal differed for TNG300. Here the total signal agrees better with the KiDS+GAMA observations than the matched sub-sample, whose curves lie within the error bars for radial scales between the central satellite galaxy term and the central bump. Notwithstanding differences in the selection functions, this demonstrates that the models are capable to provide accurate (average) lensing profiles for satellites in some groups, and thus the problem for the red galaxies might be narrowed down to the colour and the precise group definition.

Lastly, we investigate the baryonic imprint on this group lens sample for the centrals, especially using TNG300. TNG100 and Illustris suffer from poor statistics, which affect the 1-to-2 halo transition regime at $r \sim 1-2\,h^{-1}\,\text{Mpc}$, where nearby and line-of-sight contributions from surrounding structures produces large scatter (which can amount to a factor 10 or more for the lowest mass bin) for the two lowest mass bins with $M_\ast < 10^{10.6}\,M_\odot$. Owing to its larger volume, TNG300 does not suffer to the same extent from this scatter, but it is partly visible for the lowest mass bin where the signal is computed for $\sim 20$\,galaxies ($f_\text{sat} = 98\,\%$). From $M_\ast > 10^{10.6}\,M_\odot$, the deformations are very similar to what was shown in Fig.~\ref{fig:barEffectsIllustrisTNGEAGLE}. We note a slightly weaker trend with decreasing deformation at the very massive end, observed for both the stellar mass only centrals in Fig.~\ref{fig:barEffectsIllustrisTNGEAGLE} and in Fig.~\ref{fig:tng300LBGMstarSelectionRedBlue} for red centrals. This can be due to the wider mass bin in Fig.~\ref{fig:barEffectsIllustrisTNGEAGLE}, and our restriction to galaxy groups where the AGN feedback is the strongest. The results from Illustris, EAGLE and TNG100 are also very similar to their stellar mass-only central counterparts. Again, TNG100 suggests a shallower deformation than TNG300 by a few percent. For the two most massive bins, the maximum suppression for EAGLE (by about $\sim 10\,\%$) lies more in line with TNG100 than TNG300.

\section{Discussion}\label{sec:discussion}

We have carried out a comparison between different SAMs as well as 
IllustrisTNG for different galaxy-galaxy lensing and galaxy clustering
datasets and found broadly satisfactory agreement across several of them.

Starting with the TNG300 results, the simulation can achieve excellent lensing as well as clustering predictions, passing yet another milestone in physical modelling, for several of the datasets covered, especially for stellar-mass only ($\chi^2 = 1.81$) and group lens samples and LBG lensing, for all ($\chi^2 = 3.80$) and red galaxies $(\chi^2 = 3.06)$, and blue if one neglects the most massive bin. In addition, it is interesting to note that this simulation can help to ameliorate the construction and parameter choices of semi-analytical models, suggesting a future mutually beneficial relationship between the two frameworks to arrive at realistic synthetic catalogues for Gigaparsec volumes. The dependency of the (red) galaxy clustering predictions on the stellar-mass resolution correction, shown in Fig.~\ref{fig:clusteringHenriques17ExtremeVaryingBin2RedBlue}, highlights the future requirement to construct hydrodynamical simulations which feature e.g. resolution-independent recipes for star formation  \citep[cf. the discussion in][regarding weak vs. strong numerical convergence for parameter values in the calibration steps]{2015MNRAS.446..521S, 2018MNRAS.475..648P} to obtain the proper volume scaling relations, or to directly calibrate for large simulation runs, as was performed for e.g. the BAHAMAS suite \citep{2017MNRAS.465.2936M, 2018MNRAS.476.2999M}.

A residual challenge is the modelling of the SDSS-DR7 stellar mass + colour sample, both for TNG300 and the different SAMs, 
where we cannot reach sufficient agreement even for extreme model
parameter variations in the SAM. Specifically, the tension we observe is that, for $10^{9.4} < M_\ast \left [ M_\odot \right ] < 10^{11.0}$, TNG300 and TNG100 (as well as rTNG300) predict a $\Delta\Sigma(r)$ for red galaxies which is generically above the SDSS data from \citet{2016MNRAS.457.4360Z} and \citet{2016MNRAS.457.3200M} for $r > 0.1-0.3 \, h^{-1}\,\text{Mpc}$ by up to a factor of two (i.e. several sigma). In contrast, the signal at $r < 0.1 \,h^{-1}\,\text{Mpc}$ as well as for blue galaxies at all distances is within roughly 1 sigma of the data. This mass range is the most problematic for the simulation, and at higher masses (across the whole radial range below $3\,h^{-1}\text{Mpc}$) the agreement with observations is better. The excess red signal at this mass range could be due to internal processes (i.e. the TNG feedback model at these mass scales) or external processes (i.e. an overestimated environmental impact of stripping due to too-large background gas densities). This issue propagates into the total lensing signal, where the model suggests a signal increase from $z = 0.3$ to $z = 0$, especially in the $r > 0.1 \, h^{-1}\,\text{Mpc}$ region, which is disfavoured by the data. To solve this will probably require major work on the physical processes involved in the
quenching of satellite galaxies for signal agreement around $r \sim 1\,h^{-1}\,\text{Mpc}$, as seen in Fig.~\ref{fig:zmRedGGLComponentSeparated}. Traditionally, a way to boost
galaxy clustering has been to populate massive galaxies with more
satellite galaxies and we see that this might influence the lensing. The lensing data instead suggest that one should preferentially populate lower mass host haloes with red satellites. If we compare to the iHODs from
\citet{2016MNRAS.457.3200M}, they have no trouble getting the red
lensing signal right, although we are able to produce equally well or
slightly superior predictions at the extreme low mass end. We have
also checked the halo occupation distributions for our SAMs compared
to those shown in \citet{2015MNRAS.454.1161Z} for the iHOD setup for
stellar masses at $\log_{10} M_\ast = 10 \, \pm \, 0.1\left [M_\odot \right
]$ and $\log_{10} M_\ast = 11 \, \pm \, 0.1\left [M_\odot \right ]$, and we
find adequate agreement for the central galaxies, but for
satellites the distribution has tails of the order of $0.5\,$dex
towards lower and higher masses. If we remove the orphans these two
tails disappear, and it is mostly the massive tail which dominates the
lensing signal. Hence, we deduce that the physical processes setting the stellar masses and colours of satellites 
need refinements in future simulations and SAMs to eliminate these errors. Still, the  agreement for the satellite lensing signals from the galaxy groups indicate that most of the machinery is correct. 

Our best fit LBG SAM, the $ (0.5\, \alpha_\text{dyn}, \, 0.2\,k_\text{AGN} )$ model, gives good predictions for the
\citet{2017MNRAS.471.2856V} central galaxy signals. As an additional validation step we have computed its SMF up to
$z = 3$ and its predicted red galaxy fractions with respect to the
fiducial \citetalias{2015MNRAS.451.2663H} model. As pointed
out in Section~\ref{sec:abundanceCorrections}, the new SMF is slightly too
high above the knee at $z = 0.11$ and this also applies to $z = 1$ but
given the stellar mass uncertainties, we argue that this is still consistent with the observations (and IllustrisTNG). We are thus
able to retain the good agreement to the SMF to $z = 3$, which is well beyond the depth of any near future lensing survey, meaning that
this simplified study is compatible with the fiducial MCMC constraints
used to tune \citetalias{2015MNRAS.451.2663H}. Yet, if we examine the red fractions, the new
model shifts the distribution away from the observational data
points. Still, the division is very sensitive to the actual shape of
the colour distribution, especially for stellar masses between $9.5 <
\log_{10} M_\ast \left [ h^{-2}\, M_\odot \right ] < 10.5$ where the
transition between blue and red is rapid. Thus, we do not put equal
weight on matching the colour cuts. For $z = 3$ we are in excellent
agreement with \citetalias{2015MNRAS.451.2663H}. Future endeavours
should focus on the incorporation of these lensing constraints into
the MCMC chains themselves for the model selection, but, as we have
shown, the modification of a few pertinent parameters is sufficient to
produce an acceptable improvement. We have also examined the SMF evolution of our best fit $0.1\,k_\text{AGN}$ model for stellar mass-only lenses, and it is also acceptable with respect to the data with the excess above the knee still present at $z = 1$, but within error bars. In \citet{2016MNRAS.456.2301W} it was argued that abundance
corrections could be used to bring the signal into agreement. We also
find that this is the case, but the effects are largest at the
high mass end where the uncertainties are considerable.

The approaching era of
precision cosmology requires a more profound understanding of
systematic effects such as the influence of baryons. With the baryonic
feedback prescriptions offered by TNG300, we find that we are still safe from
their impact for the current datasets by restricting the analyses to
scales $r > 30\,h^{-1}\,\text{kpc}$ to avoid the impact of the stellar
term. Furthermore, large scale cosmological analyses are mostly safe from baryonic processes if one restrict the scales to $r > 1-2 \,h^{-1}\,\text{Mpc}$, unless one considers more extreme formation models such as Illustris. That the baryonic effects on these scales are consistent for different feedback recipes is encouraging for analyses with explicit component modelling \citep[e.g.][]{2015JCAP...12..049S, 2019JCAP...03..020S, 2020MNRAS.495.4800A}, fitting formulae based measured simulation bias or extensions to the halo model \citep[e.g.][]{2015MNRAS.450.1212H, 2015MNRAS.454.1958M, 2020arXiv200500009M} or with principal components \citep[e.g.][]{2015MNRAS.454.2451E, 2019MNRAS.488.1652H} and other libraries \citep[e.g.][]{2020MNRAS.491.2424V} \citep[see e.g.][for a recent review of baryonic modelling]{2019OJAp....2E...4C}. Future progress should be directed towards understanding
the amplitude and scope of the suppression on intermediate radial
scales, which is especially important for group scale systems where
the effect of AGN feedback is the strongest. We have shown that this suppression both depends on the formation model, as well as the simulation volume. This is particularly important for further simulation campaigns where one should ensure that the baryonic  deformations for a given galaxy formation recipe are consistent. We also observed that the suppression depends on the colour of the galaxy sample, and is less pronounced for blue (central) galaxies. This could be important for galaxy surveys where one treats the systematics differently depending on colour. 

Further developments could also to be made to include additional 2-pt
statistics in the analysis, such as cosmic shear which has been shown
to offer interesting galaxy formation constraints
\citep[e.g.][]{2016MNRAS.463.3326F}, as well as lensing 3-pt statistics \citep[e.g.][]{ 2017A&A...601A..98S, 2020A&A...640A..59L}, where the \citetalias{2015MNRAS.451.2663H} SAM has been able to yield viable predictions in contrast to other SAMs.

\section{Conclusions}\label{sec:conclusions}

\begin{table*}
	\centering
	\begin{tabular}{ l l l l }
	\hline
		{\bf Dataset} & {\bf Reference} & {\bf Best model} & {\bf$\chi^2$}\\
		\hline
		 Stellar mass-only lensing & \citet{2016MNRAS.459.3251V} &  $0.1\,k_\text{AGN}$ & 1.67\\
		 LBG lensing (all) & \citet{2016MNRAS.456.2301W} & $(0.5 \,\alpha_\text{dyn}, \,0.2\,k_\text{AGN})$ & 1.59 \\
		 LBG lensing (red) & \citet{2016MNRAS.457.3200M} & $(0.5 \,\alpha_\text{dyn}, \,0.5\,k_\text{AGN}, \, 0.5 \,\epsilon_\text{reheat})$ & 1.12\\
		 LBG lensing (blue) & \citet{2016MNRAS.457.3200M} & $(0.3 \,\alpha_\text{dyn}, \,0.1\,k_\text{AGN})$ & 1.67\\
		 Clustering (all) & \citetalias{Guo:2010ap} & $0.1\,k_\text{AGN}$ & 0.72\\
		 Clustering (red) & \citet{2016MNRAS.457.4360Z} & $(0.3 \,\alpha_\text{dyn}, \,0.2\,k_\text{AGN})$ & 4.05\\
		 Clustering (blue) & \citet{2016MNRAS.457.4360Z} & $(0.3 \,\alpha_\text{dyn}, \,0.1\,k_\text{AGN})$ & 2.61\\
		\hline
	\end{tabular}
	\caption{List of the best SAMs for the constraints presented in this Paper.}
	\label{tab:samModelsBest}
\end{table*}

In this Paper we have analysed predictions from different semi-analytical
models of galaxy formation based on \textsc{L-Galaxies} and IllustrisTNG, TNG300 and TNG100, for galaxy-galaxy lensing and galaxy clustering datasets. We summarise the SAM configurations which performed the best for the different datasets in Table~\ref{tab:samModelsBest} and list the most important results below:
\begin{itemize}
    \item TNG300 is able to produce very good lensing predictions, especially for stellar mass-only selected lenses ($\chi^2 = 1.81$) (Section~\ref{sec:baryons}), but also for locally brightest galaxies (LBGs, Section~\ref{sec:lbgLensingTNG}) ($\chi^2 = 3.80$), whereas the \citetalias{2015MNRAS.451.2663H} SAM requires slight modifications to achieve the same level of agreement with fiducial performance at $\chi^2 = 7.79$ (Section~\ref{sec:mstarOnlyKiDSGAMALG}) and $\chi^2 = 5.01$ (Section~\ref{sec:lbgLensingSAMs}). One
can adjust the merger times and AGN feedback parameters of the SAMs to
bring about agreement at the high mass end, which improves the stellar-mass only lensing (Section~\ref{sec:mstarOnlyKiDSGAMALG}) and LBG lensing signals (Section~\ref{sec:lbgLensingSAMs}).
    \item TNG100 and TNG300 also produce consistent predictions for galaxy group lensing from \citet{2017MNRAS.471.2856V}, but have minor issues with matching the central galaxy signals for $M_\ast > 10^{11.2}\,M_\odot$ at $r \sim 1\, h^{-1}\,\text{Mpc}$ (Section~\ref{sec:groupCriteriaTNG}).
    \item We have found a tension (Section~\ref{sec:sdssColour}) for the red lensing
signal with excessive predictions approximately ranging from $50\,\%$ to a factor of two at $r\approx 0.6\,h^{-1}\,\text{Mpc}$ for $10.2 < \log_{10} M_\ast [h^{-2}\,M_\odot] < 11.0$ (for the SAMs and in both TNG100 and TNG300, where the tension at $r\approx 0.6\,h^{-1}\,\text{Mpc}$ also is present for $9.4 < \log_{10} M_\ast [h^{-2}\,M_\odot] < 10.2$).  Further work is required to model physics responsible for the (potentially stripped) satellite subhalo lensing signal in
both hydrodynamical simulations and in SAMs, to achieve the right colours and stellar masses in lower host mass haloes.

\item We observe that baryons reduce the lensing signal on intermediate radial scales for $0.1 < r [h^{-1}\,\text{Mpc}]\lesssim 1-2$ by 10-15\,\% depending on the galaxy evolution model and simulation volume used compared to gravity-only runs (Section~\ref{sec:baryons}). This reduction, however, is too small to explain why the hydrodynamical simulations (TNG100, TNG300, EAGLE and Illustris) perform better than the \citetalias{2015MNRAS.451.2663H} SAM for the stellar mass-only lenses for the \citet{2016MNRAS.459.3251V} dataset. We also find that the deformation is different for red and blue central galaxies at the same stellar mass (Section~\ref{sec:lbgLensingTNG}), with the suppression more prominent for red galaxies at $\approx15\,\%$ for $10.4 < \log_{10} M_\ast [M_\odot] < 11.6$ compared to a few percent for blue galaxies, and that this reduces the inferred host halo bimodality, both in TNG300 and TNG100.

\item With the dust model from  \citet{2018MNRAS.475..624N} applied to the colours, the observed tension with SDSS-data for red galaxies in TNG300 with $9.5 < \log_{10} M_\ast [h^{-2}\,M_\odot] < 10$ in \citet{2018MNRAS.475..676S} is reduced (Section~\ref{sec:clustering}). By comparing the clustering signal from the \citetalias{Guo:2010ap} and \citetalias{2015MNRAS.451.2663H} SAMs with and without dust, we find that for red galaxies with $M_\ast < 10^{11}\,h^{-2}\,M_\odot$ dust reduces the signal $30-40\,\%$
whereas the effect for the 2-halo term is $\sim 10-20\,\%$ depending
on the galaxy formation model. The effect on the lensing signal is more modest, with the dusty red signal from \citetalias{2015MNRAS.451.2663H} suppressed with
at most $\approx 15\,\%$ for $10.6 < \log_{10} M_\ast \left [
  h^{-2}\,M_\odot \right ] < 11.0$ (Section~\ref{sec:sdssColour}).

\item By combining lensing data across different stellar masses and by
adding clustering information, we identify a few viable parameter
combinations in the SAM for the central galaxy signal which we have verified to be consistent with the external group membership datasets from
\citet{2017MNRAS.471.2856V} (Section~\ref{sec:lgalaxiesGroupLensing}). Our best fit models suggest a weaker AGN radio-mode
feedback by $80-90\,\%$ from the fiducial MCMC parameter value (Section~\ref{sec:mstarOnlyKiDSGAMALG}), and for central-dominated samples also a shorter dynamical friction merger time multiplier by approximately $80\,\%$ from the value in
the fiducial \citetalias{2015MNRAS.451.2663H} model (Section~\ref{sec:lbgLensingSAMs}), and they retain
the good agreement with the SMF at $z = 0$ up to $z = 3$ (Section~\ref{sec:discussion}), making them
suitable for future lensing and clustering surveys. 
\end{itemize}
Thus we conclude
that joint 2-pt statistics analyses are powerful, promising
approaches to constrain galaxy formation.

\section*{Acknowledgements}

The Authors would like to thank the referee whose comments have improved the Paper, and in addition, Edo van Uitert, Wenting Wang, Ying Zu
and Rachel Mandelbaum for sharing their data. We thank Matthieu Schaller for sharing subhalo matching catalogues for the EAGLE simulation. In addition we thank Patrick Simon for providing useful comments on the draft. MR and
SH acknowledge support by the DFG cluster of excellence \lq{}Origin
and Structure of the Universe\rq{}
(\href{http://www.universe-cluster.de}{\texttt{www.universe-cluster.de}})
and would like to thank the Max Planck Institute
for Astrophysics and the Max Planck Computing and Data Facility for
computational resources. MR acknowledges additional support from the US DOE contract DE-AC02-06CH11357. BMBH acknowledges support from a Zwicky Prize fellowship. MV acknowledges support through an MIT RSC award, a Kavli Research Investment Fund, NASA ATP grant NNX17AG29G, and NSF grants AST-1814053 and AST-1814259. REA acknowledges support of the European Research Council through grant number ERC-StG/716151.

\section*{Data availability}

Some of the data products underlying this article are available at \url{www.tng-project.org/renneby20} and the rest are available upon request.



\bibliographystyle{mnras}
\bibliography{references}




\bsp	
\label{lastpage}
\end{document}